\documentclass[fleqn,usenatbib]{mnras}

\usepackage{newtxtext,newtxmath}

\usepackage[T1]{fontenc}

\DeclareRobustCommand{\VAN}[3]{#2}
\let\VANthebibliography\thebibliography
\def\thebibliography{\DeclareRobustCommand{\VAN}[3]{##3}\VANthebibliography}

\usepackage{graphicx}	
\usepackage{amsmath}	
 \usepackage{multirow}

\title[NLR warm-ionised outflows in NGC\;1068 and NGC\;4151]{Outflow densities and ionisation mechanisms in the NLRs of the prototypical Seyfert galaxies NGC\;1068 and NGC\;4151}

\author[L.R. Holden et al.]{
Luke R. Holden,$^{1}$\thanks{E-mail: lholden2@sheffield.ac.uk}
Clive N. Tadhunter,$^{1}$
\\
$^{1}$Department of Physics $\&$ Astronomy, University of Sheffield, S6 3TG Sheffield, UK. \\
}

\date{Accepted XXX. Received YYY; in original form ZZZ}

\pubyear{2015}

\begin{document}
\label{firstpage}
\pagerange{\pageref{firstpage}--\pageref{lastpage}}
\maketitle

\begin{abstract}
Despite being thought to play an important role in galaxy evolution, the true impact of outflows driven by active galactic nuclei (AGN) on their host galaxies is unclear. In part, this may be because electron densities of outflowing gas are often underestimated: recent studies that use alternative diagnostics have measured much higher densities than those from commonly used techniques, and consequently find modest outflow masses and kinetic powers. Furthermore, outflow ionisation mechanisms --- which are often used to probe acceleration mechanisms --- are also uncertain. To address these issues, we have analysed archival HST/STIS spectra of the inner regions (r\;\textless\;160\;pc) of the nearby prototypical Seyfert galaxies NGC\;1068 and NGC\;4151, which show evidence of warm-ionised outflows driven by the central AGN. We derive high electron densities ($10^{3.6}$\;\textless\;$n_e$\;\textless\;$10^{4.8}$\;cm$^{-3}$) using the transauroral [OII] and [SII] emission lines ratios for the first time with spatially-resolved observations. Moreover, we find evidence that the gas along the radio axis in NGC\;1068 has a significant AGN-photoionised matter-bounded component, and there is evidence for shock-ionisation and/or radiation-bounded AGN-photoionisation along the radio axis in NGC\;4151. We also note that the outflow extents are similar to those of the radio structures, consistent with acceleration by jet-induced shocks. Taken together, our investigation demonstrates the diversity of physical and ionisation conditions in the narrow line regions of Seyfert galaxies, and hence reinforces the need for robust diagnostics of outflowing gas densities and ionisation mechanisms.
\end{abstract}




\section{Introduction}
\label{section: intro}

Active galactic nuclei (AGN) can drive gas outflows through radiation-pressure driven winds from their accretion disks \citep{DiMatteo2005, Hopkins2010} and/or radio jets \citep{Axon1998, Wagner2011, Mukherjee2018}. These outflows, as well as the heating and ionising of near-nuclear gas, may constitute an important part of `AGN feedback', which now routinely plays a crucial role in theoretical models of galaxy evolution. AGN-feedback is required to explain observed galaxy properties (e.g. \citealt{DiMatteo2005, Somerville2008, Schaye2015, Dubois2016, Dave2019} and empirical scaling relations between super massive black holes and host galaxy properties (e.g. \citealt{Magorrian1998, Silk1998, Fabian1999, Gebhardt2000, Ferrarese2000}). Models often require that the kinetic power ($\dot{\mathrm{E}}_\mathrm{kin}$) of the outflowing gas is above a certain fraction of the AGN bolometric luminosity (L$_\mathrm{bol}$): this is characterised by a ratio known as the `coupling factor' ($\epsilon_f$=$\dot{\mathrm{E}}_\mathrm{kin}$/L$_\mathrm{bol}$), and is typically required to be in the range 0.5\;\textless$\epsilon_\mathrm{f}$\;\textless10\;per\;cent \citep{DiMatteo2005, Springel2005, Hopkins2010}.

Observational studies commonly attempt to quantify the impact of outflows on their host galaxies by comparing measured coupling efficiencies to those required by models (e.g. \citealt{Liu2013, Cicone2014, Harrison2014, Rose2018, Riffel2021b}). However, many key outflow properties are highly uncertain, leading to a wide range of observationally-derived coupling efficiencies \citep{Harrison2018}. For the warm ionised outflow phase (i.e. traced by [OIII] and H$\mathrm{\beta}$; 10,000\;\textless\;$T_e$\;\textless\;25,000\;K), the largest source of uncertainty is likely to be the electron density of the outflowing gas, which is often estimated or assumed to be in the range $n_e \sim$100--1000\;cm$^{-3}$ (e.g. \citealt{Kraemer2000II, Nesvadba2006, Fiore2017}). This is because the commonly used `traditional' density diagnostics --- the [SII](6717/6731) and [OII](3726/3729) emission-line doublet ratios --- are only sensitive up to $n_\mathrm{e}\sim10^{3.5}$\;cm$^{-3}$, and are often blended in the case of complex outflow kinematics. However, in recent years, alternative density diagnostics have been developed and used, such as detailed photoionisation modelling that makes use of a wide range of emission lines \citep{Collins2009, Crenshaw2015, Revalski2021, Revalski2022}, and a technique involving ionisation parameter measurements with infrared estimates of outflow radii \citep{Baron2019b}. Such methods have measured higher electron densities for the warm ionised phase than commonly-used traditional techniques, up to $n_e\sim10^{5.5}$\;cm$^{-3}$. Studies which make use of the higher critical density `transauroral' (`TR'; \citealt{Boyce1933}) [OII](3726 + 3729)/(7319 + 7331) and [SII](4068 + 4076)/(6717 + 6731) diagnostic ratios have similarly found densities in the range of \mbox{10$^3$\;\textless\;$n_\mathrm{e}$\;\textless\;$10^{5.5}$\;cm$^{-3}$} \citep{Holt2011, Rose2018, Santoro2018, Spence2018, RamosAlmeida2019, Santoro2020, Davies2020, Speranza2022, Holden2022}. Considering that the derived outflow kinetic power is inversely proportional to the electron density, if electron densities are truly orders of magnitude higher than are commonly assumed or estimated, the resulting kinetic powers and coupling factors for the warm ionised phase will be orders of magnitude lower. This could significantly change our understanding of the importance of AGN feedback in galaxy evolution.

Moreover, where possible, it is important to use spatially-resolved observations when deriving electron densities, since global electron densities may significantly underestimate the values at small radial distances from the nucleus, where the outflows are the most extreme (\citealt{Revalski2022}; but see also \citealt{Kakkad2018}). Thus, detailed spatially-resolved observations are needed to robustly assess electron densities in different types of AGN, as well as to compare and verify different density diagnostics.

Investigations into the impact of outflows on their host galaxies are further complicated by the fact that the dominant acceleration and ionisation mechanisms are unclear: while it is thought that outflows may be accelerated by radiation pressure from the AGN (either `in situ': e.g. \citealt{Crenshaw2015, Fischer2017, Revalski2018, Meena2023}, or from the nucleus: e.g. \citealt{Hopkins2010, Meena2021}), a study of a large sample of local AGN found a link between intermediate radio power AGN (L$_\mathrm{1.4\;GHz}=10^{23-25}$\;W\;Hz$^{-1}$) and outflow kinematics \citep{Mullaney2013}, suggesting that feedback from jets is also important in AGN that are classified as radio-quiet. Indeed, hydrodynamic simulations have shown that jets interacting with the ISM on kpc-scales can explain observed gas kinematics in some objects (e.g. \citealt{Mukherjee2018, Audibert2023}), and may have both a positive and negative impact on local star formation rates \citep{Mandal2021}. Therefore, determining dominant acceleration mechanisms is crucial for facilitating proper comparisons between observations and predictions from theoretical modelling, which are needed to interpret the role of outflows in AGN feedback.

The ionisation and excitation mechanisms of the outflowing gas may provide clues as to the acceleration mechanism(s) present. For example, shock-ionised gas must have passed through (and been accelerated by) a shock. However, AGN-photoionised gas may have been previously accelerated by another mechanism, and reionised by photons from the AGN after cooling \citep{Holden2022}. Hence, the true nature of the relationship between outflow acceleration and ionisation mechanisms is complex, and requires further careful analysis.

Regardless of how outflows are accelerated, understanding the dominant ionisation mechanisms impacts our ability to extract key diagnostic information for the warm outflowing gas. Specifically, the techniques presented by \citet{Holt2011} and \citet{Revalski2021} (see also \citealt{Collins2009}) both rely on photoionisation models, and the transauroral lines (in the case of the \citealt{Holt2011} method) cannot be emitted by a matter-bounded component. If, in reality, a gas outflow is shock-ionised or has a large contribution from a matter-bounded component, this may have a significant impact on the validity of these methods. Thus, it is important to investigate the ionisation mechanisms present in active galaxies for which these techniques have been applied in the past, as well the potential impact of matter-bounded components or shock ionisation on derived densities.

In order to address these issues, we are undertaking detailed spatially-resolved studies of nearby AGN that show clear evidence of outflows on pc to kpc scales. In \citet{Holden2022}, we presented a detailed study of the central regions of the nearby Seyfert 2 IC\;5063 using \mbox{Very Large Telescope (VLT) / Xshooter} ultraviolet (UV), optical and near-infrared (NIR) spectroscopy: we found electron densities just above the critical density of the traditional [SII] ratio, and evidence for a post-shock cooling sequence and reionisation via AGN photoionisation. 

There is a clear need to determine whether the conditions found in the narrow line region (NLR) of IC\;5063 are similar in other Seyfert galaxies, specifically to further investigate the true outflow gas density, kinetic powers, and ionisation mechanisms present on different spatial scales. Therefore, here we analyse archival Hubble Space Telescope (HST) / Space Telescope Imaging Spectrograph (STIS) spectra of the inner NLRs ($r$\;\textless\;160\;kpc) of the prototypical Seyfert galaxies NGC\;1068 and NGC\;4151, and apply and expand upon many of the techniques presented in \citet{Holden2022}. We take the distances to NGC\;1068 and NGC\;4151 to be $D = 13.0$\;Mpc \citep{Revalski2021} and $D = 15.8$\;Mpc \citep{Yuan2020}, respectively, which correspond to spatial scales of 0.067\;kpc/arcseconds for NGC\;1068 and 0.078\;kpc/arcseconds for NGC\;4151.

The structure of the paper is as follows: in Section \ref{section: introduction_seyferts}, we introduce the prototypical Seyfert galaxies NGC\;1068 and NGC\;4151; in Section \ref{section: observations_reduction}, we detail the archival HST/STIS observations and our data reduction and handling processes; in Section \ref{section: analysis}, we present our analysis of the STIS data; in Section \ref{section: discussion}, we discuss the implications of our findings, and in Section \ref{section: conclusions} we give our conclusions.

\section{Two prototypical Seyferts: NGC\;1068 \& NGC\;4151}
\label{section: introduction_seyferts}

NGC\;1068 and NGC\;4151 appeared in Carl Seyfert's original paper that established the Seyfert class \citep{Seyfert1943}, and are respectively the prototypical Seyfert 2 (Sey2) and Seyfert 1 (Sey1) galaxies. In consequence, they are perhaps the most well-studied AGN of their respective types. Their close proximity to Earth, and the previous, extensive multi-wavelength studies of their properties, make them ideal objects for our project: the outflows in their central regions can be spatially resolved, and we can compare our results to those obtained using other methods. Principally, this allows us to assess the validity of the different density diagnostic techniques, as well as investigate the ionisation of the gas.

\subsection{NGC\;1068}
\label{section: ngc1068}

NGC\;1068 is one of the closest and brightest (in terms of observed flux) Seyfert 2 galaxies, allowing detailed spatially-resolved observations, and thus making it the target for extensive studies that cover a range of spatial scales in the optical (e.g. \citealt{Cecil1990, Evans1991, Axon1998, Crenshaw2000a, Kraemer2000III, Das2006}), NIR (e.g. \citealt{Raban2009, MUllerSanchez2009, May2017}) and radio (e.g. \citealt{Wilson1983, Gallimore1996, GarciaBurillo2014, GarciaBurillo2019}). NGC\;1068 has a radio luminosity of L$_\mathrm{1.4\;GHz}=2.3\times10^{23}$\;W\;Hz$^{-1}$ \citep{Ulvestad1984}, placing it in the upper end of the radio luminosity range for Seyfert galaxies, and its high bolometric luminosity (\mbox{0.4\;\textless\;$L_\mathrm{bol}$\textless\;4.7$\times10^{38}$\;W}: \citealt{Woo2002, AlonsoHerrero2011, LopezRodriguez2018, Gravity2020}) is close to the lower boundary of the luminosity range for quasars (L$_\mathrm{bol}$\;\textgreater\;10$^{38}$\;W). The galaxy also has an important historical role, as it was the first object used to verify the orientation-based unified scheme for AGN \citep{Antonucci1985}.

The NLR of NGC\;1068 presents as an `hourglass'-shaped bicone \citep{Riffel2014, Barbosa2014, May2017} with an opening angle of $\theta \sim 40^\circ$ along PA$=30\pm2^\circ$ at an inclination of $i=5^\circ$, placing the bicone axis close to the plane of the sky and inclined $\sim45^\circ$ out of the galaxy's disk (\citealt{Das2006}; but see also \citealt{Crenshaw2000a}). Outflows of warm-ionised gas with velocities up to $\sim$1500\;km\;s$^{-1}$ have been detected in the bicone \citep{Crenshaw2000_N1068, Das2006}. In the NE cone, the radio axis is closely aligned with the bicone axis --- interpreted as a radio jet propagating within the hollowed-out cone --- with a radio lobe that extends just beyond the maximum extent of the cone (\citealt{WilsonUlvestad1987}; shown in Figure \ref{fig: seyferts_ifu_slit}). Lower velocity cold molecular CO(3-2) outflows have been detected at this position, indicating that the lobe may represent the termination of the AGN-driven outflows \citet{GarciaBurillo2014}.

The outflows in the NLR of NGC\;1068 have been argued to be radiatively accelerated by some authors \citep{Kraemer2000III, Das2006, Revalski2021, Meena2023}, while others have proposed they are driven by jet-induced shocks \citep{Capetti1997, Axon1998}. \citet{May2017} propose a scenario in which the radio jet impacts molecular clouds on small radial scales near the central AGN, accelerating high-velocity `bullets' of gas that propagate within the bicone but constitute only a small fraction of the total outflowing mass.

\subsection{NGC\;4151}
\label{section: ngc4151}

NGC\;4151 is the prototypical Seyfert 1 (Sey1) galaxy\footnote{NGC\;4151 was later classified as an intermediate `Seyfert 1.5' \citep{OsterbrockKoski1976, Robinson1994}.} and is also one of the closest and brightest (in terms of observed flux) of its class, leading to its NLR outflows being the target of extensive studies of the coronal (e.g. \citealt{Storchi-Bergmann2009, Storchi-Bergmann2010}), warm ionised (e.g. \citealt{Winge1997, Hutchings1999, Crenshaw2000_N4151, Das2005, May2020}) and warm molecular (H$_\mathrm{2}$; $T\sim$2000\;K, e.g. \citealt{May2020}) gas phases, which have distinct flux distributions \citep{Storchi-Bergmann2009}. Similar to NGC\;1068, the bicone-shaped NLR also has an hourglass morphology \citep{May2020}, with PA=22$^\circ$ at an inclination of $i=21^\circ$ (\citealt{Pedlar1992}; 36$^\circ$ to the galactic disk) and an opening angle of 33$^\circ$ \citep{Das2005}. However, the bolometric luminosity of the AGN in NGC\;4151 ($L_\mathrm{bol}=1.4\times10^{37}$\;W) is approximately an order of magnitude below that of NGC\;1068.

The radio source (of luminosity L$_\mathrm{1.4\;GHZ}$=1.6$\times10^{22}$\;W\;Hz$^{-1}$; \citealt{Ulvestad1984}) consists of a double sided jet ($\mathrm{PA}\sim77^\circ$) originating from the nucleus. High-resolution radio imaging \citep{Carral1990, Pedlar1993, Williams2017} shows several radio knots along this structure within the central few arcseconds, whereas lower-resolution radio observations \citep{Johnston1982, Pedlar1993} reveal a larger-scale lower surface brightness structure with a radio lobe in the NE cone extending to 6.3\;arcseconds from the nucleus along the radio axis. It has been argued that the radio jet has little connection to the NLR outflow kinematics in NGC\;4151 \citep{Hutchings1999, Crenshaw2000_N4151, Das2005}. However, enhanced line fluxes from the warm ionised gas, high electron temperatures (T$_\mathrm{e}$\;\textgreater16,000\;K) and high [FeII]/[PII] ratios have been spatially associated with the radio structure \citep{Mundell2003, Storchi-Bergmann2009, Storchi-Bergmann2010}, indicating that jet-ISM interactions may still drive shocks into the gas at certain locations within the bicone (see also \citealt{Wang2011b, Wang2011, Williams2017}).

\citet{May2020} propose a similar model as they proposed for NGC\;1068 \citep{May2017} --- albeit on smaller spatial scales with less extreme kinematics --- to explain the NLR and outflow structure in NGC\;4151: the radio jet impacts a molecular cloud near the nucleus (potentially due to misalignment between the jet and torus/disk: \citealt{Storchi-Bergmann2010, May2020}), driving fragmented, shock-accelerated gas into the cones and contributing to the NLR morphology.  

\subsection{Previous photoionisation modelling of NGC\;1068 and NGC\;4151}

\citet{Crenshaw2015} and \citet{Revalski2021, Revalski2022} performed detailed, multi-ionisation component photoionisation modelling of the warm ionised outflows in NGC\;1068 and NGC\;4151, finding densities in the range \mbox{$10^{3.0}$\;cm$^{-3}$\;\textless\;$n_e$\;\textless\;$10^{7.2}$\;cm$^{-3}$} for the NLR gas in both objects, and coupling efficiencies above the lower limit required by galaxy evolution models (0.5\;per cent: \citealt{Hopkins2010}) in the case of NGC\;1068. In order to further investigate the electron densities of the outflowing gas in the NLR of these two important objects, and to attempt to clarify the uncertainties regarding the acceleration and ionisation mechanisms of the gas, we require high spatial resolution, wide wavelength-coverage long-slit spectroscopy with the slit aligned along the radio axes (which is approximately along the bicone axes).

\section{Observations and data reduction}
\label{section: observations_reduction}

\subsection{Archival HST/STIS observations}
\label{section: stis_spectra}
\begin{figure*}
    \centering
    \includegraphics[width=1\linewidth]{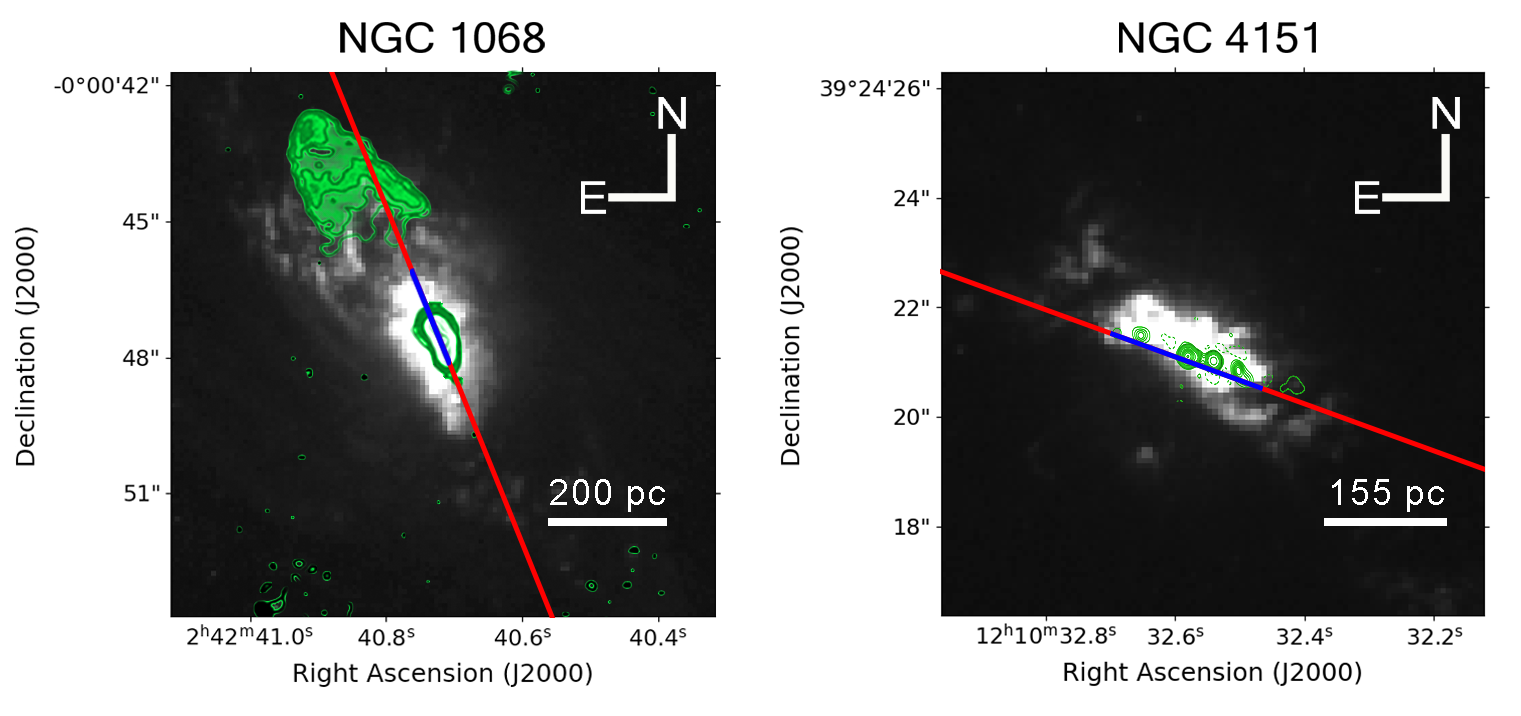}
    \caption{The STIS slits of our archival observations (red) shown plotted over archival HST/WFPC2 [OIII] emission-line images of the inner regions of NGC\;1068 and NGC\;4151, taken with the F502N filter (NGC\;1068: GTO:5754, PI Ford; NGC\;4151: GTO:5124; PI Ford). The extents of our apertures (Section \ref{section: apertures}) along the slits are shown in blue. \textbf{Left:} The STIS slit shown over the [OIII] emission-line image of the near-nuclear regions of NGC\;1068. VLA 22\;GHz contours from \citet{Gallimore1996} are presented in green, showing the radio structure near the core and an extended lobe to the NE. \textbf{Right:} the STIS slit shown over the [OIII] emission-line image of the near-nuclear regions of NGC\;4151; the green contours are from high-resolution eMERLIN 1.5\;GHz imaging presented by \citet{Williams2017}, and show a string of radio knots near the nucleus. We note that, while the narrow-band images are not continuum-subtracted, the brighter parts of the NLR emission are dominated by [OIII] emission in the filter bandpass, and so the images provide a good representation of the main NLR structures.}
    \label{fig: seyferts_ifu_slit}
\end{figure*}

To achieve our science goals, suitable archival HST/STIS long-slit spectra were downloaded from the Hubble Legacy Archive (\url{https://hla.stsci.edu/hlaview.html}). We required data taken using both the G430L and G750L gratings in order to ensure sufficient wavelength coverage, namely that the spectra contained the blue [SII]$\lambda\lambda4068,4076$ and red [OII]$\lambda\lambda7319,7331$ transauroral doublets. Both gratings have a spatial pixel scale of 0.051\;arcseconds per pixel, and the dispersions of the two gratings are 2.72\;\AA/pixel (G430L; 2900--5700\;{\AA}) and 4.92\;\AA/pixel (G750L; 5240--10270\;{\AA}). We also required that these data were taken along (or close to) the PA of the radio/bicone structures to ensure we are tracing the the gas that is impacted most by the jet. The data for NGC\;1068 were taken as part of the Cycle 7 HST Proposal GTO:7573 (PI Kraemer), with a 52$\times$0.1\;arcsecond slit along PA=202$^\circ$, centred on a bright emission-line knot close (\textless0.4$^{\prime\prime}$) to the nucleus (see \citealt{Crenshaw2000b} and \citealt{Kraemer2000II}). Data for NGC\;4151 were taken with a 52$\times$0.1\;arcsecond slit along PA=70$^\circ$, offset to the south by 0.1\;arcsecond to reduce contamination from the bright Sey1 nucleus, and were taken in Cycle 7 as part of HST Proposal GTO:7569 (PI Hutchings) --- a full description of the NGC\;4151 observations is given by \citet{Nelson2000}. We show the positions of the STIS slits over the central regions of the two Seyferts in Figure \ref{fig: seyferts_ifu_slit}.

\subsection{Reduction and handling of STIS data}
\label{section: data_handling}
\subsubsection{Data reduction}
\label{section: data_reduction}

The first step in the data reduction was performed with the standard \mbox{\textsc{CALSTIS}} pipeline. For the NGC\;1068, only a single exposure for each grating was available, while for NGC\;4151 we took the average of two exposures for each grating using \textsc{Python} scripts which made use of the \textsc{Numpy} \citep{Harris2020} and \mbox{\textsc{AstroPy}} \citep{AstropyCollaboration2013, AstropyCollaboration2018} modules. In order to ensure that the individual exposures for each grating were aligned, we first extracted spatial slices along the slit direction in a line-free region of the continuum covering the wavelength range 5480--5600\;{\AA} for the G430L grating and 6795---6890\;{\AA} for the G750L grating. The centroids of the spatial peaks --- determined with Gaussian profile fits --- were consistent within better than 0.4\;pixels, confirming that each exposure was taken with the same telescope pointing within 0.02\;arcseconds. We also checked that the spectra taken with the G430L and G750L gratings for the each object were aligned, using the same method of Gaussian fits to the spatial flux profiles. Again, the spatial positions of the peak flux between gratings were consistent to within better than 0.4\;pixels, indicating that the observations with different gratings were closely spatially aligned.

Residual hot pixels and cosmic rays were removed from the spectra using the \textsc{CLEAN} command from the \textsc{STARLINK FIGARO} software package \citep{Currie2014}. We then corrected for extinction due to dust in the Milky Way using the Galactic extinction maps presented by \citet{Schlegel1998} and recalibrated by \citet{Schlafly2011}. Using the NASA/IPAC Infrared Science Archive reddening lookup tool (\url{https://irsa.ipac.caltech.edu/applications/DUST/}) with these maps, we find that there are mean colour excesses in the directions of NGC\;1068 and NGC\;4151 of \mbox{E(B-V)$_\mathrm{mean}=0.0289\pm0.0004$} and \mbox{E(B-V)$_\mathrm{mean}=0.0237\pm0.0011$} respectively. The $R_v=3.1$ extinction law presented by \citet{Cardelli1989} (hereafter CCM89) was then used to correct for Galactic extinction.

\subsubsection{Aperture selection and extraction}
\label{section: apertures}

\begin{figure*}
	\includegraphics[width=\linewidth]{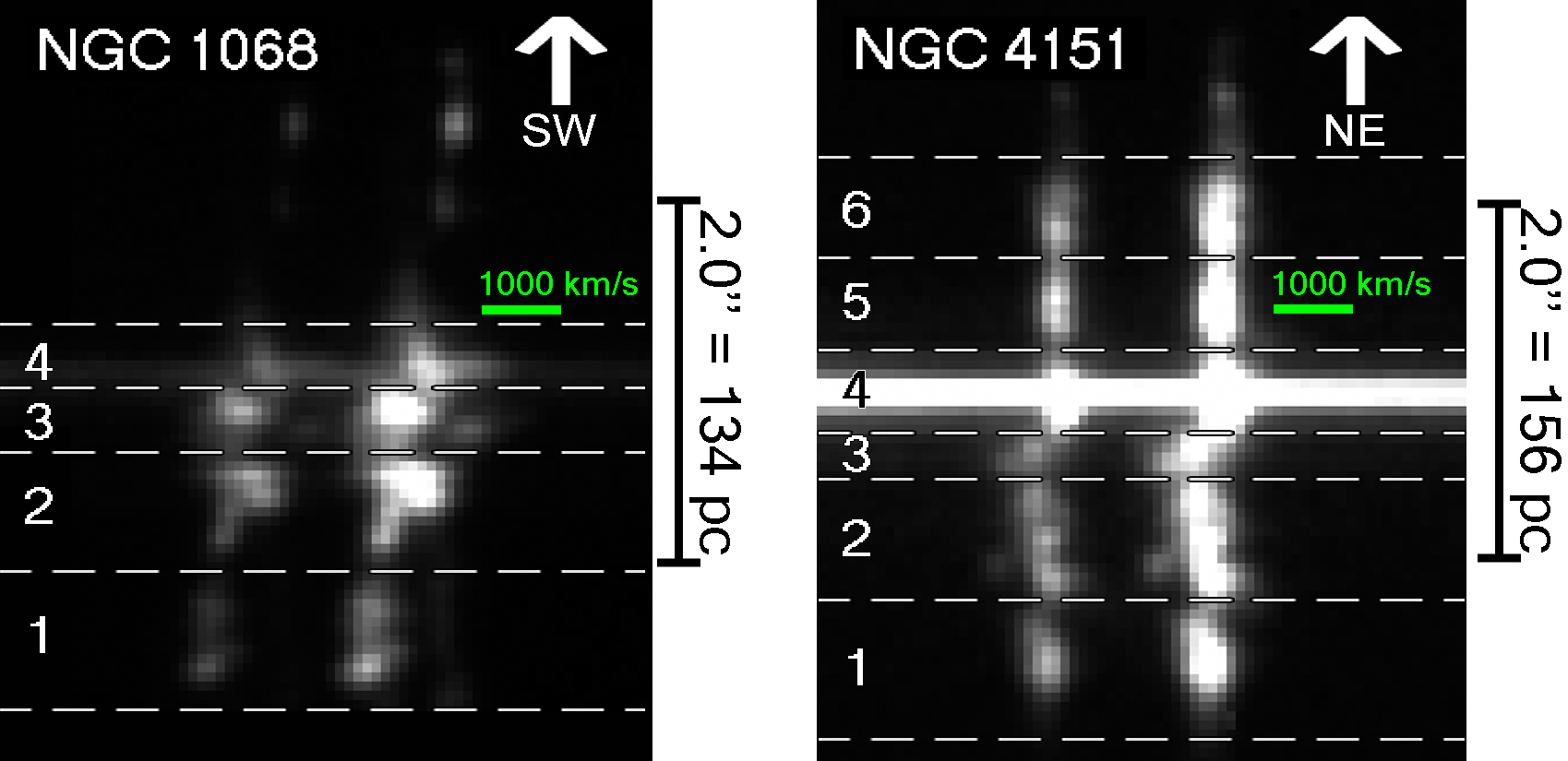}
	\caption{Selected apertures for NGC\;1068 (left) and NGC\;4151 (right), positioned over the [OIII]$\lambda\lambda$4959,5007 doublet in the two-dimensional STIS G430L spectra. The spectral direction is horizontal (left = bluewards; right = redwards) and the vertical direction is spatial along the slit (with the direction shown by the labelled arrows); velocity scale bars are shown in green, and the spatial extents in arcseconds and parsecs are shown to the right of each spectrum. The apertures are shown as regions bounded by dashed lines, and are labelled on the left of each image --- they were chosen to contain enough signal for the measurement of faint lines in distinct kinematic regions within the central few hundred parsecs of each galaxy.}
	\label{fig: apertures}
\end{figure*}

\begin{figure*}
    \centering
    \includegraphics[width=\linewidth]{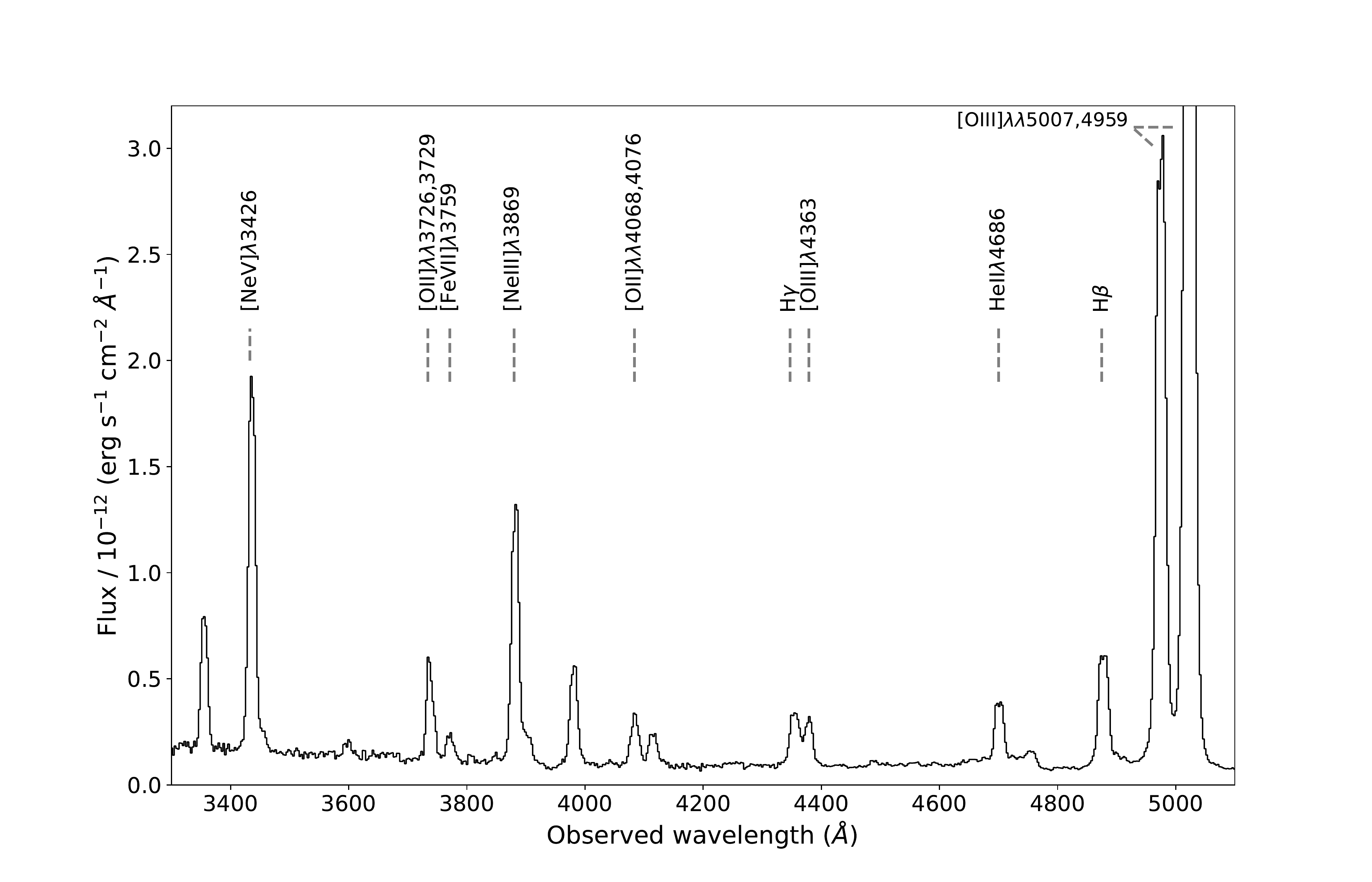}
    \caption{The G430L grating spectrum for Aperture 2 of NGC\;1068 (Figure \ref{fig: apertures}). Key emission lines that are used in our analysis are labelled with dotted lines. Note that, for presentation reasons, the limit on the flux axis has been chosen so that fainter lines can be clearly seen; as a result, the peak of the [OIII]$\lambda$5007 line is not visible.}
    \label{fig: ngc1068_g430l_ap2}
\end{figure*}

The STIS long-slit spectra of NGC\;1068 and NGC\;4151 show disturbed kinematics (indicating outflows) and several bright emission-line knots in the central few hundred parsecs, as noted by previous studies \citep{Crenshaw2000a, Kraemer2000II, Das2005, Das2006, Meena2023}. We extracted several apertures (integrated groupings of pixel rows) from the two-dimensional G430L and G750L spectra, with each aperture forming an integrated one-dimensional spectrum that corresponds to a certain spatial position along the slit. We selected the apertures to cover the locations of the bright emission knots seen in our two-dimensional spectra (Figure \ref{fig: apertures}). The widths of the apertures (6--15\;pixels; 0.3--0.8\;arcseconds) were set to contain sufficient signal in the fainter emission lines that are used for diagnostics in our analysis, namely the fainter transauroral [OII]$\lambda\lambda$7319,7331 and [SII]$\lambda\lambda$4068,4076 doublets. We extracted the same apertures from the G430L and G750L spectra for each object, as we previously determined that the spectra were closely spatially aligned \mbox{(Section \ref{section: data_reduction})}. Flux errors were determined by adding the flux errors from individual pixel rows (which constitute a given aperture) in quadrature. As an example, we present part of the spectrum of Aperture 2 for NGC\;1068 in Figure \ref{fig: ngc1068_g430l_ap2}. The chosen apertures extended out to a maximum radial distance of 139\;pc for NGC\;1068, and 151\;pc in the case of NGC\;4151.

Aperture 3 for NGC\;1068 was placed over a bright emission knot that corresponds to a previously detected radio source at the likely position of the galaxy's nucleus (see discussion in \citealt{Kraemer2000II}), while Aperture 4 for NGC\;4151 corresponds to the location along the slit that is closest to the nucleus. We note that the spectra for NGC\;4151 do not directly cover the nucleus, due to the 0.1\;arcsecond slit offset to the south to avoid nuclear contamination. Unfortunately, the south-west part of the slit for NGC\;1068 (seen above Aperture 4 in Figure \ref{fig: apertures}) did not contain enough signal for the measurement of the faint [OII]$\lambda\lambda7319,7331$ transauroral doublet, even when integrated as a single aperture. Therefore, we omit this region from our analysis. 

Following aperture extraction, we ensured that the flux calibration was consistent between the two gratings for each aperture by overplotting the spectra in the region where the wavelength ranges of the gratings overlap (5275--5705\;{\AA}). We found that all apertures for NGC\;1068 are closely matched in flux. However, for apertures 2 and 4 of NGC\;4151, the flux in the overlap region was \textgreater8\;per cent higher in the G430L grating than the G750L grating, potentially due to internal reflections within the instrument caused by the bright Type 1 nucleus (see \citealt{Nelson2000}). Therefore, we do not use these apertures in further analysis. 

\subsubsection{The contribution of stellar continua to the spectra}
\label{section: stellar_continua}

We did not model and subtract the underlying stellar continuum in detail using a template-fitting approach (as was done for similar analyses of other objects by \citealt{Rose2018} and \citealt{Holden2022}) for various reasons. First, our archival STIS G430L and G750L spectra did not have sufficient spectral resolution to clearly resolve absorption features that could be used to verify the robustness of the continuum fits. Second, there may be substantial contamination by direct and scattered AGN continuum \citep{Antonucci1985} and nebular continuum \citep{Tadhunter2016} that precludes accurate stellar continuum modelling. Finally, the emission lines in our spectra have relatively high equivalent widths, which fill in various stellar absorption features.

In order to verify whether stellar continuum modelling was needed in this study, we measured the equivalent widths (EWs) for the H$\mathrm{\beta}$ recombination line. We find \mbox{36\;\textless\;EW\;\textless\;148\;{\AA}} in our NGC\;1068 apertures and \mbox{30\;\textless\;EW\;\textless\;151\;{\AA}} in the NGC\;4151 apertures. The lowest emission-line equivalent width we measure (EW$=30$\;{\AA} for Aperture 3 in NGC\;4151) is a factor of three higher than that of the H$\mathrm{\beta}$ absorption feature as modelled for a $\sim$400\;Myr old stellar population (which gives the highest EWs in modelling by \citealt{GonzalezDelgado1999}). Thus, underlying stellar absorption features may affect our measured H$\mathrm{\beta}$ luminosities by a maximum factor of 1.3 (for a stellar EW$=10$\;{\AA}). However, this is very much an upper limit since we do not detect a Balmer break in the continuum in any of our apertures, as would be expected for intermediate age stellar populations that have strong Balmer absorption lines.

\subsubsection{Fits to key emission lines}
\label{section: oiii_models}

The NLR kinematics in NGC\;1068 and NGC\;4151 are complex, and have been previously modelled in detail as biconical outflows based on higher resolution STIS spectra than those used here (\citealt{Das2006} and \citealt{Das2005} respectively; but see also \citealt{Crenshaw2000_N1068} and \citealt{Crenshaw2000_N4151}). In those studies, the [OIII]$\lambda\lambda$4959,5007 doublet line profiles were fit with multiple Gaussians for each pixel row of the 2D spectra. Here, we perform a similar procedure for our extracted apertures by simultaneously fitting a 1st or 2nd order polynomial to the continuum surrounding the [OIII]$\lambda\lambda$4959,5007 doublet, and one or two Gaussian profiles to each of the lines in the doublet itself. We set the wavelength separation of the lines in the doublet, as well as the intensity ratio of the lines (1:2.99), to those defined by atomic physics \citep{Osterbrock2006}. Furthermore, we constrained the widths of a given Gaussian component to be the same for each line in the doublet. We present the model parameters for each aperture in Table \ref{tab: oiii_models}.

Once we had established [OIII]$\lambda\lambda$4959,5007 doublet fits in each aperture, we calculated the difference between the mean wavelength of each Gaussian component and the rest [OIII] wavelength in the reference frame of the galaxy, using redshifts\footnote{21\;cm redshifts from the NASA/IPAC Extragalactic Database (\url{https://ned.ipac.caltech.edu/}).} of $z=0.00381$ for NGC\;1068 and $z=0.003262$ for NGC\;4151. We also determined the intrinsic width of each component by subtracting the instrumental width of the STIS G430L grating in quadrature from the measured widths. According to the STIS manual, for a slit of width 0.1\;arcseconds, the instrumental broadening in the spectral direction is in the range \mbox{2\;\textless\;FWHM\;\textless\;3} pixels, corresponding to \mbox{5.5\;\textless\;FWHM\;\textless\;8.2\;{\AA}} for the G430L grating and \mbox{9.8\;\textless\;FWHM\;\textless\;14.8\;{\AA}} for the G750L grating. By fitting single Gaussians to the [OIII]$\lambda\lambda$4959,5007 emission-line doublet at a radial distance of 4\;arcseconds from the nucleus of NGC\;4151 in the G430L spectra (where the lowest line widths are measured), we measure a line width of FWHM$_\mathrm{inst}=6.0\pm0.4$\;{\AA}; similarly, measuring the [SII]$\lambda$9531 line in the G750L spectra with this method resulted in a line width of FWHM$_\mathrm{inst}=12.3\pm2.4$\;{\AA}. Thus, we adopt instrumental widths of FWHM$_\mathrm{inst}=6.0$\;{\AA} (360\;km\;s$^{-1}$ at 5007\;{\AA}) and FWHM$_\mathrm{inst}=12.3$\;{\AA} (560\;km\;s$^{-1}$ at 6575\;{\AA}) for the G430L and G750L gratings, respectively.

In subsequent analysis, we only consider \textit{total} line fluxes --- including all Gaussians components used --- rather than fluxes from individual components (i.e. potentially representing outflowing and quiescent gas). This was done because of the low spectral resolutions of the G430L and G750L gratings, which made it challenging to separate different kinematic components in cases where lines are heavily blended. Nonetheless, in order to improve the accuracy of the fits to the weaker emission lines and blends in the spectra, we used the kinematics (velocity shifts and widths) derived from fits to the [OIII] doublet in each aperture to constrain the fits to the other key diagnostic lines used in our analysis, such as H$\mathrm{\beta}$, H$\mathrm{\gamma}$, [OIII]$\lambda$4363, [OII]$\lambda$3726,3729, [OII]$\lambda\lambda$7319,7331, [SII]$\lambda\lambda$4068,4076, [SII]$\lambda\lambda$6717,6731, [ArIV]$\lambda\lambda$4711,4740 and HeII$\lambda$4686. We found that this procedure produced acceptable fits to these lines, including the transauroral [SII]$\lambda\lambda$4068,4076 and [OII]$\lambda\lambda$7319,7331 doublets. However, for closely spaced doublets such as [OII]$\lambda$3726,3729, the low spectral resolution meant that we did not resolve individual lines, and so we modelled the \textit{total} doublet profile as a single emission line during the fitting process.

\begin{table*}
\def\arraystretch{1.5}
\centering
\begin{tabular}{cccccccccc}
\hline
Aperture & \parbox{1.5cm}{Distance \\ (arcseconds)} &  \parbox{1.5cm}{Distance \\ (pc)}  & \parbox{1cm}{$v_\mathrm{c, a}$ \\ (km\;s$^{-1}$)} & \parbox{1cm}{FWHM$_\mathrm{c, a}$ \\ (km\;s$^{-1}$)} & \parbox{1cm}{$v_\mathrm{c, b}$ \\ (km\;s$^{-1}$)} & \parbox{1cm}{FWHM$_\mathrm{c, b}$ \\ (km\;s$^{-1}$)} & log$_{10}(n_e$[cm$^{-3}$]) & E(B-V)$_\mathrm{TR}$ & $T_e$ (K)    \\ \hline
\multicolumn{10}{c}{NGC 1068} \\ \hline
1 & $-1.45$ & $-97$ & $-828\pm$4 & 572$\pm$25 & 295$\pm$40 & 1078$\pm$96 & 4.06$^{+0.05}_{-0.06}$ & 0.16$^{+0.04}_{-0.05}$ & 14300$^{+2100}_{-1300}$ \\
2 & $-0.74$ & $-50$ & $-184\pm$7 & 1017$\pm$31 & --- & --- & 4.65$^{+0.05}_{-0.04}$ & 0.05$^{+0.04}_{-0.05}$ & 14400$^{+1500}_{-1100}$ \\
3 & $-0.23$ & $-15$ & $-306\pm$3 & 662$\pm$26 & $-5\pm$20 & 1770$\pm$43 & 4.74$^{+0.05}_{-0.05}$ & 0.16$^{+0.05}_{-0.04}$ & 16100$^{+1400}_{-1000}$ \\
4 & $0.10$ & $7$ & $95\pm$3 & 367$\pm$26 & 235$\pm$8 & 1684$\pm$34 & 4.45$^{+0.09}_{-0.09}$ & 0.17$^{+0.08}_{-0.08}$ & 16000$^{+1600}_{-1100}$ \\ \hline
\multicolumn{10}{c}{NGC 4151} \\ \hline
1 & $-1.58$ & $-123$ & $-172\pm$2 & 420$\pm$25 & $-263\pm$29 & 1261$\pm$102 & 3.68$^{+0.08}_{-0.10}$ & 0.11$^{+0.05}_{-0.07}$ & 16300$^{+3400}_{-1800}$ \\
3 & $-0.38$ & $-30$ & $-392\pm$6 & 0$\pm28$\;$^a$ & $-356\pm$3 & 1065$\pm$26 & 4.04$^{+0.07}_{-0.15}$ & 0.13$^{+0.05}_{-0.06}$ & 21000$^{+3300}_{-2100}$ \\
5 & 0.48 & $37$ & $34\pm$1 & 234$\pm$24 & 121$\pm$11 & 1013$\pm$44 & 3.94$^{+0.10}_{-0.10}$ & 0.15$^{+0.08}_{-0.08}$ & 17300$^{+5200}_{-2400}$ \\
6 & 1.02 & $80$ & $40\pm$1 & 307$\pm$25 & 126$\pm$8 & 768$\pm$35  & 3.75$^{+0.8}_{-0.10}$  & 0.23$^{+0.07}_{-0.07}$ & 15200$^{+3400}_{-1800}$ \\ \hline
\end{tabular} \\
$^a$ The measured width of the \textit{component} is consistent with the instrumental width, and hence unresolved. \\
\caption{[OIII] model parameters (galaxy rest-frame component velocity shift: $v_\mathrm{c}$; instrumentally-corrected component velocity width: FWHM$_\mathrm{c}$), distances from the nucleus (in arcseconds and pc), electron densities, reddening values and electron temperatures for each of our apertures for NGC\;1068 and NGC\;4151. In apertures where there are multiple Gaussian components for the [OIII] models, we label the kinematic parameters for the two components with the subscripts `a' and `b'. The densities and reddenings were determined simultaneously using the transauroral line technique (Section \ref{section: tr_diagnostics}; Figure \ref{fig: tr_ddd}), and the temperatures were determined using the [OIII](5007+4959)/4363 ratio (Section \ref{section: electron_temperatures}).}
\label{tab: oiii_models}
\end{table*}

\section{Analysis of the STIS spectra}
\label{section: analysis}

\subsection{Transauroral line diagnostics}
\label{section: tr_diagnostics}

In order to provide estimates of the electron densities of the warm ionised gas in NGC\;1068 and NGC\;4151, we make use of a technique first described by \citet{Holt2011} which requires measurement of the transauroral [SII] and [OII] ratios:
\begin{align*}
TR([OII]) = F(3726 + 3729) / F(7319 + 7331),
\end{align*}
\begin{align*}
TR([SII]) = F(4068 + 4076) / F(6717 + 6731).
\end{align*} 
In this technique, measured TR([OII]) and TR([SII]) ratios are compared to those expected from photoionisation modelling in order to simultaneously derive electron densities and reddenings. This has several important advantages as a density diagnostic over commonly-used, traditional methods. First, these lines have higher critical densities (Appendix \ref{section: appendix_critical_densities}), meaning that the TR ratios are sensitive to higher electron densities ($n_\mathrm{e}\sim10^{5.5}$\;cm$^{-3}$) than the traditional [OII](3726/3729) and [SII](6717/6731) density diagnostics, which are only sensitive up to $n_\mathrm{e}\sim10^{3.5}$\;cm$^{-3}$. Furthermore, the TR method uses the ratios of the \textit{total} line fluxes of widely-separated emission-line doublets, unlike the traditional [SII] and [OII] techniques, which rely on the flux ratios of lines \textit{within} the doublets. This means that the TR ratios are less susceptible to uncertainties from fit degeneracy resulting from the larger velocity widths (as often seen for outflowing gas) and low spectral resolutions (as for our STIS spectra) that lead to blending of line profiles within the doublets. 

\begin{figure}
    \includegraphics[width=\linewidth]{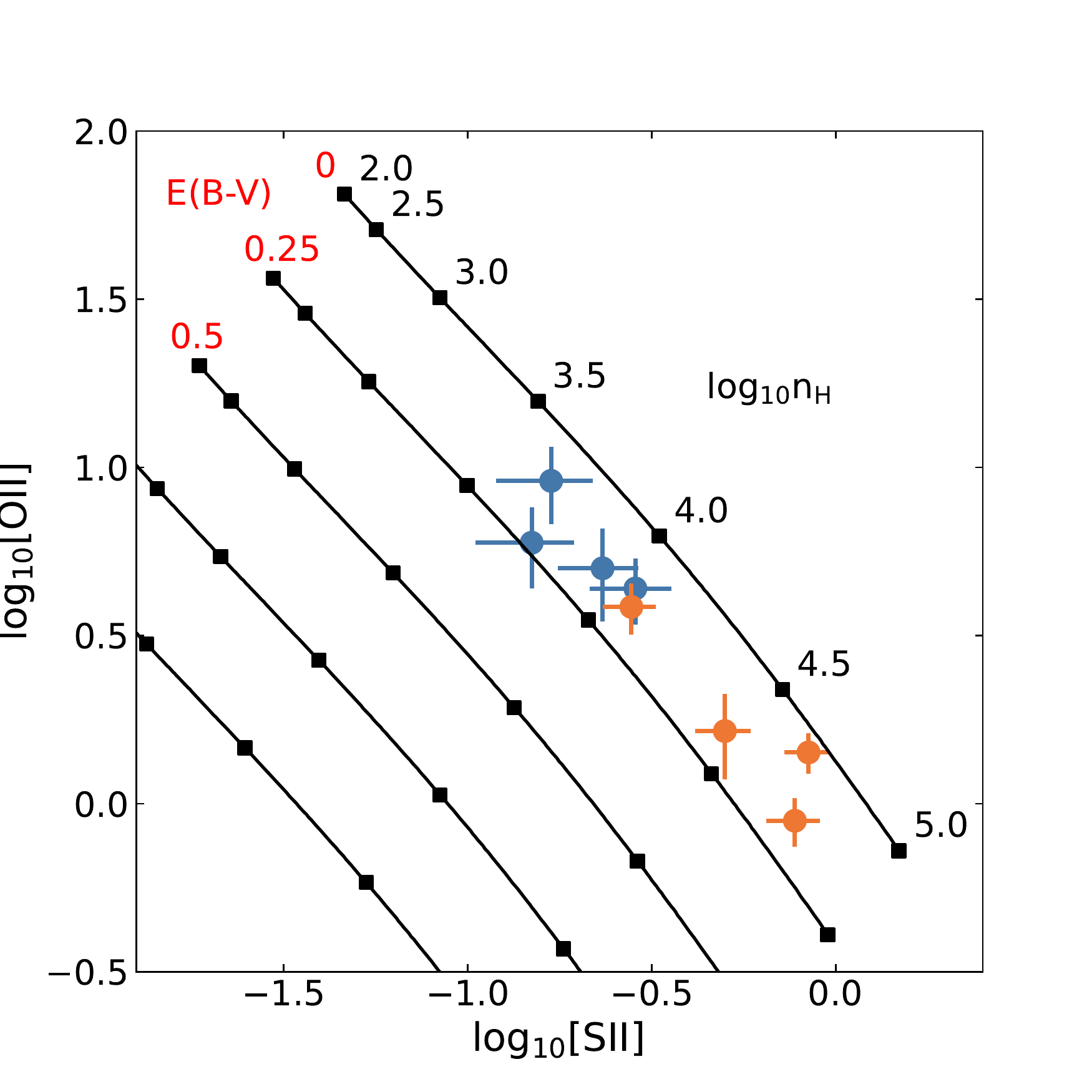}
    \caption{Grid of modelled transauroral (TR) [SII] and [OII] line ratios for radiation-bounded gas at different electron densities and reddenings, (black joined squares; as modelled with the \textsc{Cloudy} code and the CCM89 extinction curve) and measured line ratios for NGC\;1068 (orange circles) and NGC\;4151 (blue circles).}
    \label{fig: tr_ddd}
\end{figure}

We used the [OIII] model fits to the TR lines to measure line fluxes, which were then used to calculate measured TR ratios. The \textsc{CLOUDY} code (version C17.02: \citealt{Ferland2017}) was then used to generate plane-parallel, single-slab, radiation-bounded models of solar-composition gas with no dust depletion, photoionised by a central source. We set the ionising continuum of this source to follow a power-law of shape $F_v \propto v^{-\alpha}$ between 10\;{\textmu}m and 50\;keV, with a spectral index of $\alpha=1.5$. This is close to the average optical to X-ray spectral index measured in radio-quiet AGN \citep{Zamorani1981, Miller2011}, and is consistent with photoionisation modelling of the emission-line ratios of the extended and nuclear NLRs in various samples of AGN (e.g. \citealt{Ferland1983}, \citealt{Robinson1987}). We note, however, that the TR ratios are relatively insensitive to the shape of the ionising continuum (see Appendix B in \citealt{Santoro2020}). We selected an ionisation parameter of log\;$U=-3$ (the highest value that reproduced the measured TR ratios) and varied the electron density of the modelled gas in 0.01\;dex steps between \mbox{2.00\;\textless\;log$_{10}(n_e$ [cm$^{-3}$])\;\textless\;5.00}. We then reddened the modelled TR ratios produced for each electron density value with the $R_v=3.1$ CCM89 law, producing a grid of values that we compared to our measured ratios in order to provide simultaneous values of electron density and reddening. The resulting TR grid is shown in Figure \ref{fig: tr_ddd}, and the derived values are given in Table {\ref{tab: oiii_models}}.

The electron densities measured in this way for NGC\;1068 have values in the range \mbox{4.00\;\textless\;log$_{10}(n_e$ [cm$^{-3}$])\;\textless\;4.75}, while those for NGC\;4151 are approximately an order of magnitude lower (\mbox{3.60\;\textless\;log$_{10}(n_e$ [cm$^{-3}$])\;\textless\;4.10}). This is the first time that densities above $n_e=10^{3.5}$\;cm$^{-3}$ have been found using the transauroral lines with \textit{spatially-resolved} observations, and agree with similarly high electron densities derived using this technique for non-spatially resolved observations of other AGN (e.g. \citealt{Holt2011, Rose2018, Santoro2018, Spence2018, Davies2020, Speranza2022}). Importantly, the densities we find here are above the critical densities of the traditional [OII](3726/3729) and [SII](6717/6731) line ratios (Appendix \ref{section: appendix_critical_densities}), and since we do not separate broad (outflowing) and narrow (quiescent; non-outflowing) components, are likely to be underestimates for the outflowing gas (which is expected to be denser than the quiescent gas: e.g. \citealt{VillarMartin1999, Holden2022}). The reddenings that we measure are relatively modest and in the range \mbox{0.05\;\textless\;E(B-V)$_\mathrm{TR}$\;\textless\;0.25} for both objects --- these values were used to deredden our spectra for all further analysis.

\subsection{Ionisation states and mechanisms of the warm gas}
\label{section: mechanisms}

The relatively low-ionisation transauroral lines must be emitted by radiation-bounded clouds. Therefore it is uncertain how well densities derived from the transauroral ratios would represent the densities of clouds or cloud complexes that have been shock-ionised or have significant matter-bounded components. Furthermore, the model used in the transauroral ratio method assumes radiation-bounded AGN-photoionised clouds, with no contribution from a matter-bounded component or shock-ionisation. Similarly, the multi-component ionisation modelling by \citet{Revalski2021} --- which has previously been applied to NGC\;1068 and NGC\;4151 --- uses AGN photoionisation models. Therefore, it is important to investigate the ionisation mechanisms for the gas detected in our STIS slits, which potentially can also give information regarding the outflow acceleration mechanism(s) present.

\subsubsection{Electron temperatures}
\label{section: electron_temperatures}

Electron temperatures of the warm ionised phase are expected to be higher for shocked gas than AGN-photoionised gas (e.g. \citealt{Fosbury1978, VillarMartin1999}). Therefore, to provide a first indication of the ionisation mechanisms of the warm ionised gas observed in our apertures, we measured electron temperatures using the (dereddened) [OIII](5007+4959)/4363 emission-line ratio and the \textsc{PyNeb Python} module \citep{Luridiana2015}, taking the electron densities for the apertures to be those derived using the transauroral line technique for both objects (3.75\;\textless\;log$_{10}(n_e$[cm$^{-3}$])\;\textless\;4.75: see Table \ref{tab: oiii_models} and Section \ref{section: tr_diagnostics}). We present the measured electron temperatures in Table \ref{tab: oiii_models}, which are found to be high (14,300\;\textless\;T$_\mathrm{e}$\;\textless\;21,000\;K) for every aperture in both objects, with particularly high temperatures (up to T$_\mathrm{e}=21,000$\;K) being found in the central apertures of NGC\;4151. The high electron temperatures that we find in our apertures for both objects may not be fully explainable as being due to AGN-photoionisation of radiation-bounded gas \citep{Fosbury1978, Binette1996, VillarMartin1999, Holden2022}.

\subsubsection{Shock-ionisation vs matter-bounded AGN photoionisation}
\label{section: shock_vs_agn}

\begin{figure}
    \includegraphics[width=\linewidth]{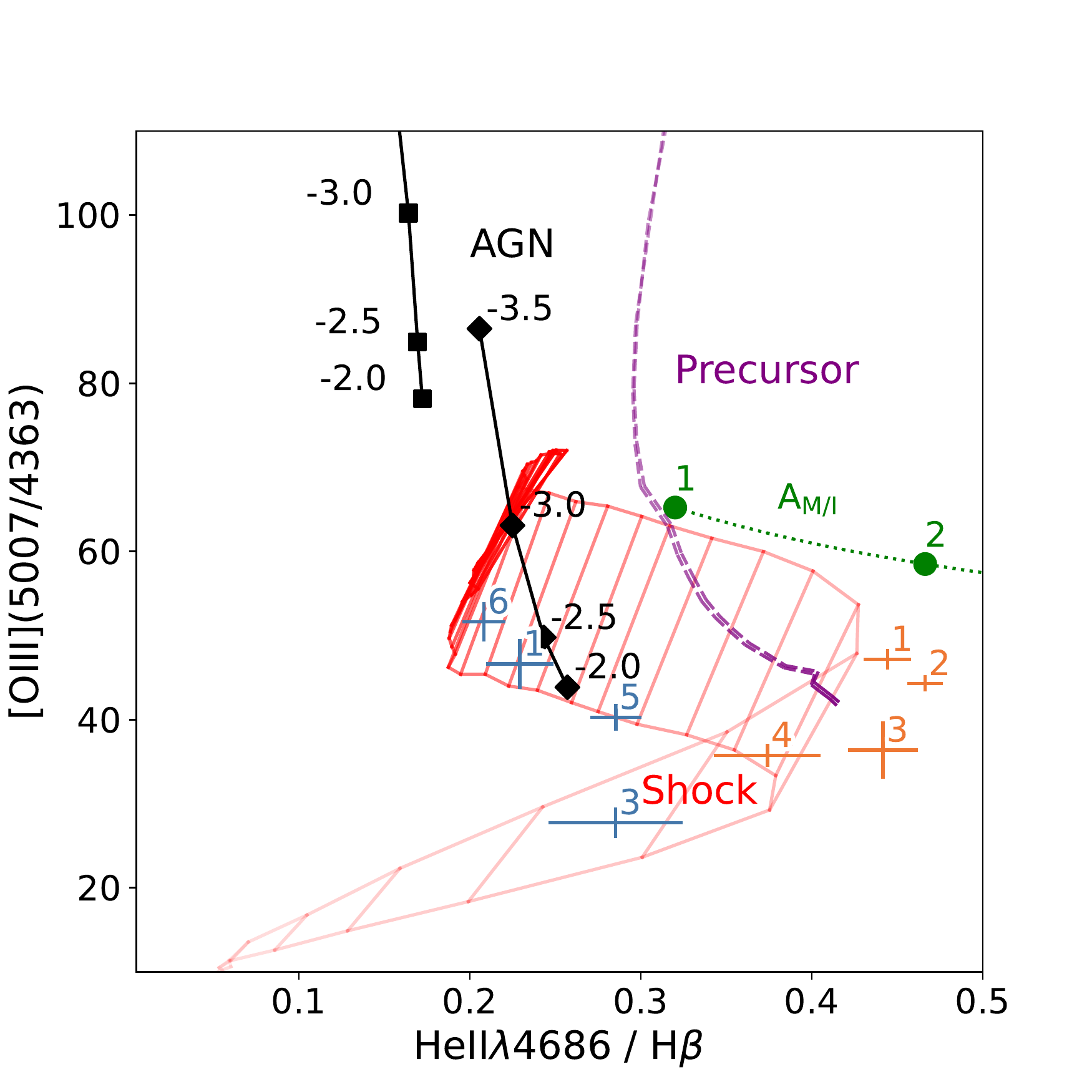}
    \caption{[OIII](5007/4363) vs HeII/H$\mathrm{\beta}$ diagnostic diagram \citep{VillarMartin1999}, used to distinguish between radiation-bounded AGN-photoionisation, matter-bounded AGN-photoionisation and shock-ionisation. The black markers show the predicted line ratios from radiation-bounded \textsc{Cloudy} modelling (see Section \ref{section: tr_diagnostics}) for solar-composition gas with a density of \mbox{$n_e$ = 10$^{4}$\;cm$^{-3} $} and varying ionisation parameters (log\;$U$; labelled) and spectral indices (squares: $\alpha$=1.5; diamonds: $\alpha$=1.0). The solid red grid shows the line ratios predicted from shock modelling \citep{Allen2008} for solar-composition gas with a pre-shock density of \mbox{$n$ = 10$^{2}$\;cm$^{-3}$} and magnetic parameters of \mbox{$B/\sqrt{n}$ = 2,4\;$\mu$G\;cm$^{3/2}$}, with lighter regions on the plot corresponding to lower shock velocities. The purple dashed lines show the predicted emission from the precursor gas, which has not yet passed through (but is photoionised by) the shock, and the dotted green line shows the line ratios expected for different ratios of matter-bounded and radiation-bounded clouds (A$_\mathrm{M/I}$, labelled and marked with green circles) from modelling by \citet{Binette1996}. Observed line ratios for each aperture are shown in orange for NGC\;1068 and blue for NGC\;4151, with the aperture number annotated.}
    \label{fig: oiii_heii_hbeta_stis}
\end{figure}

In order to investigate the cause of the high electron temperatures further, we produced the [OIII](5007/4363) vs HeII/H$\mathrm{\beta}$ diagnostic diagram developed by \citet{VillarMartin1999}, as shown in Figure \ref{fig: oiii_heii_hbeta_stis}. The radiation-bounded photoionisation models shown here are the same as those used for the TR ratio grid in Section \ref{section: tr_diagnostics} (Figure \ref{fig: tr_ddd}), albeit for an electron density of $n_e=10^{4}$\;cm$^{-3}$, varying ionisation parameters (between -3.5\;\textless\;log$_{10}U$\;\textless\;-2.0), and two values of spectral index ($\alpha=1.0$, 1.5). The pure shock and precursor (pre-shock) models are taken from the \textsc{MAPPINGS III} library presented by \citet{Allen2008}, with varying shock velocities in the range 100\;\textless\;$v_\mathrm{shock}$\;\textless\;1000\;km\;s$^{-1}$ and magnetic parameters of \mbox{$B/\sqrt{n}$ = 2,4\;$\mu$G\;cm$^{3/2}$} for a solar-composition pre-shock gas with a density of \mbox{$n$ = 10$^{2}$\;cm$^{-3}$}. The magnetic parameters were chosen to cover a reasonable range of values expected in the ISM \citep{Dopita1995}, in addition to being close to the magnetic parameters near equipartition (\mbox{$B/\sqrt{n}\sim3.23\;\mu$G\;cm$^{3/2}$}: \citealt{Allen2008}). Note that we do not use the standard `BPT' diagrams \citep{Baldwin1981} to investigate the ionisation of the gas, because some of the lines involved in those diagrams (such as H$\mathrm{\alpha}$ and [NII]$\lambda\lambda$6548,6583) are strongly blended in our apertures due to the outflow kinematics and relatively low spectral resolution, and therefore are affected by major fit degeneracies.

In Figure \ref{fig: oiii_heii_hbeta_stis}, we also plot [OIII](5007/4363) and HeII/H$\mathrm{\beta}$ as functions of A$_\mathrm{M/I}$: the ratio of the solid angles subtended by matter-bounded clouds and radiation-bounded clouds, from modelling by \citet{Binette1996}. This ratio allows us to estimate the relative contribution of matter-bounded clouds and radiation-bounded clouds in our apertures. The modelling by \citet{Binette1996} assumes solar-metallicity gas, with an ionising source spectral index of $\alpha=-1.3$, an ionisation parameter of log\;$U=-1.4$, and a density of $n_\mathrm{MB}=50$\;cm$^{-3}$. The radiation-bounded clouds are ionised by UV photons which have passed through the matter-bounded component, thus the shape of the ionising spectrum reaching the radiation-bounded clouds has changed relative to that from the source --- the parameters of the radiation-bounded clouds are determined using the resulting ionising spectrum and by assuming that the clouds have fixed pressures.

Due to the continuum underlying the H$\mathrm{\beta}$, [HeII]$\lambda$4686, and [OIII]$\lambda$4363 lines being more complex than that which underlies the [OIII]$\lambda\lambda$4959,5007 doublet and transauroral lines, we used a MCMC (Markov Chain Monte Carlo) fitting routine to fit the lines involved in the HeII$\lambda$4686/H$\mathrm{\beta}$ and [OIII](5007/4363) ratios in each aperture for both objects --- this was done to ensure that we were not significantly overestimating line flux uncertainties due to blending of spectral lines and the continuum. We used the results of the Gaussian fits described in Section \ref{section: oiii_models} (determined using least squares optimisation) to these lines as initial starting points for the MCMC routine, which fit the same models (namely one or two Gaussians and a low order polynomial) to the spectra --- taking into account the observational the flux uncertainty of the HST data --- with priors chosen to ensure the resulting models were physical (i.e. the line fluxes, mean wavelengths, and line widths must have been positive). For each fit, we initialised 500 walkers in a Gaussian distribution around the starting parameters, and used a total of 5000 iterations (including a 1000 iteration `burn-in' phase). The MCMC fits themselves were run using the \textsc{emcee} \textsc{Python} module \citep{FormanMackey2013}.

From Figure \ref{fig: oiii_heii_hbeta_stis}, we find clear evidence for significant matter-bounded emission in Apertures 1, 2, and 3 in NGC\;1068, implied by high electron temperatures and HeII/H$\mathrm{\beta}$\;\textgreater\;0.4 (similar ratio values were also measured by \citealt{Kraemer2000III}); the approximate ratio of matter-bounded to radiation-bounded clouds is A$_\mathrm{M/I}\sim2$. The difference between the [OIII](5007/4363) ratios measured in the NGC\;1068 apertures and those predicted from the \citet{Binette1996} modelling can be explained as due to the models only representing one combination of parameters: it is possible for matter-bounded clouds with different parameters to have similar [OIII](5007/4363 ratios to those found for NGC\;1068. Specifically, this ratio would be smaller for higher electron densities than the low density assumed by \citet{Binette1996}. Moreover, the presence of matter-bounded emission in these apertures is supported by the strength of high-ionisation emission lines (E$_\mathrm{ion}$\;\textgreater\;100\;eV), such as [NeV]$\lambda$3426, [FeVII]$\lambda$3759, and [FeVII]$\lambda$6087, relative to lower-ionisation lines (such as [OIII]) in our STIS spectra. These and other high-ionisation lines were previously identified in the same dataset by \citet{Kraemer2000II}. For Aperture 4 of NGC\;1068 (centered slightly above the nucleus: Figure \ref{fig: apertures}), we measure HeII/H$\mathrm{\beta}$ ratios consistent with both matter-bounded AGN-photoionisation and shock-ionisation. 

\begin{figure}
    \includegraphics[width=\linewidth]{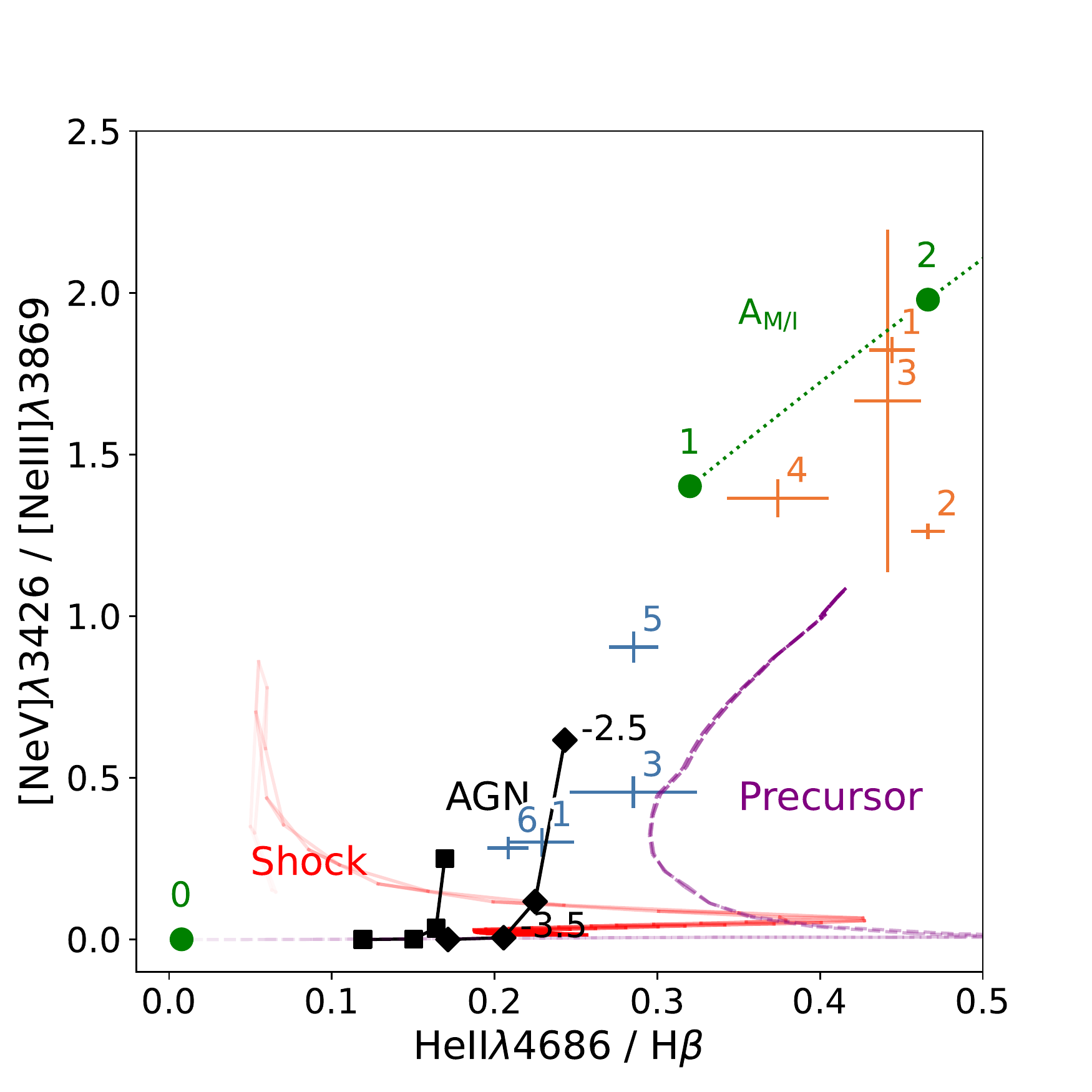}
    \caption{[NeV]3426/[NeIII]3869 vs HeII/H$\mathrm{\beta}$ diagnostic diagram --- both ratios are sensitive to the presence of significant matter-bounded components. The line and marker scheme is the same as Figure \ref{fig: oiii_heii_hbeta_stis}. The line ratios measured in our NGC\;1068 apertures are located in the matter-bounded photoionisation region of the diagram (corresponding to \mbox{1\;\textless\;A$_\mathrm{M/I}$\;\textless\;2}; consistent with Figure \ref{fig: nev_neiii_heii_hbeta_stis}) whereas the NGC\;4151 line ratios fall in the shock/precursor/radiation-bounded AGN photoionisation region.}
    \label{fig: nev_neiii_heii_hbeta_stis}
\end{figure}

To further probe the ionisation mechanism of the gas, we also measured the [NeV]$\lambda$3426/[NeIII]$\lambda$3869 ratio --- which is sensitive to higher ionisation gas  --- using the same MCMC fitting routine described earlier. We produced a diagnostic diagram of [NeV]$\lambda$3426/[NeIII]$\lambda$3869 vs HeII/H$\mathrm{\beta}$ using the same radiation-bounded photoionisation, matter-bounded photoionisation, and shock-ionisation models as used for the [OIII](5007/4363) and HeII/H$\mathrm{\beta}$ diagram (Figure \ref{fig: oiii_heii_hbeta_stis}), and present this in Figure \ref{fig: nev_neiii_heii_hbeta_stis}. We find that the values for all of the NGC\;1068 apertures are consistent with matter-bounded AGN-photoionisation with  1\;\textless\;A$_\mathrm{M/I}$\;\textless\;2. This further indicates that the gas in these apertures is matter-bounded and AGN-photoionised, including Aperture 4.

With the exception of Aperture 3, the [OIII](5007/4363) vs HeII/H$\mathrm{\beta}$ ratios measured in our NGC\;4151 apertures (Figure \ref{fig: oiii_heii_hbeta_stis}), are consistent with both shock ionisation and radiation-bounded AGN photoionisation (assuming a relatively flat spectral index of $\alpha=1.0$ and log\;$U\sim-2.0$). However, from the [NeV]$\lambda$3426/[NeIII]$\lambda$3869 vs HeII/H$\mathrm{\beta}$ diagram (Figure \ref{fig: nev_neiii_heii_hbeta_stis}), it can be seen that the measured ratios for NGC\;4151 are not consistent with pure shock-ionisation alone: if the gas is shock ionised, then a contribution from the precursor component is required. Alternatively, the gas in these apertures may have pure radiation-bounded AGN photoionisation, however we highlight that this requires a relatively flat spectral index ($\alpha=1.0$), and/or higher ionisation parameters ($-3.0$\;\textless\;log\;$U$\;\textless\;$-2.0$) and densities ($n_e$\;\textgreater$10^5$\;cm$^{-3}$) than can explain our transauroral line ratios (Section \ref{section: tr_diagnostics}). Ultimately, it is not possible to determine unambiguously the true, dominant ionisation mechanism of the gas in our NGC\;4151 apertures with the diagnostic features that are available in our data.

\subsubsection{The viability of shock-ionisation}
\label{section: shock_viability}

In order to further investigate the viability of shocks as the dominant ionisation mechanism along our slits for NGC\;1068 and NGC\;4151, we compared our measured H$\mathrm{\beta}$ fluxes to those expected from shock models --- a technique presented by \citet{Baron2017}. First, we converted our measured (and dereddened) H$\mathrm{\beta}$ fluxes ($F_\mathrm{H\beta}$) into H$\mathrm{\beta}$ luminosities using the luminosity distances ($D_L$) for each galaxy. The resulting luminosities were then converted into luminosities per surface area using the aperture sizes in arcseconds (i.e. the aperture width multiplied by the slit width) and the spatial scales for each object (0.067\;kpc/arcsecond and 0.078\;kpc/arcsecond, respectively). 

We then compared the measured luminosities per surface area to those expected from the \textsc{MAPPINGS III} shock models of pre-shock density $n=10^2$\;cm$^{-3}$ (corresponding the densities measured in our apertures, assuming a compression factor of 100: \citealt{Sutherland2017}) and magnetic parameters $B/\sqrt{n}=2,4$\;$\mathrm{\mu}$G\;cm$^{3/2}$. From this comparison, we find that the H$\mathrm{\beta}$ luminosities per surface area, as measured in each aperture for NGC\;1068 \mbox{($4.9\times10^{-3}$\;\textless\;L$_\mathrm{H\beta}$\;\textless\;$2.2\times10^{-2}$\;erg\;s$^{-1}$\;cm$^{-2}$)} and NGC\;4151 \mbox{($2.2\times10^{-3}$\;\textless\;L$_\mathrm{H\beta}$\;\textless\;$4.6\times10^{-3}$\;erg\;s$^{-1}$\;cm$^{-2}$)}, can be accounted for by shocks with velocities $v_\mathrm{shock}$\;\textgreater\;425\;km\;s$^{-1}$ and  $v_\mathrm{shock}$\;\textgreater\;225\;km\;s$^{-1}$ respectively. In both cases, the outflow velocities for our apertures (Section \ref{section: kinematics}; Table \ref{tab: energetics}) are above these required velocities. This demonstrates that shock-ionisation \textit{could} feasibly produce the recombination line fluxes measured in both objects, however this alone does not necessarily confirm the ionisation mechanism.

Note that here we assumed a gas covering factor of unity relative to the shock (i.e. that the emitting-gas covers the entire area of the shock within each aperture), which may not be the case in reality. If this covering factor is in fact much lower than unity, then a larger shock area or higher shock velocities would be needed to produce the same $\mathrm{H\beta}$ luminosity.

\subsection{The high-ionisation gas in NGC\;1068}
\label{section: high_ionisation}

The relative strengths of the high ionisation (E$_\mathrm{ion}$\;\textgreater\;$100$\;eV) lines detected in several of our apertures for NGC\;1068 indicate the presence of matter-bounded clouds, and therefore may play an important role in the structure of the cloud complexes present in the NLR. Determining the physical conditions of this high-ionisation component is therefore necessary. To this end, we measured the [FeVII](6087/3759) and [NeV]$\lambda$3426/[FeVII]$\lambda$6086 emission-line ratios, which are sensitive to the density and ionisation parameter of the high-ionisation gas. These ratios were calculated using the measured line fluxes of the lines in the ratios, which were themselves determined using the same MCMC fitting method described in Section \ref{section: mechanisms}.

\begin{figure}
    \includegraphics[width=\linewidth]{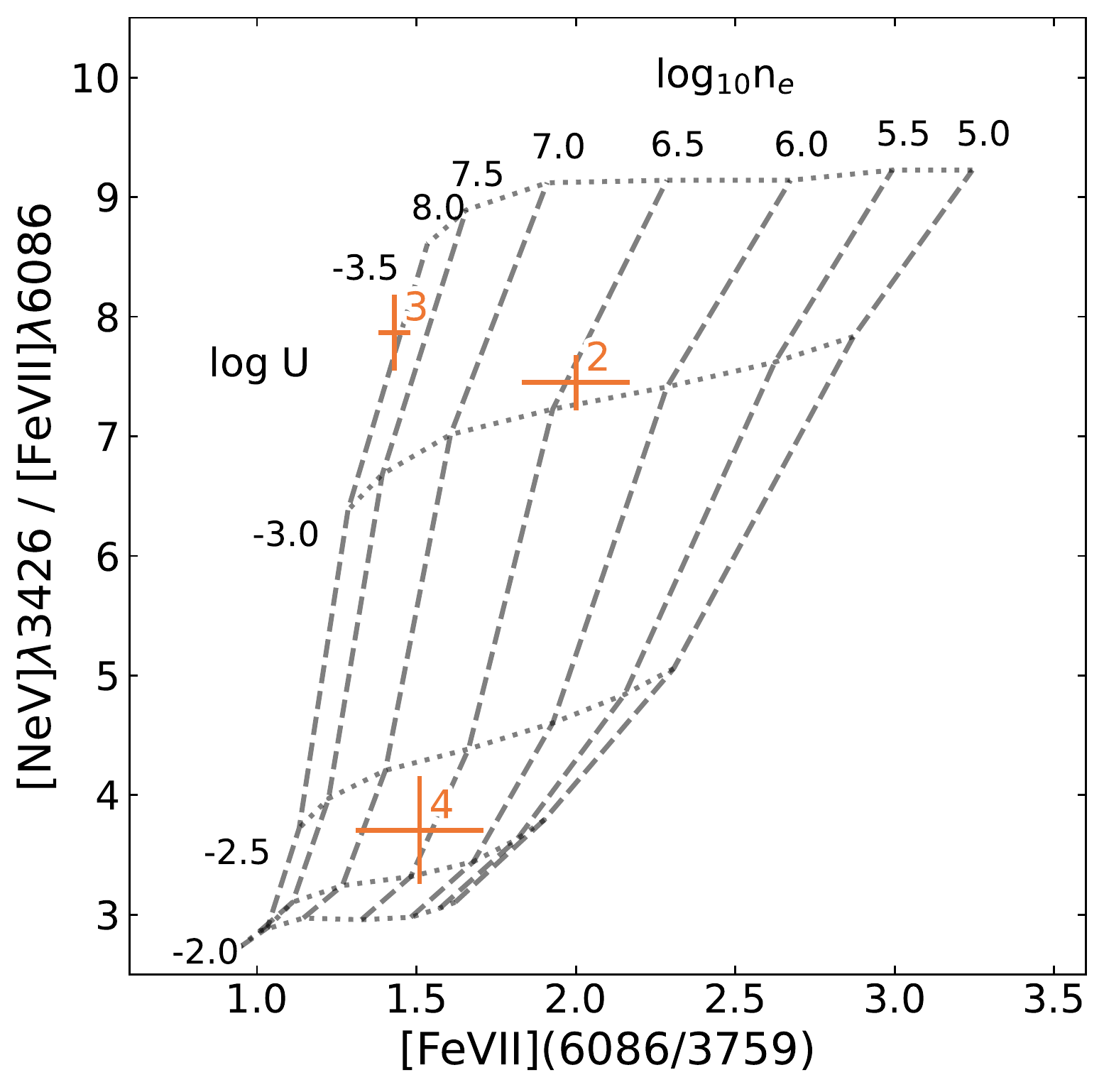}
    \caption{[FeVII](6087/3759) vs [NeV]$\lambda$3426/[FeVII]$\lambda$6086 diagnostic diagram, sensitive to the spectral index, ionisation parameter, and electron density of the gas. The grey grid was generated using \textsc{CLOUDY} for radiation-bounded AGN-photoionisation with a spectral index of $\alpha=1.5$ (see Appendix \ref{section: appendix_nev_fevii}) and varying electron densities and ionisation parameters: dashed lines connect points of constant density (labelled), and dotted lines connect points of constant ionisation parameter (also labelled). The measured ratio values in our NGC\;1068 apertures are shown in orange and labelled. From this diagram, the high ionisation gas is found to have densities in the range 6.45\;\textless\;log$_{10}$($n_e$[cm$^{-3}$])\;\textless\;8.00.}
    \label{fig: nev_fevii}
\end{figure}

We present the [FeVII](6087/3759) vs [NeV]$\lambda$3426/[FeVII]$\lambda$6086 diagnostic diagram (see \citealt{Rose2011}) with our measured line ratios for the NGC\;1068 apertures in Figure \ref{fig: nev_fevii}; a \textsc{CLOUDY} radiation-bounded photoionisation grid for a solar metallicity, plane-parallel, single-slab cloud of varying ionisation parameters (-3.5\;\textless\;log\;$U$\;\textless\;2.0) and electron densities (5.0\;\textless\;log($n_e$[cm$^{-3}$])\;\textless\;8.0), and a central ionising source with spectral index $\alpha=1.5$ (see Appendix \ref{section: appendix_nev_fevii}), is shown. From this grid, we determine the densities of the high-ionisation gas to be in the range 6.45\;\textless\;log$_{10}$($n_e$[cm$^{-3}$])\;\textless\;8.00: several orders of magnitude higher than the gas traced by the lower critical-density [OII] and [SII] lines. We discuss the implications of this for the gas structures within our apertures in Section \ref{section: disc-density-mb}.

\subsection{Energetics of the outflowing gas}
\label{section: energetics}

\subsubsection{Outflow kinematics}
\label{section: kinematics}

In order to determine the mass outflow rates, kinetic powers and coupling efficiencies of the gas outflows detected in our STIS spectra, we required measurements of the kinematics of the outflowing gas\footnote{We do not use kinematics derived from our [OIII] models due to the relatively low spectral resolution and high instrumental widths of our spectra.}. For this purpose, we used the results from detailed kinematic modelling (based the same HST/STIS spectra used here) of NGC\;1068 and NGC\;4151 presented by \citet{Crenshaw2000_N1068} and \citet{Crenshaw2000_N4151} (hereafter CKN1068 and CKN4151), respectively. We note that, due to the different PAs used and the fact that the outflow geometry likely depends greatly on PA, we do not use the updated kinematic models from \citet{Das2005} and \citet{Das2006}. 

To calculate deprojected velocities, we first derived a universal `deprojection factor' by dividing the maximum observed velocities (located at the velocity `turnover' position - see \citealt{Crenshaw2000_N1068} and \citealt{Crenshaw2000_N4151}) by the maximum model-deprojected velocities from the CKN1068 and CKN4151 bicone models. We then took the the highest observed (projected) velocity at the position of each aperture, and divided these velocities by our determined deprojection factor to give the maximum deprojected outflow velocity in each aperture. We label the deprojected outflow velocities as $v_\mathrm{out}$, and give their values in Table \ref{tab: energetics}.

\subsubsection{Mass outflow rates, kinetic powers and coupling efficiencies}
\label{section: mout_ekin_fkin}

We used the H$\mathrm{\beta}$ luminosities to determine masses for the warm ionised gas in each aperture with
\begin{equation}
    M_\mathrm{ion} = \frac{L(H\mathrm{\beta})m_\mathrm{p}}{\alpha^\mathrm{eff}_{\mathrm{H\beta}}hv_{\mathrm{H\beta}}n_\mathrm{e}},
    \label{eq: mion}
\end{equation}
where $M_\mathrm{ion}$ is the total mass of the warm ionised gas, $m_\mathrm{p}$ is the proton mass, $\alpha^{eff}_{H\beta}$ is the Case B recombination coefficient for H$\mathrm{\beta}$ (taken to be 1.61$\times10^{-14}$\;cm$^3$s$^{-1}$ for a gas of density $n_\mathrm{e}=10^4$\;cm$^{-3}$ and temperature $T_\mathrm{e}=20,000$\;K; \citealt{Osterbrock2006}) and $v_\mathrm{H\beta}$ is the frequency of the H$\mathrm{\beta}$ line.

Assuming that the derived masses (estimated using the total line fluxes) are dominated by outflowing gas, we combined them with the aperture crossing time to calculate mass outflow rates
\begin{equation}
    \dot{M}_\mathrm{out} = \frac{M_\mathrm{ion}v_\mathrm{out}}{\Delta R},
    \label{eq: mion_rate}
\end{equation}
where $v_\mathrm{out}$ is the outflow velocity from the CKN1068 and CKN4151 models, and $\Delta R$ is the aperture width.

Kinetic powers were estimated from the mass outflow rates using
\begin{equation}
    \dot{E}_\mathrm{kin} = \frac{1}{2}M_\mathrm{out}v^2_\mathrm{out}.
    \label{eq: ekin}
\end{equation}

Finally, the ratio of the kinetic power to the bolometric AGN luminosity ($L_\mathrm{bol}$) was taken to estimate coupling efficiencies for each aperture:
\begin{equation}
    \epsilon_\mathrm{f} = \frac{\dot{E}_\mathrm{kin}}{L_\mathrm{bol}}.
    \label{eq: fkin}
\end{equation}
NGC\;1068 is estimated to have a bolometric luminosity in the range 0.4\;\textless\;$L_\mathrm{bol}$\textless\;4.7$\times10^{38}$\;W \citep{Woo2002, AlonsoHerrero2011, LopezRodriguez2018, Gravity2020}, of which we take the lowest value to ensure higher estimates of coupling efficiencies and thus determine the maximum potential impact of the outflowing gas on the host galaxy. For NGC\;4151, we took the bolometric luminosity to be $L_\mathrm{bol}=1.4\times10^{37}$\;W \citep{Kraemer2020}.

We present our derived mass outflow rates, kinetic powers and coupling efficiencies for both cases in Table \ref{tab: energetics}. For NGC\;1068, our estimates are less than the maximum values determined from photoionisation modelling by \citet{Revalski2021} ($\dot{M}_\mathrm{out}=9.0\pm1.13$\;M$_\odot$yr$^{-1}$; $\dot{E}_\mathrm{kin}=(5.4\pm0.5)\times10^{35}$\;W, $\epsilon_\mathrm{f}=0.54\pm0.05$\;per\;cent)\footnote{\citet{Crenshaw2015} and \citet{Revalski2021} assume bolometric luminosities of L$\mathrm{bol}=1\times10^{38}$\;W for NGC\;1068 and L$\mathrm{bol}=7.9\times10^{36}$\;W for NGC\;4151 when calculating coupling efficiencies.}. For NGC\;4151, our derived values are similar to the results of photoionisation modelling by \citet{Crenshaw2015} ($\dot{M}_\mathrm{out}\sim3.01\pm0.45$\;M$_\odot$yr$^{-1}$; $\dot{E}_\mathrm{kin}=(4.3\pm1.0)\times10^{34}$\;W,  $\epsilon_\mathrm{f}=0.54\pm0.11$\;per\;cent). Our calculated mass outflow rates for NGC\;4151 are also consistent with previous values derived for the warm ionised phase by \citet{Storchi-Bergmann2010} ($M_\mathrm{out}\approx2.4$\;M$_\odot$) and the X-ray emitting gas ($M_\mathrm{out}\approx2$\;M$_\odot$yr$^{-1}$: \citealt{Wang2011} and \citealt{Kraemer2020}).

For NGC\;1068, the mass outflow rates for the warm-ionised phase are much below that of the cold molecular gas at a similar extent from the nucleus (i.e. traced by CO, HCN; $T\sim$100\;K): \citet{GarciaBurillo2014} derive a mass outflow rate of $\dot{M}_\mathrm{out}=63^{+21}_{-37}$\;M$_\odot$yr$^{-1}$ within the $r\sim$200\;pc circumnuclear disk (CND) of NGC\;1068. This indicates that most of the outflowing mass may be present in the colder gas phases, as has been found for other objects (see \citealt{RamosAlmeida2019} and \citealt{Holden2022}).

\begin{table*}
\begin{tabular}{cccccc}
\hline
Aperture & $v_\mathrm{out}$ (km\;s$^{-1}$) & M$_\mathrm{out}$ (M$_\odot$) &  $\dot{M}_\mathrm{out}$ (M$_\odot$yr$^{-1}$) & $\dot{E}_\mathrm{kin}$ (W) & $\epsilon_\mathrm{f}$ (per\;cent) \\ \hline
\multicolumn{6}{c}{NGC 1068} \\ \hline
1 & -1300 & ($1.7\pm0.3)\times10^5$ & $3.7\pm0.6$ & $(2.0\pm0.3)\times10^{35}$ & $(4.9\pm0.7)\times10^{-1}$    \\
2 & -1100 & ($6.3\pm0.8)\times10^4$ & $1.6\pm0.2$ & $(6.1\pm0.7)\times10^{34}$ & $(1.5\pm0.2)\times10^{-1}$ \\
3 & -450 & $(7.3\pm0.9)\times10^4$ & $1.2\pm0.1$ & $(7.6\pm0.9)\times10^{33}$ & $(1.9\pm0.2)\times10^{-2}$    \\
4 & -150 & $(9.4\pm2.2)\times10^4$ & $(6.0\pm1.4)\times10^{-1}$ & $(4.2\pm1.0)\times10^{32}$ & $(1.1\pm0.2)\times10^{-3}$    \\ \hline
\multicolumn{6}{c}{NGC 4151}    \\ \hline
1 & -700 & $(2.5\pm0.6)\times10^5$ & $3.7\pm1.0$ & $(5.8\pm0.2)\times10^{34}$ & $(4.1\pm1.1)\times10^{-1}$    \\
3 & -800 & $(7.2\pm3.0)\times10^4$ & $3.4\pm1.4$ & $(6.9\pm2.9)\times10^{34}$ & $(4.9\pm2.0)\times10^{-1}$    \\
5 & 800 & $(1.7\pm0.4)\times10^5$ & $4.5\pm1.2$ & $(9.2\pm2.4)\times10^{34}$ & $(6.5\pm1.7)\times10^{-1}$ \\
6 & 800 & $(2.9\pm0.7)\times10^5$ & $6.9\pm1.8$ & $(1.4\pm0.4)\times10^{35}$ & $(9.9\pm2.6)\times10^{-1}$    \\ \hline
\end{tabular}
\caption{Outflow velocities, outflow masses, mass outflow rates, kinetic powers and coupling efficiencies for the apertures of our STIS spectra of NGC\;1068 and NGC\;4151. The outflow velocities used to calculate the mass outflow rates and kinetic powers presented here are from the CKN1068 and CKN4151 models (see Section \ref{section: kinematics}).}
\label{tab: energetics}
\end{table*}

\section{Discussion}
\label{section: discussion}

From our analysis of archival STIS spectra of the central regions ($r$\;\textless\;160\;pc) of NGC\;1068 and NGC\;4151, we find evidence for dense (10$^{3.6}$\;cm$^{-3}$\textless\;$n_e$\;\textless\;10$^{4.8}$\;cm$^{-3}$) gas that shows line ratios consistent with matter-bounded AGN-photoionisation in the case of NGC\;1068, and shock-ionisation (with precursor gas ionisation) or radiation-bounded AGN-photoionisation in the case of NGC\;4151. Furthermore, we find that the measured H$\mathrm{\beta}$ luminosities could be explained as being due to shock-ionisation for both objects, assuming a shock covering factor of unity. In both objects, we find coupling efficiencies that are close to the lowest value required by models of galaxy evolution, however these are likely underestimates. In this section, we discuss the implication of these results on the dominant ionisation and acceleration mechanisms of the gas seen in our slits, compare our results to past work on these two well-studied objects, and investigate the impact on the density diagnostic techniques used. Finally, we place our results in a broader context by comparing with those from a similar study of the nearby Seyfert 2 galaxy IC\;5063.

\subsection{The outflow ionisation and acceleration mechanisms in the NLRs of NGC\;1068 and NGC\;4151}
\label{section: disc-ionisation}

To determine the true impact of the outflowing gas on the host galaxies, quantitative comparison of observations to theoretical modelling is needed. However, both modelling of jet-ISM interactions (e.g. \citealt{Mukherjee2018, Audibert2023}) and AGN radiation-pressure-driven outflows (e.g. \citealt{Crenshaw2000_N1068, Meena2023}) is able to explain outflow kinematics in different objects. In order to enable accurate future comparisons to theoretical models and therefore accurately quantify the impact of the outflows in NGC\;1068 and NGC\;4151 --- which have been conversely argued to be radiatively-accelerated \citep{Crenshaw2000_N4151, Crenshaw2000_N1068, Das2005, Das2006, Revalski2021, Meena2023} and jet-accelerated \citep{Capetti1997, Axon1998, May2017, May2020} --- the dominant outflow acceleration mechanisms in these objects need to be robustly identified.

\subsubsection{Matter-bounded ionisation and the acceleration mechanism in NGC\;1068}

It has been previously proposed that the outflows in NGC\;1068 are driven via radiation pressure \citep{Kraemer2000II, Das2006, Revalski2021, Meena2023}, instead of via shocks induced by the radio jet colliding with the ISM within the bicone. While we do not separate the outflowing gas from the quiescent gas in this work, our results are consistent with this mechanism: we find evidence for matter-bounded AGN-photoionisation of the warm-ionised gas in the form of simultaneous high [OIII] temperatures (Table \ref{tab: oiii_models}: $T_e \sim$15,000\;K; corresponding to [OIII](5007/4363)\;\textless\;60) and line ratios (\mbox{HeII$\lambda$4686\;/\;H$\mathrm{\beta}$\;\textgreater\;0.4}: Figure \ref{fig: oiii_heii_hbeta_stis}; \mbox{[NeV]$\lambda3426$\;/\;[NeIII]$\lambda$3869\;\textgreater\;1.0}: Figure \ref{fig: nev_neiii_heii_hbeta_stis}) within a 134\;pc radius from the nucleus in the NE cone along the radio axis, consistent with radiative acceleration. However, it is \textit{possible} that the outflowing gas has been shock-ionised and accelerated by the jet, but has subsequently cooled and then been reionised by the AGN (e.g. as in \citealt{Holden2022}). Spatially-resolved, high spectral-resolution observations are needed to further investigate this situation by separating the emission from the outflowing and quiescent gas, and then determining the ionisation and excitation mechanisms of each kinematic component. In addition, comparing the electron densities of the outflowing and non-outflowing gas may reveal signs of shock compression, which is expected to be a factor of $\sim$4--100 \citep{Sutherland2017}.

We note that the outflowing gas appears to be spatially confined to the extent of the radio structure: the broad (FWHM$_\mathrm{v}$\;\textgreater\;250\;km\;s$^{-1}$) [OIII]$\lambda\lambda$4959,5007 emission in our spectra is seen to a maximum radius of $\sim4.8$\;arcseconds from the nucleus in the NE cone (as measured from the line profiles of the [OIII] emission that extends beyond the regions covered by our apertures), similar to the maximum radial extent of the NE radio lobe (6.18\;arcseconds; 420\;pc) measured from radio imaging (e.g. 15\;GHz: \citealt{WilsonUlvestad1987} 5\;GHz: \citealt{Wilson1983}, \citealt{Gallimore1996}; 22\;GHz: \citealt{Gallimore1996}; 1.4\;GHz: \citealt{Gallimore1996}, \citealt{GarciaBurillo2014}). This is also in agreement with ground-based Fabry-Pérot integral field spectroscopy by \citet{Cecil1990} --- which finds no significant velocity deviation from the systematic velocity beyond the radio lobe --- and kinematic modelling by \citet{Crenshaw2000_N1068}, \citet{Das2006} and \citet{Meena2023}, which find outflows extended up to $\sim$5.1\;arcseconds from the nucleus\footnote{An [OIII] emission knot in the NLR of NGC\;1068, labelled `A' by \citet{Meena2023} and located 7.3\;arcseconds from the nucleus (i.e. beyond the radio source), has outflow-like kinematics (200\;\textless\;FWHM\;\textless\;1000\;kms$^{-3}$; v$_\mathrm{out}=863$\;kms$^{-3}$). As noted by \citet{Meena2023}, this knot lies beyond the expected extent of radiatively-driven outflows. Regardless, we highlight that the vast majority of the outflows along the radio axis are located at lower radii than the maximum extent of the NE radio lobe.}. Furthermore, VLT/MUSE spectroscopy presented by \citet{Venturi2021} shows that the measured [OIII] W70 velocity parameter\footnote{W70 is defined as the difference between the velocities that contain 85\;per\;cent and 15\;per\;cent of the total flux of the fits to the line profile (see \citealt{Venturi2021}).} has high values \mbox{(400\;km\;s$^{-1}$\;\textless\;[OIII]\;W70\;\textless\;1200\;km\;s$^{-1}$)} between the nucleus and the lobe, out to a radius of 3.6\;arcseconds along the bicone axis. Moreover, the NLR molecular CO(3--2) outflows (as seen in ALMA imaging by \citealt{GarciaBurillo2014}) decelerate within the radio lobe, at a distance of $\sim400$\;pc ($\sim$5.7\;arcseconds) from the nucleus. Taken together, this shows that the NE cone outflows have a similar extent to the NE radio lobe. This is evidence for the outflows being accelerated by the radio jet, although it does not entirely rule out radiative acceleration.

\subsubsection{Shock-ionisation and acceleration in NGC\;4151}

Our results for NGC\;4151 indicate that the near-nuclear gas along the radio axis may be shock-ionised, since the measured [OIII](5007/4363), HeII/H$\mathrm{\beta}$, and [NeV]$\lambda$3426/[NeIII]$\lambda$3869 line ratios and H$\mathrm{\beta}$ luminosities are consistent with those expected from a mixture of shock and shock-precursor ionisation (Figures \ref{fig: oiii_heii_hbeta_stis} and \ref{fig: nev_neiii_heii_hbeta_stis}; Section \ref{section: shock_viability}).

The radio structure in the NLR of NGC\;4151, as seen in low-resolution 1.5--5\;GHz VLA radio imaging by \citet{Johnston1982}, has a lobe-like component with a centroid 6.43\;arcseconds from the nucleus along the radio axis in the NE cone. This structure lies beyond the maximum $\sim$4\;arcseconds extent of the warm-ionised outflows (\citealt{Meena2023}; see also \citealt{Das2005}), and --- as we have argued for the situation in NGC\;1068 --- is consistent with the outflows being launched by the radio jet. From HST/PC + HST/WFPC2 imaging, \citet{Williams2017} found higher [OIII]/H$\mathrm{\alpha}$ ratios close to the string of radio knots that are seen in their higher-resolution 1.51\;GHz observations (shown here in Figure \ref{fig: seyferts_ifu_slit}), with the values of this ratio decreasing beyond $\sim$4\;arcseconds from the nucleus along the radio axis. The authors interpreted this as the radio jet having a contribution to the ionisation of the gas close to the nucleus, but AGN-photoionisation being dominant further out. This is also in agreement with the results from X-ray and optical imaging by \citet{Wang2011b}, who propose a mixture of shock-ionisation and AGN-photoionisation in the NLR of NGC\;4151. 

Taken together with the findings of these previous investigations, the results presented here may indicate that the outflows in NGC\;4151 have been shock-accelerated and then re-ionised by photons from the AGN, with AGN-photoionisation being dominant further from the nucleus.

\subsection{The effect of ionisation mechanisms on density diagnostics}
\label{section: disc-density-effects}

The ionisation mechanisms (Section \ref{section: mechanisms}), electron temperatures (Section \ref{section: electron_temperatures}), and densities (Sections \ref{section: tr_diagnostics} and \ref{section: high_ionisation}) of the warm gas detected in our STIS slits allows us to investigate the structures and conditions of the line-emitting clouds, and therefore verify the origin of different emission lines and thus the precision of diagnostics which make use of them. For example, the TR density diagnostic (Section \ref{section: tr_diagnostics}) relies on AGN-photoionisation being dominant, with no significant contribution from a matter-bounded component or shock-ionisation. Since we find evidence for matter-bounded emission in NGC\;1068 and potential shock-ionisation in NGC\;4151, it is important to investigate the effect of this on derived densities.

\subsubsection{The impact of matter-bounded photoionisation}
\label{section: disc-density-mb}

If the higher ionisation lines are indeed emitted by matter-bounded gas structures in the outflow (as shown by the [OIII] temperatures, HeII/H$\mathrm{\beta}$ ratios, and [FeVII](6086/3759) vs [NeV]$\lambda$3426/[FeVII]$\lambda$6086 diagram: Sections \ref{section: mechanisms} and \ref{section: high_ionisation}), then the transauroral lines cannot be emitted by the same  structures. However, it is possible that they are emitted by different clouds within the same cloud complexes, considering that we see these lines with similar profiles in each of our apertures. Alternatively, or perhaps in addition, it is possible that the outer layers of a single cloud are matter-bounded, while the denser core is radiation-bounded (one of the scenarios presented by \citealt{Binette1996}). In this scenario, the matter-bounded layers may represent lower density gas that was driven away from the ionisation front by the increase in pressure that occurred when the gas structure was first photoionised by the AGN. However, this is not consistent with our findings: in Section \ref{section: high_ionisation}, we use the [FeVII](6087/3759) and [NeV]$\lambda$3426/[FeVII]$\lambda$6086 emission-line ratios to determine high-ionisation gas densities of 6.45\;\textless\;log$_{10}$($n_e$[cm$^{-3}$])\;\textless\;8.00 in our NGC\;1068 apertures: significantly higher than that of the lower-ionisation gas. A potential explanation is that the gas that is emitting the high-ionisation [FeVII] and [NeV] lines represents dense fragments of the expanding matter-bounded component: since these lines have high critical densities (7.1\;\textless\;log$_{10}(n_\mathrm{crit}$[cm$^{-3}$])\;\textless\;8.5; Appendix \ref{section: appendix_critical_densities}), they would only be emitted strongly by such dense cloud components.

Therefore, given the ionisation energies of the lines (Appendix \ref{section: appendix_critical_densities}), we propose that the [FeVII] and [NeV] lines trace matter-bounded, higher ionisation clouds within the complexes (or edges of individual clouds), and the [OII] and [SII] lines are emitted from radiation-bounded clouds (or cores of individual clouds). In this scenario, much of the [OIII] emission must arise from the matter-bounded regime in order to explain the high electron temperatures that we measure in our NGC\;1068 apertures (Section \ref{section: electron_temperatures}). Hence, given the high density of the high-ionisation gas, it is likely that the gas emitting the [OIII] lines is denser than the gas that is emitting the transauroral lines. This reinforces the need for outflow diagnostics that are sensitive to high (\textgreater10$^{3.5}$\;cm$^{-3}$) densities.

\subsubsection{The impact of shock-ionisation}

Since the gas in our NGC\;4151 apertures may be shock-ionised, it is essential to quantify the effect of this on the transauroral ratio density diagnostic. In Appendix \ref{section: appendix_tr_shock}, we plot the TR ratios from shock models over the TR photoionisation diagnostic grid used in Section \ref{section: tr_diagnostics}, and quantify the impact of shock-ionisation on the TR electron density and reddening values derived from the photoionisation grid. We find that, overall, the effect on the derived density is $\pm$0.38 orders of magnitude, and the effect on derived reddenings is E(B-V)$\pm$0.13. Crucially, we note that this is much less than the impact of using lower-critical-density techniques (such as the [SII](6717/6731) ratio), and is similar to the effect of varying the parameters of the photoionisation model (\citealt{Santoro2020}: log($n_e$[cm$^{-3}$])$\pm$(0.1--0.7); E(B-V)$\pm$(0.1--0.2)). In summary, while using the transauroral line method presented by \citet{Holt2011} as a density and reddening diagnostic for shock-ionised gas does incur some uncertainty on the derived densities, the derived densities are still likely more accurate than those derived from commonly used, traditional methods.

Gas that has been shock-ionised by jet-ISM interactions presents a problem for the photoionisation modelling method used by \citet{Revalski2021}, as the technique relies on assuming that the material at a given distance from the nucleus is being photoionised by the central AGN engine. In the case of shock-ionisation, the outflows are instead being shock-ionised \textit{locally} by the jet within the bicone at any given distance from the nucleus, and so any electron densities derived using an assumed ionisation parameter and distance will be incorrect. \citet{Revalski2021} used the standard BPT diagrams \citep{Baldwin1981} in an attempt to ensure all of the measured line ratios were consistent with AGN-photoionisation. However, the regions of AGN shock and photoionisation in these diagrams overlap considerably, thus further diagnostics should also be used in order to disentangle the contribution from shocks and photoionisation, such as the [OIII](5007/4363) vs HeII$\lambda$4686/H$\mathrm{\beta}$ \citep{VillarMartin1999} and [FeII]$\lambda$12570/Pa$\mathrm{\beta}$ vs H$_2\lambda$21218/Br$\mathrm{\gamma}$ \citep{Rodriguez-Ardila2005, Riffel2013a, Colina2015, Riffel2021, Holden2022} diagnostic diagrams, and/or the three-dimensional diagram (which makes use of line ratios and velocity dispersion) presented by \citet{DAgostino2019}.

Overall, despite the challenges that shock-ionisation and significant matter-bounded photoionisation components present to the transauroral line technique and the \citet{Revalski2021} photoionisation modelling, we argue that these methods are nonetheless more robust density diagnostics than the commonly used [SII](6717/6731) and [OII](3726/3729) ratios. In the case of matter-bounded photoionisation, the [SII]$\lambda\lambda$6717,6731 and [OII]$\lambda\lambda$3726,3729 lines arise from the same part of the ionisation structure of the cloud as the transauroral lines, meaning they face the same issues as the TR method, while the \citet{Revalski2021} modelling allows for higher-ionisation components, and therefore is a more accurate diagnostic of the overall cloud density. Furthermore, we have established here that using radiation-bounded photoionisation grids to measure the TR densities of shock-ionised gas incurs an uncertainty on the overall density that is much less than using lower-critical density line ratios for high-density ($n_e$\;\textgreater\;10$^{3}$) gas: in the case of NGC\;4151 (where there may be some contribution from shock-ionisation), the TR-derived densities are similar to those reported by \citet{Revalski2022} (see also \citealt{Crenshaw2015}), indicating that both methods still give more precise density determinations than traditional methods, despite some of their underlying assumptions potentially being incorrect.

\subsection{Comparison of the TR electron densities to other techniques}
\label{section: disc-density-tr}

Using the TR method, we find high electron densities in both objects: 4.00\;\textless\;log($n_e$[cm$^{-3}$])\;\textless\;4.75 in NGC\;1068 and 3.50\;\textless\;log($n_e$[cm$^{-3}$])\;\textless\;4.10 in NGC\;4151 (Section \ref{section: tr_diagnostics}; Table \ref{tab: oiii_models}). This agrees with the similarly high densities (\textgreater\;10$^3$\;cm$^{-3}$) derived from multi-component photoionisation modelling of both objects presented in \citet{Crenshaw2015} and \citet{Revalski2021} (see also \citet{Collins2009} and \citealt{Revalski2022}). Crucially, the derived densities from both techniques lie above the sensitivity range of the traditional [SII](6717/6731) and [OII](3726/2739) techniques, which are commonly used (either directly or as a basis for assumption) to derive electron densities in studies of the warm-ionised phase (e.g. \citealt{Nesvadba2006, Liu2013, Harrison2014, Fiore2017}), thus further supporting the need for robust warm-ionised gas electron density diagnostics such as the transauroral line technique and multi-component photoionisation modelling.

Considering the traditional [SII](6717/6731) ratio, \citet{Kraemer2000II} (using the same STIS dataset as used in this work) and \citet{Kakkad2018} and \citet{Mingozzi2019} (using IFU data) derived electron densities of $n_e\sim10^3$\;cm$^{-3}$ for the outflows in the NLR of NGC\;1068. These [SII]-derived densities are 1--1.5 orders of magnitude lower than those we find using the TR method, and are close to the upper limit of the density range for the [SII] ratio technique (Appendix \ref{section: appendix_critical_densities}: $n_\mathrm{crit}\sim10^{3.5}$\;cm$^{-3}$). This provides further evidence that, for gas of electron density $n_e$\;\textgreater\;10$^{3.5}$\;cm$^{-3}$, the [SII](6717/6731) ratio may underestimate the true electron density by more than an order of magnitude.

\subsection{The impact of the outflowing gas on the host galaxies}

Using densities derived from the transauroral line ratios, reddening-corrected recombination line fluxes and kinematics taken from previous modelling, we find mass outflow rates in the range 0.6\;\textless$\dot{M}_\mathrm{out}$\textless\;6.9\;M$_\odot$yr$^{-1}$, and coupling efficiencies in the range 1.1$\times10^{-3}$\;\textless$\epsilon_\mathrm{kin}$\;\textless\;0.99\;per\;cent (Table \ref{tab: energetics}). In many cases, our calculated coupling efficiencies are just above the lower limit required by models of the co-evolution of galaxies and their supermassive black holes (e.g. $\sim$0.5--10\;per\;cent: \citealt{DiMatteo2005, Springel2005, Hopkins2010}). It is important to note that there is likely more outflowing material within the bicones that is not covered by our slits (which are only 0.1\;arcseconds wide), and that comparisons between coupling efficiencies from models and observations are not straightforward (see \citealt{Harrison2018} for further discussion). To properly account for the impact of the warm ionised outflows, detailed studies that make use of robust density diagnostics, separate emission from the outflowing and quiescent gas and, importantly, cover the entire NLRs of both objects, are needed. Moreover, we highlight that assessments of \textit{all} gas phases --- not just the warm ionised phase --- are needed to robustly assess the \textit{total} impact of the AGN-driven outflows \citep{Cicone2018}, as the warm ionised gas may represent just a fraction of the total outflowing gas mass at a given radius (e.g. \citealt{RamosAlmeida2019, Holden2022}). Therefore, it is likely the the true coupling efficiencies of the total NLR outflows in NGC\;1068 and NGC\;4151 are higher than we calculate here.

\subsection{A tale of three Seyferts: NGC\;1068, NGC\;4151 and IC\;5063}

\begin{table*}
\begin{tabular}{llllll}
\hline
Object     & L$_\mathrm{bol}$ (W)    & L$_\mathrm{1.4\;GHz}$ (W\;Hz$^{-1}$) & P$_\mathrm{jet}$ (W)  & $\theta_\mathrm{jet}$ & Ionisation mechanism$^a$ \\ \hline
NGC\;1068 & 0.4--4.7$\times10^{38}$ & 2.3$\times10^{23}$   & 1.8$\times10^{36}$ \citep{GarciaBurillo2014}    & $\sim45^\circ$ & Matter-bounded photoionisation \\
NGC\;4151 & 1.4$\times10^{37}$ & 1.6$\times$10$^{22}$  & $\sim10^{35}$ \citep{Wang2011}    & $\sim36^\circ$ &  Photo- and/or shock-ionisation \\
IC\;5063  & 7.6$\times 10^{37}$     & 3$\times10^{23}$ & 10$^{37-38}$ \citep{Mukherjee2018} & $\sim5^\circ$ & Radiation-bounded photoionisation    \\ \hline
\end{tabular} \\
$^a$Determined with line ratios detected in slits along PA=202$^\circ$ (NGC\;1068), PA=70$^\circ$ (NGC\;4151) and PA=115$^\circ$ (IC\;5063). \\
\caption{Bolometric luminosities, 1.4\;GHz radio luminosities, jet powers (P$_\mathrm{jet}$), jet orientations with respect to the disk ($\theta_\mathrm{jet}$) and ionisation mechanisms detected along the radio axis for NGC\;1068, NGC\;4151 and IC\;5063.}
\label{tab: three_seyferts}
\end{table*}

Finally, using the results for the nearby Seyfert 2 IC\;5063 presented in \citet{Holden2022} along with the results for NGC\;1068 and NGC\;4151 that we present here, we can begin to construct a sample of nearby Seyferts with spatially-resolved, detailed studies of their NLR outflows.

IC\;5063 is a nearby ($z=0.01131$) early-type Seyfert 2 galaxy that is seen close to edge-on, with a radio jet propagating almost in the plane of the disk which drives fast (\mbox{$v_\mathrm{out}$\;\textgreater\;700\;km\,s$^{-1}$}) outflows \citep{Morganti1998, Oosterloo2000, Morganti2015, Mukherjee2018, Holden2022}. These outflows are seen in multiple gas phases, including warm ionised \citep{Morganti2007, Sharp2010, Congiu2017, Venturi2021, Holden2022}; neutral \citep{Morganti1998, Oosterloo2000}; warm molecular \citep{Tadhunter2014, Holden2022} and cold molecular \citep{Morganti2013, Morganti2015, Dasyra2016, Oosterloo2017}.

\begin{figure}
    \includegraphics[width=\linewidth]{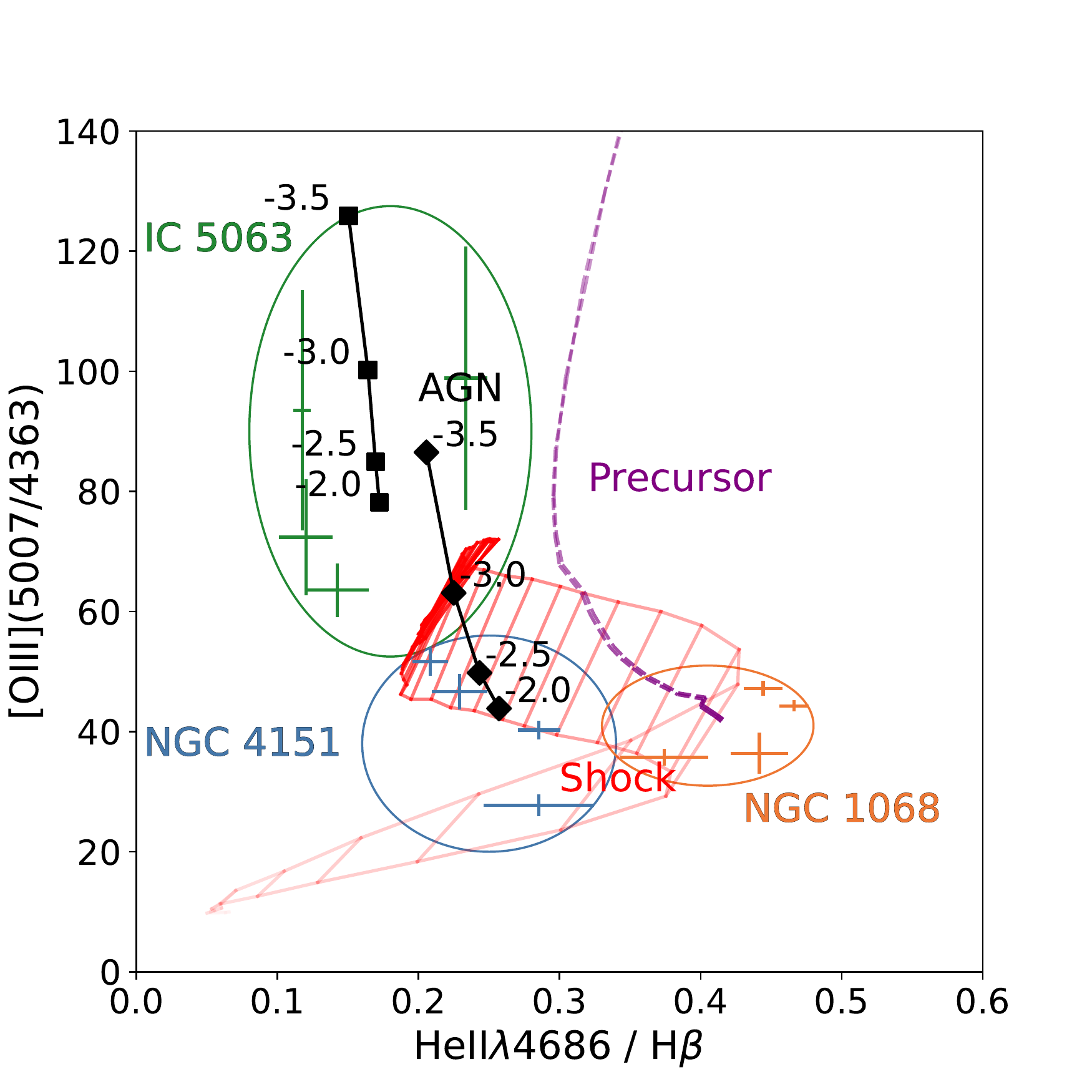}
    \caption{[OIII](5007/4363) vs HeII4686/H$\mathrm{\beta}$ diagnostic diagram (as in Figure \ref{fig: oiii_heii_hbeta_stis}) with line ratios measured from the STIS spectra of NGC\;1068 and NGC\;4151 (presented in this work) and from Xshooter spectra of IC\;5063 (presented in \citealt{Holden2022}). The AGN, shock and precursor models are the same as those described in Section \ref{section: mechanisms}. The three objects each show distinct ionisation conditions, hinting at the complex nature of the NLRs of Seyfert galaxies.}
    \label{fig: oiii_heii_hb_seyferts}
\end{figure}

In \citet{Holden2022}, we presented evidence that both the outflowing and quiescent warm ionised gas in IC\;5063 has dominant AGN-photoionisation --- even though the outflows show clear signatures of shock acceleration --- and that the different outflow phases may represent a post-shock cooling sequence. We interpreted this situation as the pre-shock gas being AGN-photoionised, and the closest post-shock gas to the AGN kept in an ionised state by photoionisation. In Figure \ref{fig: oiii_heii_hb_seyferts}, we add the [OIII](5007/4363) and HeII4686/H$\mathrm{\beta}$ ratios for IC\;5063 from \citet{Holden2022} to the diagnostic diagram presented in this work (Figure \ref{fig: oiii_heii_hbeta_stis}). Furthermore, we present [OII](7319+7331)/[OIII]$\lambda$5007 and [SII](4068+4076)/H$\mathrm{\beta}$ ratios for IC\;5063 (alongside NGC\;1068 and NGC\;4151) in Appendix \ref{appendix: tr-origin} (Figure \ref{fig: tr_oiii_hbeta}) --- determined using the dataset from \citet{Holden2022} --- and find that they are consistent with radiation-bounded AGN-photoionisation with gas densities \mbox{10$^3$\;cm$^{-3}$\;\textless\;$n_e$\;\textless\;10$^4$\;cm$^{-3}$} and ionisation parameters in the range \mbox{$-3$\;\textless\;log\;$U$\;\textless\;$-2$}, in agreement with the values determined in \citet{Holden2022}. 

It is interesting that the overall differences in ionisation conditions \textit{between} the three galaxies are significantly larger than the range of ionisation conditions \textit{within} the galaxies. Our small sample thus shows three distinct cases in the three objects: radiation-bounded AGN-photoionisation in IC\;5063, matter-bounded AGN-photoionisation in NGC\;1068, and shock-ionisation or radiation-bounded AGN photoionisation with a relatively flat spectral index and higher ionisation parameters in NGC\;4151 --- despite all being classified as Seyferts, the details of the ionisation mechanisms in the objects vary greatly. This is particularly interesting considering that in all three objects, the outflows detected along the radio axes appear to be spatially-confined to the radio structures (i.e. the outflows do not extend beyond the radio lobes in the NLRs). As argued in Section \ref{section: disc-ionisation}, this is consistent with shock-acceleration, although it does not rule out radiative-acceleration. If the outflows in IC\;5063, NGC\;1068 and NGC\;4151 are shock-accelerated, then this would highlight the importance of not deriving information regarding the outflow acceleration mechanisms based solely on the ionisation/excitation mechanisms or kinematics of the gas in NLRs: a full account, involving detailed multi-wavelength observations with multiple diagnostics, is required to properly evaluate the relative contributions of different mechanisms. 

Despite evidence that the outflows in all three objects are being driven by the radio jet, the densities of the outflowing gas differ by more than an order of magnitude: for IC\;5063, we found that the outflowing gas has densities in the range 3.17\;\textless\;log($n_e$[cm$^{-3}$])\;\textless\;3.43, while for NGC\;1068 and NGC\;4151 we find densities in the ranges 4.00\;\textless\;log($n_e$[cm$^{-3}$])\;\textless\;4.75 and 3.50\;\textless\;log($n_e$[cm$^{-3}$])\;\textless\;4.10 respectively (Table \ref{tab: oiii_models}). The reason for this may simply be due to different pre-shock gas densities in the different objects (assuming the outflows in all three are shock-accelerated): for IC\;5063 the pre-shock density is 2.1\;\textless\;log($n_e$[cm$^{-3}$])\;\textless\;2.7, however without higher velocity resolution spectra, we are unable to determine the quiescent gas densities in NGC\;1068 and NGC\;4151. In addition, the differing post-shock densities in the three Seyferts may be due to different cooling conditions behind the shock front. Standard shock-jump conditions predict a compression factor of $\sim$4, however this may be much higher ($\sim$100) if the post-shock gas has cooled in pressure equilibrium \citep{Sutherland2017, Santoro2018}.

Moreover, all three objects have low-to-intermediate radio luminosities (1.6$\times10^{22}$\;\textless\;L$_\mathrm{1.4\;GHz}$\;\textless\;3.0$\times10^{23}$\;W\;Hz$^{-1}$; Table \ref{tab: three_seyferts}) --- again, if the outflows in these Seyferts are shock-accelerated, then this would reinforce the importance of jet-driven shocks as a feedback mechanism in the inner regions of galaxies, even at lower radio luminosities, in agreement with a statistical study of nearby AGN presented by \citet{Mullaney2013}. Furthermore, the radio jets in NGC\;1068 and NGC\;4151 are oriented out of galactic disks by $\sim45^\circ$ and $\sim36^\circ$ respectively, unlike IC\;5063 in which the jet propagates almost directly into the plane of the disk. Therefore, at least within the central few hundred parsec of the AGN, this would show that inclined jets can still have an impact on the kinematics and ionisation of the NLR, as predicted by recent relativistic hydrodynamic simulations \citep{Mukherjee2018, Meenakshi2022}, which show that a jet inclined $\theta_\mathrm{jet}\sim45^\circ$ to the galaxy's disk may have a significant effect on the kinematics, density and temperature of the gas within the central few kpc (albeit less so than a jet inclined in the plane of the disk, such as is the case in IC\;5063). Similar hydrodynamic simulations, specifically tailored to the situations in NGC\;1068 and NGC\;4151, could thus be used to quantify the impact of their radio jets on the star-forming gas in their NLRs, as well as the impact of inclined kpc-scale jets in general.

Ultimately, further observations of NGC\;1068 and NGC\;4151 are required to decisively determine the outflow acceleration mechanism(s). Namely, wide wavelength coverage spectroscopy (to make available a range of diagnostics) with sufficient velocity resolution to kinematically discriminate between outflowing (post-shock?) and quiescent (pre-shock?) gas.

\section{Conclusions}
\label{section: conclusions}

By analysing archival HST/STIS spectra taken along the radio axes of the inner few hundred parsecs of the NLR of the prototypical Seyfert galaxies NGC\;1068 and NGC\;4151, we have found the following.

\begin{itemize}
    \item Using the transauroral line ratio technique, we derive spatially-resolved electron densities of 4.00\;\textless\;log$_{10}n_e$[cm$^{-3}$])\;\textless\;4.75 for NGC\;1068 and 3.60\;\textless\;log$_{10}n_e$[cm$^{-3}$])\;\textless\;4.10 for NGC\;4151. These values are an order of magnitude above those commonly reported and assumed based on traditional density estimates, but are in agreement with the results from alternative diagnostics such as multi-component photoionisation modelling. Overall, our results provide further motivation for the use of the transauroral lines in deriving electron densities of AGN-driven outflows.
    \item The measured emission-line ratios for the warm ionised gas are consistent with the dominant ionisation mechanisms being matter-bounded AGN-photoionisation in NGC\;1068, and shock-ionisation and/or radiation-bounded AGN-photoionisation with a relatively flat spectral index (and/or higher ionisation parameters and lower metallicities) in NGC\;4151.
    \item Along the radio axes, the outflows in the northeastern cones of both objects have similar spatial extents to the radio structures --- this is consistent with the outflows in their NLRs being shock-accelerated by the radio jets and reionised by radiation from the AGN, although it does not rule out radiative acceleration.
    \item Applying the transauroral line technique to gas that has dominant shock-ionisation may incur an uncertainty on the derived electron densities by up to $\pm0.38$ orders of magnitude, which is still far below the potential order-of-magnitude error incurred when using techniques which are not sensitive to higher density gas. However, care must still be taken when using detailed density diagnostic techniques, as the ionisation mechanism of the gas may alter the results. Therefore, robust ionisation-mechanism diagnostics should be used to verify the validity of the density measurements.
    \item Finally, by combining our findings with those for the nearby Seyfert 2 galaxy IC\;5063, we find that the ionisation mechanisms and outflow conditions along the radio axes in the central few hundred parsecs vary significantly between the different objects. Thus overall, our study highlights the necessity of care when deriving information about outflow acceleration mechanisms from the ionisation of the gas, and the need for robust ionisation-mechanism diagnostics with detailed observations.
\end{itemize}

\section*{Acknowledgements}
We thank the anonymous referee for their helpful comments and suggestions, which improved the clarity of this manuscript. LRH and CNT acknowledge support from STFC. Based on observations made with the NASA/ESA Hubble Space Telescope, and obtained from the Hubble Legacy Archive, which is a collaboration between the Space Telescope Science Institute (STScI/NASA), the Space Telescope European Coordinating Facility (ST-ECF/ESA) and the Canadian Astronomy Data Centre (CADC/NRC/CSA). This research has made use of the NASA/IPAC Infrared Science Archive, which is funded by the National Aeronautics and Space Administration and operated by the California Institute of Technology. This work makes use of the Starlink software \citep{Currie2014}, which is currently supported by the East Asian Observatory. For the purposes of open access, the authors have applied a Creative Commons Attribution (CC BY) licence to any Author Accepted Manuscript Arising.

\section*{Data Availability}

The data used in this report is available from the Hubble Legacy Archive (HLA) (\url{https://hla.stsci.edu/hlaview.html}) with proposal IDs GTO:5754 (PI Ford) and GTO:5124 (PI Ford) for the HST/WFPC2 [OIII] imaging, and proposal IDs GTO:7573 (PI Kraemer) and GTO:7569 (PI Hutchings) for the HST/STIS spectra.



\bibliographystyle{mnras}
\bibliography{stis_seyferts} 

\begin{thebibliography}{}
\makeatletter
\relax
\def\mn@urlcharsother{\let\do\@makeother \do\$\do\&\do\#\do\^\do\_\do\%\do\~}
\def\mn@doi{\begingroup\mn@urlcharsother \@ifnextchar [ {\mn@doi@}
  {\mn@doi@[]}}
\def\mn@doi@[#1]#2{\def\@tempa{#1}\ifx\@tempa\@empty \href
  {http://dx.doi.org/#2} {doi:#2}\else \href {http://dx.doi.org/#2} {#1}\fi
  \endgroup}
\def\mn@eprint#1#2{\mn@eprint@#1:#2::\@nil}
\def\mn@eprint@arXiv#1{\href {http://arxiv.org/abs/#1} {{\tt arXiv:#1}}}
\def\mn@eprint@dblp#1{\href {http://dblp.uni-trier.de/rec/bibtex/#1.xml}
  {dblp:#1}}
\def\mn@eprint@#1:#2:#3:#4\@nil{\def\@tempa {#1}\def\@tempb {#2}\def\@tempc
  {#3}\ifx \@tempc \@empty \let \@tempc \@tempb \let \@tempb \@tempa \fi \ifx
  \@tempb \@empty \def\@tempb {arXiv}\fi \@ifundefined
  {mn@eprint@\@tempb}{\@tempb:\@tempc}{\expandafter \expandafter \csname
  mn@eprint@\@tempb\endcsname \expandafter{\@tempc}}}

\bibitem[\protect\citeauthoryear{{Allen}, {Groves}, {Dopita}, {Sutherland}  \&
  {Kewley}}{{Allen} et~al.}{2008}]{Allen2008}
{Allen} M.~G.,  {Groves} B.~A.,  {Dopita} M.~A.,  {Sutherland} R.~S.,
  {Kewley} L.~J.,  2008, \mn@doi [\apjs] {10.1086/589652}, \href
  {https://ui.adsabs.harvard.edu/abs/2008ApJS..178...20A} {178, 20}

\bibitem[\protect\citeauthoryear{{Alonso-Herrero} et~al.,}{{Alonso-Herrero}
  et~al.}{2011}]{AlonsoHerrero2011}
{Alonso-Herrero} A.,  et~al., 2011, \mn@doi [\apj]
  {10.1088/0004-637X/736/2/82}, \href
  {https://ui.adsabs.harvard.edu/abs/2011ApJ...736...82A} {736, 82}

\bibitem[\protect\citeauthoryear{{Antonucci} \& {Miller}}{{Antonucci} \&
  {Miller}}{1985}]{Antonucci1985}
{Antonucci} R.~R.~J.,  {Miller} J.~S.,  1985, \mn@doi [\apj] {10.1086/163559},
  \href {https://ui.adsabs.harvard.edu/abs/1985ApJ...297..621A} {297, 621}

\bibitem[\protect\citeauthoryear{{Astropy Collaboration} et~al.,}{{Astropy
  Collaboration} et~al.}{2013}]{AstropyCollaboration2013}
{Astropy Collaboration} et~al., 2013, \mn@doi [A\&A]
  {10.1051/0004-6361/201322068}, \href
  {https://ui.adsabs.harvard.edu/abs/2013A&A...558A..33A} {558, A33}

\bibitem[\protect\citeauthoryear{{Astropy Collaboration} et~al.,}{{Astropy
  Collaboration} et~al.}{2018}]{AstropyCollaboration2018}
{Astropy Collaboration} et~al., 2018, \mn@doi [ApJ] {10.3847/1538-3881/aabc4f},
  \href {https://ui.adsabs.harvard.edu/abs/2018AJ....156..123A} {156, 123}

\bibitem[\protect\citeauthoryear{{Audibert} et~al.,}{{Audibert}
  et~al.}{2023}]{Audibert2023}
{Audibert} A.,  et~al., 2023, \mn@doi [\aap] {10.1051/0004-6361/202345964},
  \href {https://ui.adsabs.harvard.edu/abs/2023A&A...671L..12A} {671, L12}

\bibitem[\protect\citeauthoryear{{Axon}, {Marconi}, {Capetti}, {Macchetto},
  {Schreier}  \& {Robinson}}{{Axon} et~al.}{1998}]{Axon1998}
{Axon} D.~J.,  {Marconi} A.,  {Capetti} A.,  {Macchetto} F.~D.,  {Schreier} E.,
    {Robinson} A.,  1998, \mn@doi [\apjl] {10.1086/311249}, \href
  {https://ui.adsabs.harvard.edu/abs/1998ApJ...496L..75A} {496, L75}

\bibitem[\protect\citeauthoryear{{Baldwin}, {Phillips}  \&
  {Terlevich}}{{Baldwin} et~al.}{1981}]{Baldwin1981}
{Baldwin} J.~A.,  {Phillips} M.~M.,   {Terlevich} R.,  1981, \mn@doi [\pasp]
  {10.1086/130766}, \href
  {https://ui.adsabs.harvard.edu/abs/1981PASP...93....5B} {93, 5}

\bibitem[\protect\citeauthoryear{{Barbosa}, {Storchi-Bergmann}, {McGregor},
  {Vale}  \& {Rogemar Riffel}}{{Barbosa} et~al.}{2014}]{Barbosa2014}
{Barbosa} F.~K.~B.,  {Storchi-Bergmann} T.,  {McGregor} P.,  {Vale} T.~B.,
  {Rogemar Riffel} A.,  2014, \mn@doi [\mnras] {10.1093/mnras/stu1637}, \href
  {https://ui.adsabs.harvard.edu/abs/2014MNRAS.445.2353B} {445, 2353}

\bibitem[\protect\citeauthoryear{{Baron} \& {Netzer}}{{Baron} \&
  {Netzer}}{2019}]{Baron2019b}
{Baron} D.,  {Netzer} H.,  2019, \mn@doi [MNRAS] {10.1093/mnras/stz1070}, \href
  {https://ui.adsabs.harvard.edu/abs/2019MNRAS.486.4290B} {486, 4290}

\bibitem[\protect\citeauthoryear{{Baron}, {Netzer}, {Poznanski}, {Prochaska}
  \& {F{\"o}rster Schreiber}}{{Baron} et~al.}{2017}]{Baron2017}
{Baron} D.,  {Netzer} H.,  {Poznanski} D.,  {Prochaska} J.~X.,   {F{\"o}rster
  Schreiber} N.~M.,  2017, \mn@doi [\mnras] {10.1093/mnras/stx1329}, \href
  {https://ui.adsabs.harvard.edu/abs/2017MNRAS.470.1687B} {470, 1687}

\bibitem[\protect\citeauthoryear{{Binette}, {Wilson}  \&
  {Storchi-Bergmann}}{{Binette} et~al.}{1996}]{Binette1996}
{Binette} L.,  {Wilson} A.~S.,   {Storchi-Bergmann} T.,  1996, \aap, \href
  {https://ui.adsabs.harvard.edu/abs/1996A&A...312..365B} {312, 365}

\bibitem[\protect\citeauthoryear{{Boyce}, {Menzel}  \& {Payne}}{{Boyce}
  et~al.}{1933}]{Boyce1933}
{Boyce} J.~C.,  {Menzel} D.~H.,   {Payne} C.~H.,  1933, \mn@doi [Proceedings of
  the National Academy of Science] {10.1073/pnas.19.6.581}, \href
  {https://ui.adsabs.harvard.edu/abs/1933PNAS...19..581B} {19, 581}

\bibitem[\protect\citeauthoryear{{Capetti}, {Macchetto}  \&
  {Lattanzi}}{{Capetti} et~al.}{1997}]{Capetti1997}
{Capetti} A.,  {Macchetto} F.~D.,   {Lattanzi} M.~G.,  1997, \mn@doi [\apjl]
  {10.1086/310498}, \href
  {https://ui.adsabs.harvard.edu/abs/1997ApJ...476L..67C} {476, L67}

\bibitem[\protect\citeauthoryear{{Cardelli}, {Clayton}  \& {Mathis}}{{Cardelli}
  et~al.}{1989}]{Cardelli1989}
{Cardelli} J.~A.,  {Clayton} G.~C.,   {Mathis} J.~S.,  1989, \mn@doi [ApJ]
  {10.1086/167900}, \href
  {https://ui.adsabs.harvard.edu/abs/1989ApJ...345..245C} {345, 245}

\bibitem[\protect\citeauthoryear{{Carral}, {Turner}  \& {Ho}}{{Carral}
  et~al.}{1990}]{Carral1990}
{Carral} P.,  {Turner} J.~L.,   {Ho} P. T.~P.,  1990, \mn@doi [\apj]
  {10.1086/169280}, \href
  {https://ui.adsabs.harvard.edu/abs/1990ApJ...362..434C} {362, 434}

\bibitem[\protect\citeauthoryear{{Cecil}, {Bland}  \& {Tully}}{{Cecil}
  et~al.}{1990}]{Cecil1990}
{Cecil} G.,  {Bland} J.,   {Tully} R.~B.,  1990, \mn@doi [ApJ]
  {10.1086/168742}, \href
  {https://ui.adsabs.harvard.edu/abs/1990ApJ...355...70C} {355, 70}

\bibitem[\protect\citeauthoryear{{Cicone} et~al.,}{{Cicone}
  et~al.}{2014}]{Cicone2014}
{Cicone} C.,  et~al., 2014, \mn@doi [A\&A] {10.1051/0004-6361/201322464}, \href
  {https://ui.adsabs.harvard.edu/abs/2014A&A...562A..21C} {562, A21}

\bibitem[\protect\citeauthoryear{{Cicone}, {Brusa}, {Ramos Almeida}, {Cresci},
  {Husemann}  \& {Mainieri}}{{Cicone} et~al.}{2018}]{Cicone2018}
{Cicone} C.,  {Brusa} M.,  {Ramos Almeida} C.,  {Cresci} G.,  {Husemann} B.,
  {Mainieri} V.,  2018, \mn@doi [Nat. Astron.] {10.1038/s41550-018-0406-3},
  \href {https://ui.adsabs.harvard.edu/abs/2018NatAs...2..176C} {2, 176}

\bibitem[\protect\citeauthoryear{{Colina} et~al.,}{{Colina}
  et~al.}{2015}]{Colina2015}
{Colina} L.,  et~al., 2015, \mn@doi [\aap] {10.1051/0004-6361/201425567}, \href
  {https://ui.adsabs.harvard.edu/abs/2015A&A...578A..48C} {578, A48}

\bibitem[\protect\citeauthoryear{{Collins}, {Kraemer}, {Crenshaw}, {Bruhweiler}
   \& {Mel{\'e}ndez}}{{Collins} et~al.}{2009}]{Collins2009}
{Collins} N.~R.,  {Kraemer} S.~B.,  {Crenshaw} D.~M.,  {Bruhweiler} F.~C.,
  {Mel{\'e}ndez} M.,  2009, \mn@doi [\apj] {10.1088/0004-637X/694/2/765}, \href
  {https://ui.adsabs.harvard.edu/abs/2009ApJ...694..765C} {694, 765}

\bibitem[\protect\citeauthoryear{{Congiu} et~al.,}{{Congiu}
  et~al.}{2017}]{Congiu2017}
{Congiu} E.,  et~al., 2017, \mn@doi [\mnras] {10.1093/mnras/stx1628}, \href
  {https://ui.adsabs.harvard.edu/abs/2017MNRAS.471..562C} {471, 562}

\bibitem[\protect\citeauthoryear{{Crenshaw} \& {Kraemer}}{{Crenshaw} \&
  {Kraemer}}{2000a}]{Crenshaw2000b}
{Crenshaw} D.~M.,  {Kraemer} S.~B.,  2000a, \mn@doi [\apj] {10.1086/308570},
  \href {https://ui.adsabs.harvard.edu/abs/2000ApJ...532..247C} {532, 247}

\bibitem[\protect\citeauthoryear{{Crenshaw} \& {Kraemer}}{{Crenshaw} \&
  {Kraemer}}{2000b}]{Crenshaw2000a}
{Crenshaw} D.~M.,  {Kraemer} S.~B.,  2000b, \mn@doi [\apjl] {10.1086/312581},
  \href {https://ui.adsabs.harvard.edu/abs/2000ApJ...532L.101C} {532, L101}

\bibitem[\protect\citeauthoryear{{Crenshaw} \& {Kraemer}}{{Crenshaw} \&
  {Kraemer}}{2000c}]{Crenshaw2000_N1068}
{Crenshaw} D.~M.,  {Kraemer} S.~B.,  2000c, \mn@doi [\apjl] {10.1086/312581},
  \href {https://ui.adsabs.harvard.edu/abs/2000ApJ...532L.101C} {532, L101}

\bibitem[\protect\citeauthoryear{{Crenshaw} et~al.,}{{Crenshaw}
  et~al.}{2000}]{Crenshaw2000_N4151}
{Crenshaw} D.~M.,  et~al., 2000, \mn@doi [\aj] {10.1086/301574}, \href
  {https://ui.adsabs.harvard.edu/abs/2000AJ....120.1731C} {120, 1731}

\bibitem[\protect\citeauthoryear{{Crenshaw}, {Fischer}, {Kraemer}  \&
  {Schmitt}}{{Crenshaw} et~al.}{2015}]{Crenshaw2015}
{Crenshaw} D.~M.,  {Fischer} T.~C.,  {Kraemer} S.~B.,   {Schmitt} H.~R.,  2015,
  \mn@doi [\apj] {10.1088/0004-637X/799/1/83}, \href
  {https://ui.adsabs.harvard.edu/abs/2015ApJ...799...83C} {799, 83}

\bibitem[\protect\citeauthoryear{{Currie}, {Berry}, {Jenness}, {Gibb}, {Bell}
  \& {Draper}}{{Currie} et~al.}{2014}]{Currie2014}
{Currie} M.~J.,  {Berry} D.~S.,  {Jenness} T.,  {Gibb} A.~G.,  {Bell} G.~S.,
  {Draper} P.~W.,  2014, in {Manset} N.,  {Forshay} P.,  eds,  Astronomical
  Society of the Pacific Conference Series Vol. 485, Astronomical Data Analysis
  Software and Systems XXIII. p.~391

\bibitem[\protect\citeauthoryear{{D'Agostino}, {Kewley}, {Groves}, {Medling},
  {Dopita}  \& {Thomas}}{{D'Agostino} et~al.}{2019}]{DAgostino2019}
{D'Agostino} J.~J.,  {Kewley} L.~J.,  {Groves} B.~A.,  {Medling} A.,  {Dopita}
  M.~A.,   {Thomas} A.~D.,  2019, \mn@doi [\mnras] {10.1093/mnrasl/slz028},
  \href {https://ui.adsabs.harvard.edu/abs/2019MNRAS.485L..38D} {485, L38}

\bibitem[\protect\citeauthoryear{{Das} et~al.,}{{Das} et~al.}{2005}]{Das2005}
{Das} V.,  et~al., 2005, \mn@doi [\aj] {10.1086/432255}, \href
  {https://ui.adsabs.harvard.edu/abs/2005AJ....130..945D} {130, 945}

\bibitem[\protect\citeauthoryear{{Das}, {Crenshaw}, {Kraemer}  \& {Deo}}{{Das}
  et~al.}{2006}]{Das2006}
{Das} V.,  {Crenshaw} D.~M.,  {Kraemer} S.~B.,   {Deo} R.~P.,  2006, \mn@doi
  [\aj] {10.1086/504899}, \href
  {https://ui.adsabs.harvard.edu/abs/2006AJ....132..620D} {132, 620}

\bibitem[\protect\citeauthoryear{{Dasyra}, {Combes}, {Oosterloo}, {Oonk},
  {Morganti}, {Salom{\'e}}  \& {Vlahakis}}{{Dasyra} et~al.}{2016}]{Dasyra2016}
{Dasyra} K.~M.,  {Combes} F.,  {Oosterloo} T.,  {Oonk} J.~B.~R.,  {Morganti}
  R.,  {Salom{\'e}} P.,   {Vlahakis} N.,  2016, \mn@doi [A\&A]
  {10.1051/0004-6361/201629689}, \href
  {https://ui.adsabs.harvard.edu/abs/2016A&A...595L...7D} {595, L7}

\bibitem[\protect\citeauthoryear{{Dav{\'e}}, {Angl{\'e}s-Alc{\'a}zar},
  {Narayanan}, {Li}, {Rafieferantsoa}  \& {Appleby}}{{Dav{\'e}}
  et~al.}{2019}]{Dave2019}
{Dav{\'e}} R.,  {Angl{\'e}s-Alc{\'a}zar} D.,  {Narayanan} D.,  {Li} Q.,
  {Rafieferantsoa} M.~H.,   {Appleby} S.,  2019, \mn@doi [\mnras]
  {10.1093/mnras/stz937}, \href
  {https://ui.adsabs.harvard.edu/abs/2019MNRAS.486.2827D} {486, 2827}

\bibitem[\protect\citeauthoryear{{Davies} et~al.,}{{Davies}
  et~al.}{2020}]{Davies2020}
{Davies} R.,  et~al., 2020, \mn@doi [MNRAS] {10.1093/mnras/staa2413}, \href
  {https://ui.adsabs.harvard.edu/abs/2020MNRAS.498.4150D} {498, 4150}

\bibitem[\protect\citeauthoryear{{Di Matteo}, {Springel}  \& {Hernquist}}{{Di
  Matteo} et~al.}{2005}]{DiMatteo2005}
{Di Matteo} T.,  {Springel} V.,   {Hernquist} L.,  2005, \mn@doi [Nat]
  {10.1038/nature03335}, \href
  {https://ui.adsabs.harvard.edu/abs/2005Natur.433..604D} {433, 604}

\bibitem[\protect\citeauthoryear{{Dopita} \& {Sutherland}}{{Dopita} \&
  {Sutherland}}{1995}]{Dopita1995}
{Dopita} M.~A.,  {Sutherland} R.~S.,  1995, \mn@doi [\apj] {10.1086/176596},
  \href {https://ui.adsabs.harvard.edu/abs/1995ApJ...455..468D} {455, 468}

\bibitem[\protect\citeauthoryear{{Dubois}, {Peirani}, {Pichon}, {Devriendt},
  {Gavazzi}, {Welker}  \& {Volonteri}}{{Dubois} et~al.}{2016}]{Dubois2016}
{Dubois} Y.,  {Peirani} S.,  {Pichon} C.,  {Devriendt} J.,  {Gavazzi} R.,
  {Welker} C.,   {Volonteri} M.,  2016, \mn@doi [\mnras]
  {10.1093/mnras/stw2265}, \href
  {https://ui.adsabs.harvard.edu/abs/2016MNRAS.463.3948D} {463, 3948}

\bibitem[\protect\citeauthoryear{{Evans}, {Ford}, {Kinney}, {Antonucci},
  {Armus}  \& {Caganoff}}{{Evans} et~al.}{1991}]{Evans1991}
{Evans} I.~N.,  {Ford} H.~C.,  {Kinney} A.~L.,  {Antonucci} R.~R.~J.,  {Armus}
  L.,   {Caganoff} S.,  1991, \mn@doi [\apjl] {10.1086/185951}, \href
  {https://ui.adsabs.harvard.edu/abs/1991ApJ...369L..27E} {369, L27}

\bibitem[\protect\citeauthoryear{{Fabian}}{{Fabian}}{1999}]{Fabian1999}
{Fabian} A.~C.,  1999, \mn@doi [MNRAS] {10.1046/j.1365-8711.1999.03017.x},
  \href {https://ui.adsabs.harvard.edu/abs/1999MNRAS.308L..39F} {308, L39}

\bibitem[\protect\citeauthoryear{{Ferland} \& {Netzer}}{{Ferland} \&
  {Netzer}}{1983}]{Ferland1983}
{Ferland} G.~J.,  {Netzer} H.,  1983, \mn@doi [\apj] {10.1086/160577}, \href
  {https://ui.adsabs.harvard.edu/abs/1983ApJ...264..105F} {264, 105}

\bibitem[\protect\citeauthoryear{{Ferland} et~al.,}{{Ferland}
  et~al.}{2017}]{Ferland2017}
{Ferland} G.~J.,  et~al., 2017, RMXAA, \href
  {https://ui.adsabs.harvard.edu/abs/2017RMxAA..53..385F} {53, 385}

\bibitem[\protect\citeauthoryear{{Ferrarese} \& {Merritt}}{{Ferrarese} \&
  {Merritt}}{2000}]{Ferrarese2000}
{Ferrarese} L.,  {Merritt} D.,  2000, \mn@doi [ApJ] {10.1086/312838}, \href
  {https://ui.adsabs.harvard.edu/abs/2000ApJ...539L...9F} {539, L9}

\bibitem[\protect\citeauthoryear{{Fiore} et~al.,}{{Fiore}
  et~al.}{2017}]{Fiore2017}
{Fiore} F.,  et~al., 2017, \mn@doi [A\&A] {10.1051/0004-6361/201629478}, \href
  {https://ui.adsabs.harvard.edu/abs/2017A&A...601A.143F} {601, A143}

\bibitem[\protect\citeauthoryear{{Fischer} et~al.,}{{Fischer}
  et~al.}{2017}]{Fischer2017}
{Fischer} T.~C.,  et~al., 2017, \mn@doi [\apj] {10.3847/1538-4357/834/1/30},
  \href {https://ui.adsabs.harvard.edu/abs/2017ApJ...834...30F} {834, 30}

\bibitem[\protect\citeauthoryear{{Foreman-Mackey}, {Hogg}, {Lang}  \&
  {Goodman}}{{Foreman-Mackey} et~al.}{2013}]{FormanMackey2013}
{Foreman-Mackey} D.,  {Hogg} D.~W.,  {Lang} D.,   {Goodman} J.,  2013, \mn@doi
  [\pasp] {10.1086/670067}, \href
  {https://ui.adsabs.harvard.edu/abs/2013PASP..125..306F} {125, 306}

\bibitem[\protect\citeauthoryear{{Fosbury}, {Mebold}, {Goss}  \&
  {Dopita}}{{Fosbury} et~al.}{1978}]{Fosbury1978}
{Fosbury} R.~A.~E.,  {Mebold} U.,  {Goss} W.~M.,   {Dopita} M.~A.,  1978,
  \mn@doi [\mnras] {10.1093/mnras/183.4.549}, \href
  {https://ui.adsabs.harvard.edu/abs/1978MNRAS.183..549F} {183, 549}

\bibitem[\protect\citeauthoryear{{Gallimore}, {Baum}, {O'Dea}  \&
  {Pedlar}}{{Gallimore} et~al.}{1996}]{Gallimore1996}
{Gallimore} J.~F.,  {Baum} S.~A.,  {O'Dea} C.~P.,   {Pedlar} A.,  1996, \mn@doi
  [\apj] {10.1086/176798}, \href
  {https://ui.adsabs.harvard.edu/abs/1996ApJ...458..136G} {458, 136}

\bibitem[\protect\citeauthoryear{{Garc{\'\i}a-Burillo}
  et~al.,}{{Garc{\'\i}a-Burillo} et~al.}{2014}]{GarciaBurillo2014}
{Garc{\'\i}a-Burillo} S.,  et~al., 2014, \mn@doi [\aap]
  {10.1051/0004-6361/201423843}, \href
  {https://ui.adsabs.harvard.edu/abs/2014A&A...567A.125G} {567, A125}

\bibitem[\protect\citeauthoryear{{Garc{\'\i}a-Burillo}
  et~al.,}{{Garc{\'\i}a-Burillo} et~al.}{2019}]{GarciaBurillo2019}
{Garc{\'\i}a-Burillo} S.,  et~al., 2019, \mn@doi [\aap]
  {10.1051/0004-6361/201936606}, \href
  {https://ui.adsabs.harvard.edu/abs/2019A&A...632A..61G} {632, A61}

\bibitem[\protect\citeauthoryear{{Gebhardt} et~al.,}{{Gebhardt}
  et~al.}{2000}]{Gebhardt2000}
{Gebhardt} K.,  et~al., 2000, \mn@doi [ApJ] {10.1086/312840}, \href
  {https://ui.adsabs.harvard.edu/abs/2000ApJ...539L..13G} {539, L13}

\bibitem[\protect\citeauthoryear{{Gonz{\'a}lez Delgado}, {Leitherer}  \&
  {Heckman}}{{Gonz{\'a}lez Delgado} et~al.}{1999}]{GonzalezDelgado1999}
{Gonz{\'a}lez Delgado} R.~M.,  {Leitherer} C.,   {Heckman} T.~M.,  1999,
  \mn@doi [\apjs] {10.1086/313285}, \href
  {https://ui.adsabs.harvard.edu/abs/1999ApJS..125..489G} {125, 489}

\bibitem[\protect\citeauthoryear{{Gravity Collaboration} et~al.,}{{Gravity
  Collaboration} et~al.}{2020}]{Gravity2020}
{Gravity Collaboration} et~al., 2020, \mn@doi [\aap]
  {10.1051/0004-6361/201936255}, \href
  {https://ui.adsabs.harvard.edu/abs/2020A&A...634A...1G} {634, A1}

\bibitem[\protect\citeauthoryear{Harris et~al.,}{Harris
  et~al.}{2020}]{Harris2020}
Harris C.~R.,  et~al., 2020, \mn@doi [Nat.] {10.1038/s41586-020-2649-2}, 585,
  357

\bibitem[\protect\citeauthoryear{{Harrison}, {Alexander}, {Mullaney}  \&
  {Swinbank}}{{Harrison} et~al.}{2014}]{Harrison2014}
{Harrison} C.~M.,  {Alexander} D.~M.,  {Mullaney} J.~R.,   {Swinbank} A.~M.,
  2014, \mn@doi [MNRAS] {10.1093/mnras/stu515}, \href
  {https://ui.adsabs.harvard.edu/abs/2014MNRAS.441.3306H} {441, 3306}

\bibitem[\protect\citeauthoryear{{Harrison}, {Costa}, {Tadhunter},
  {Fl{\"u}tsch}, {Kakkad}, {Perna}  \& {Vietri}}{{Harrison}
  et~al.}{2018}]{Harrison2018}
{Harrison} C.~M.,  {Costa} T.,  {Tadhunter} C.~N.,  {Fl{\"u}tsch} A.,  {Kakkad}
  D.,  {Perna} M.,   {Vietri} G.,  2018, \mn@doi [Nat. Astron]
  {10.1038/s41550-018-0403-6}, \href
  {https://ui.adsabs.harvard.edu/abs/2018NatAs...2..198H} {2, 198}

\bibitem[\protect\citeauthoryear{{Holden}, {Tadhunter}, {Morganti}  \&
  {Oosterloo}}{{Holden} et~al.}{2023}]{Holden2022}
{Holden} L.~R.,  {Tadhunter} C.~N.,  {Morganti} R.,   {Oosterloo} T.,  2023,
  \mn@doi [\mnras] {10.1093/mnras/stad123}, \href
  {https://ui.adsabs.harvard.edu/abs/2023MNRAS.520.1848H} {520, 1848}

\bibitem[\protect\citeauthoryear{{Holt}, {Tadhunter}, {Morganti}  \&
  {Emonts}}{{Holt} et~al.}{2011}]{Holt2011}
{Holt} J.,  {Tadhunter} C.~N.,  {Morganti} R.,   {Emonts} B.~H.~C.,  2011,
  \mn@doi [MNRAS] {10.1111/j.1365-2966.2010.17535.x}, \href
  {https://ui.adsabs.harvard.edu/abs/2011MNRAS.410.1527H} {410, 1527}

\bibitem[\protect\citeauthoryear{{Hopkins} \& {Elvis}}{{Hopkins} \&
  {Elvis}}{2010}]{Hopkins2010}
{Hopkins} P.~F.,  {Elvis} M.,  2010, \mn@doi [MNRAS]
  {10.1111/j.1365-2966.2009.15643.x}, \href
  {https://ui.adsabs.harvard.edu/abs/2010MNRAS.401....7H} {401, 7}

\bibitem[\protect\citeauthoryear{{Hutchings} et~al.,}{{Hutchings}
  et~al.}{1999}]{Hutchings1999}
{Hutchings} J.~B.,  et~al., 1999, \mn@doi [\aj] {10.1086/301076}, \href
  {https://ui.adsabs.harvard.edu/abs/1999AJ....118.2101H} {118, 2101}

\bibitem[\protect\citeauthoryear{{Johnston}, {Elvis}, {Kjer}  \&
  {Shen}}{{Johnston} et~al.}{1982}]{Johnston1982}
{Johnston} K.~J.,  {Elvis} M.,  {Kjer} D.,   {Shen} B.~S.~P.,  1982, \mn@doi
  [\apj] {10.1086/160396}, \href
  {https://ui.adsabs.harvard.edu/abs/1982ApJ...262...61J} {262, 61}

\bibitem[\protect\citeauthoryear{{Kakkad} et~al.,}{{Kakkad}
  et~al.}{2018}]{Kakkad2018}
{Kakkad} D.,  et~al., 2018, \mn@doi [\aap] {10.1051/0004-6361/201832790}, \href
  {https://ui.adsabs.harvard.edu/abs/2018A&A...618A...6K} {618, A6}

\bibitem[\protect\citeauthoryear{{Kraemer} \& {Crenshaw}}{{Kraemer} \&
  {Crenshaw}}{2000a}]{Kraemer2000II}
{Kraemer} S.~B.,  {Crenshaw} D.~M.,  2000a, \mn@doi [\apj] {10.1086/308572},
  \href {https://ui.adsabs.harvard.edu/abs/2000ApJ...532..256K} {532, 256}

\bibitem[\protect\citeauthoryear{{Kraemer} \& {Crenshaw}}{{Kraemer} \&
  {Crenshaw}}{2000b}]{Kraemer2000III}
{Kraemer} S.~B.,  {Crenshaw} D.~M.,  2000b, \mn@doi [\apj] {10.1086/317246},
  \href {https://ui.adsabs.harvard.edu/abs/2000ApJ...544..763K} {544, 763}

\bibitem[\protect\citeauthoryear{{Kraemer}, {Turner}, {Couto}, {Crenshaw},
  {Schmitt}, {Revalski}  \& {Fischer}}{{Kraemer} et~al.}{2020}]{Kraemer2020}
{Kraemer} S.~B.,  {Turner} T.~J.,  {Couto} J.~D.,  {Crenshaw} D.~M.,  {Schmitt}
  H.~R.,  {Revalski} M.,   {Fischer} T.~C.,  2020, \mn@doi [\mnras]
  {10.1093/mnras/staa428}, \href
  {https://ui.adsabs.harvard.edu/abs/2020MNRAS.493.3893K} {493, 3893}

\bibitem[\protect\citeauthoryear{{Liu}, {Zakamska}, {Greene}, {Nesvadba}  \&
  {Liu}}{{Liu} et~al.}{2013}]{Liu2013}
{Liu} G.,  {Zakamska} N.~L.,  {Greene} J.~E.,  {Nesvadba} N. P.~H.,   {Liu} X.,
   2013, \mn@doi [\mnras] {10.1093/mnras/stt1755}, \href
  {https://ui.adsabs.harvard.edu/abs/2013MNRAS.436.2576L} {436, 2576}

\bibitem[\protect\citeauthoryear{{Lopez-Rodriguez} et~al.,}{{Lopez-Rodriguez}
  et~al.}{2018}]{LopezRodriguez2018}
{Lopez-Rodriguez} E.,  et~al., 2018, \mn@doi [\apj] {10.3847/1538-4357/aabd7b},
  \href {https://ui.adsabs.harvard.edu/abs/2018ApJ...859...99L} {859, 99}

\bibitem[\protect\citeauthoryear{{Luridiana}, {Morisset}  \&
  {Shaw}}{{Luridiana} et~al.}{2015}]{Luridiana2015}
{Luridiana} V.,  {Morisset} C.,   {Shaw} R.~A.,  2015, \mn@doi [\aap]
  {10.1051/0004-6361/201323152}, \href
  {https://ui.adsabs.harvard.edu/abs/2015A&A...573A..42L} {573, A42}

\bibitem[\protect\citeauthoryear{{Magorrian} et~al.,}{{Magorrian}
  et~al.}{1998}]{Magorrian1998}
{Magorrian} J.,  et~al., 1998, \mn@doi [ApJ] {10.1086/300353}, \href
  {https://ui.adsabs.harvard.edu/abs/1998AJ....115.2285M} {115, 2285}

\bibitem[\protect\citeauthoryear{{Mandal}, {Mukherjee}, {Federrath},
  {Nesvadba}, {Bicknell}, {Wagner}  \& {Meenakshi}}{{Mandal}
  et~al.}{2021}]{Mandal2021}
{Mandal} A.,  {Mukherjee} D.,  {Federrath} C.,  {Nesvadba} N. P.~H.,
  {Bicknell} G.~V.,  {Wagner} A.~Y.,   {Meenakshi} M.,  2021, \mn@doi [\mnras]
  {10.1093/mnras/stab2822}, \href
  {https://ui.adsabs.harvard.edu/abs/2021MNRAS.508.4738M} {508, 4738}

\bibitem[\protect\citeauthoryear{{May} \& {Steiner}}{{May} \&
  {Steiner}}{2017}]{May2017}
{May} D.,  {Steiner} J.~E.,  2017, \mn@doi [\mnras] {10.1093/mnras/stx886},
  \href {https://ui.adsabs.harvard.edu/abs/2017MNRAS.469..994M} {469, 994}

\bibitem[\protect\citeauthoryear{{May}, {Steiner}, {Menezes}, {Williams}  \&
  {Wang}}{{May} et~al.}{2020}]{May2020}
{May} D.,  {Steiner} J.~E.,  {Menezes} R.~B.,  {Williams} D.~R.~A.,   {Wang}
  J.,  2020, \mn@doi [\mnras] {10.1093/mnras/staa1545}, \href
  {https://ui.adsabs.harvard.edu/abs/2020MNRAS.496.1488M} {496, 1488}

\bibitem[\protect\citeauthoryear{{Meena}, {Crenshaw}, {Schmitt}, {Revalski},
  {Fischer}, {Polack}, {Kraemer}  \& {Dashtamirova}}{{Meena}
  et~al.}{2021}]{Meena2021}
{Meena} B.,  {Crenshaw} D.~M.,  {Schmitt} H.~R.,  {Revalski} M.,  {Fischer}
  T.~C.,  {Polack} G.~E.,  {Kraemer} S.~B.,   {Dashtamirova} D.,  2021, \mn@doi
  [\apj] {10.3847/1538-4357/ac0246}, \href
  {https://ui.adsabs.harvard.edu/abs/2021ApJ...916...31M} {916, 31}

\bibitem[\protect\citeauthoryear{{Meena} et~al.,}{{Meena}
  et~al.}{2023}]{Meena2023}
{Meena} B.,  et~al., 2023, \mn@doi [\apj] {10.3847/1538-4357/aca75f}, \href
  {https://ui.adsabs.harvard.edu/abs/2023ApJ...943...98M} {943, 98}

\bibitem[\protect\citeauthoryear{{Meenakshi}, {Mukherjee}, {Wagner},
  {Nesvadba}, {Morganti}, {Janssen}  \& {Bicknell}}{{Meenakshi}
  et~al.}{2022}]{Meenakshi2022}
{Meenakshi} M.,  {Mukherjee} D.,  {Wagner} A.~Y.,  {Nesvadba} N. P.~H.,
  {Morganti} R.,  {Janssen} R. M.~J.,   {Bicknell} G.~V.,  2022, \mn@doi
  [\mnras] {10.1093/mnras/stac167}, \href
  {https://ui.adsabs.harvard.edu/abs/2022MNRAS.511.1622M} {511, 1622}

\bibitem[\protect\citeauthoryear{{Miller}, {Brandt}, {Schneider}, {Gibson},
  {Steffen}  \& {Wu}}{{Miller} et~al.}{2011}]{Miller2011}
{Miller} B.~P.,  {Brandt} W.~N.,  {Schneider} D.~P.,  {Gibson} R.~R.,
  {Steffen} A.~T.,   {Wu} J.,  2011, \mn@doi [\apj]
  {10.1088/0004-637X/726/1/20}, \href
  {https://ui.adsabs.harvard.edu/abs/2011ApJ...726...20M} {726, 20}

\bibitem[\protect\citeauthoryear{{Mingozzi} et~al.,}{{Mingozzi}
  et~al.}{2019}]{Mingozzi2019}
{Mingozzi} M.,  et~al., 2019, \mn@doi [\aap] {10.1051/0004-6361/201834372},
  \href {https://ui.adsabs.harvard.edu/abs/2019A&A...622A.146M} {622, A146}

\bibitem[\protect\citeauthoryear{{Morganti}, {Oosterloo}  \&
  {Tsvetanov}}{{Morganti} et~al.}{1998}]{Morganti1998}
{Morganti} R.,  {Oosterloo} T.,   {Tsvetanov} Z.,  1998, \mn@doi [ApJ]
  {10.1086/300236}, \href
  {https://ui.adsabs.harvard.edu/abs/1998AJ....115..915M} {115, 915}

\bibitem[\protect\citeauthoryear{{Morganti}, {Holt}, {Saripalli}, {Oosterloo}
  \& {Tadhunter}}{{Morganti} et~al.}{2007}]{Morganti2007}
{Morganti} R.,  {Holt} J.,  {Saripalli} L.,  {Oosterloo} T.~A.,   {Tadhunter}
  C.~N.,  2007, \mn@doi [A\&A] {10.1051/0004-6361:20077888}, \href
  {https://ui.adsabs.harvard.edu/abs/2007A&A...476..735M} {476, 735}

\bibitem[\protect\citeauthoryear{{Morganti}, {Frieswijk}, {Oonk}, {Oosterloo}
  \& {Tadhunter}}{{Morganti} et~al.}{2013}]{Morganti2013}
{Morganti} R.,  {Frieswijk} W.,  {Oonk} R.~J.~B.,  {Oosterloo} T.,
  {Tadhunter} C.,  2013, \mn@doi [A\&A] {10.1051/0004-6361/201220734}, \href
  {https://ui.adsabs.harvard.edu/abs/2013A&A...552L...4M} {552, L4}

\bibitem[\protect\citeauthoryear{{Morganti}, {Oosterloo}, {Oonk}, {Frieswijk}
  \& {Tadhunter}}{{Morganti} et~al.}{2015}]{Morganti2015}
{Morganti} R.,  {Oosterloo} T.,  {Oonk} J.~B.~R.,  {Frieswijk} W.,
  {Tadhunter} C.,  2015, \mn@doi [A\&A] {10.1051/0004-6361/201525860}, \href
  {https://ui.adsabs.harvard.edu/abs/2015A&A...580A...1M} {580, A1}

\bibitem[\protect\citeauthoryear{{Mukherjee}, {Wagner}, {Bicknell}, {Morganti},
  {Oosterloo}, {Nesvadba}  \& {Sutherland}}{{Mukherjee}
  et~al.}{2018}]{Mukherjee2018}
{Mukherjee} D.,  {Wagner} A.~Y.,  {Bicknell} G.~V.,  {Morganti} R.,
  {Oosterloo} T.,  {Nesvadba} N.,   {Sutherland} R.~S.,  2018, \mn@doi [MNRAS]
  {10.1093/mnras/sty067}, \href
  {https://ui.adsabs.harvard.edu/abs/2018MNRAS.476...80M} {476, 80}

\bibitem[\protect\citeauthoryear{{Mullaney}, {Alexander}, {Fine}, {Goulding},
  {Harrison}  \& {Hickox}}{{Mullaney} et~al.}{2013}]{Mullaney2013}
{Mullaney} J.~R.,  {Alexander} D.~M.,  {Fine} S.,  {Goulding} A.~D.,
  {Harrison} C.~M.,   {Hickox} R.~C.,  2013, \mn@doi [MNRAS]
  {10.1093/mnras/stt751}, \href
  {https://ui.adsabs.harvard.edu/abs/2013MNRAS.433..622M} {433, 622}

\bibitem[\protect\citeauthoryear{{M{\"u}ller S{\'a}nchez}, {Davies}, {Genzel},
  {Tacconi}, {Eisenhauer}, {Hicks}, {Friedrich}  \& {Sternberg}}{{M{\"u}ller
  S{\'a}nchez} et~al.}{2009}]{MUllerSanchez2009}
{M{\"u}ller S{\'a}nchez} F.,  {Davies} R.~I.,  {Genzel} R.,  {Tacconi} L.~J.,
  {Eisenhauer} F.,  {Hicks} E.~K.~S.,  {Friedrich} S.,   {Sternberg} A.,  2009,
  \mn@doi [\apj] {10.1088/0004-637X/691/1/749}, \href
  {https://ui.adsabs.harvard.edu/abs/2009ApJ...691..749M} {691, 749}

\bibitem[\protect\citeauthoryear{{Mundell}, {Wrobel}, {Pedlar}  \&
  {Gallimore}}{{Mundell} et~al.}{2003}]{Mundell2003}
{Mundell} C.~G.,  {Wrobel} J.~M.,  {Pedlar} A.,   {Gallimore} J.~F.,  2003,
  \mn@doi [\apj] {10.1086/345356}, \href
  {https://ui.adsabs.harvard.edu/abs/2003ApJ...583..192M} {583, 192}

\bibitem[\protect\citeauthoryear{{Nelson}, {Weistrop}, {Hutchings}, {Crenshaw},
  {Gull}, {Kaiser}, {Kraemer}  \& {Lindler}}{{Nelson}
  et~al.}{2000}]{Nelson2000}
{Nelson} C.~H.,  {Weistrop} D.,  {Hutchings} J.~B.,  {Crenshaw} D.~M.,  {Gull}
  T.~R.,  {Kaiser} M.~E.,  {Kraemer} S.~B.,   {Lindler} D.,  2000, \mn@doi
  [\apj] {10.1086/308456}, \href
  {https://ui.adsabs.harvard.edu/abs/2000ApJ...531..257N} {531, 257}

\bibitem[\protect\citeauthoryear{{Nesvadba}, {Lehnert}, {Eisenhauer},
  {Gilbert}, {Tecza}  \& {Abuter}}{{Nesvadba} et~al.}{2006}]{Nesvadba2006}
{Nesvadba} N.~P.~H.,  {Lehnert} M.~D.,  {Eisenhauer} F.,  {Gilbert} A.,
  {Tecza} M.,   {Abuter} R.,  2006, \mn@doi [ApJ] {10.1086/507266}, \href
  {https://ui.adsabs.harvard.edu/abs/2006ApJ...650..693N} {650, 693}

\bibitem[\protect\citeauthoryear{{Oosterloo}, {Morganti}, {Tzioumis},
  {Reynolds}, {King}, {McCulloch}  \& {Tsvetanov}}{{Oosterloo}
  et~al.}{2000}]{Oosterloo2000}
{Oosterloo} T.~A.,  {Morganti} R.,  {Tzioumis} A.,  {Reynolds} J.,  {King} E.,
  {McCulloch} P.,   {Tsvetanov} Z.,  2000, \mn@doi [ApJ] {10.1086/301358},
  \href {https://ui.adsabs.harvard.edu/abs/2000AJ....119.2085O} {119, 2085}

\bibitem[\protect\citeauthoryear{{Oosterloo}, {Raymond Oonk}, {Morganti},
  {Combes}, {Dasyra}, {Salom{\'e}}, {Vlahakis}  \& {Tadhunter}}{{Oosterloo}
  et~al.}{2017}]{Oosterloo2017}
{Oosterloo} T.,  {Raymond Oonk} J.~B.,  {Morganti} R.,  {Combes} F.,  {Dasyra}
  K.,  {Salom{\'e}} P.,  {Vlahakis} N.,   {Tadhunter} C.,  2017, \mn@doi [A\&A]
  {10.1051/0004-6361/201731781}, \href
  {https://ui.adsabs.harvard.edu/abs/2017A&A...608A..38O} {608, A38}

\bibitem[\protect\citeauthoryear{{Osterbrock} \& {Ferland}}{{Osterbrock} \&
  {Ferland}}{2006}]{Osterbrock2006}
{Osterbrock} D.~E.,  {Ferland} G.~J.,  2006, {Astrophysics of gaseous nebulae
  and active galactic nuclei}

\bibitem[\protect\citeauthoryear{{Osterbrock} \& {Koski}}{{Osterbrock} \&
  {Koski}}{1976}]{OsterbrockKoski1976}
{Osterbrock} D.~E.,  {Koski} A.~T.,  1976, \mn@doi [\mnras]
  {10.1093/mnras/176.1.61P}, \href
  {https://ui.adsabs.harvard.edu/abs/1976MNRAS.176P..61O} {176, 61P}

\bibitem[\protect\citeauthoryear{{Pedlar}, {Howley}, {Axon}  \&
  {Unger}}{{Pedlar} et~al.}{1992}]{Pedlar1992}
{Pedlar} A.,  {Howley} P.,  {Axon} D.~J.,   {Unger} S.~W.,  1992, \mn@doi
  [\mnras] {10.1093/mnras/259.2.369}, \href
  {https://ui.adsabs.harvard.edu/abs/1992MNRAS.259..369P} {259, 369}

\bibitem[\protect\citeauthoryear{{Pedlar}, {Kukula}, {Longley}, {Muxlow},
  {Axon}, {Baum}, {O'Dea}  \& {Unger}}{{Pedlar} et~al.}{1993}]{Pedlar1993}
{Pedlar} A.,  {Kukula} M.~J.,  {Longley} D.~P.~T.,  {Muxlow} T.~W.~B.,  {Axon}
  D.~J.,  {Baum} S.,  {O'Dea} C.,   {Unger} S.~W.,  1993, \mn@doi [\mnras]
  {10.1093/mnras/263.2.471}, \href
  {https://ui.adsabs.harvard.edu/abs/1993MNRAS.263..471P} {263, 471}

\bibitem[\protect\citeauthoryear{{Raban}, {Jaffe}, {R{\"o}ttgering},
  {Meisenheimer}  \& {Tristram}}{{Raban} et~al.}{2009}]{Raban2009}
{Raban} D.,  {Jaffe} W.,  {R{\"o}ttgering} H.,  {Meisenheimer} K.,   {Tristram}
  K. R.~W.,  2009, \mn@doi [\mnras] {10.1111/j.1365-2966.2009.14439.x}, \href
  {https://ui.adsabs.harvard.edu/abs/2009MNRAS.394.1325R} {394, 1325}

\bibitem[\protect\citeauthoryear{{Ramos Almeida}, {Acosta-Pulido}, {Tadhunter},
  {Gonz{\'a}lez-Fern{\'a}ndez}, {Cicone}  \& {Fern{\'a}ndez-Torreiro}}{{Ramos
  Almeida} et~al.}{2019}]{RamosAlmeida2019}
{Ramos Almeida} C.,  {Acosta-Pulido} J.~A.,  {Tadhunter} C.~N.,
  {Gonz{\'a}lez-Fern{\'a}ndez} C.,  {Cicone} C.,   {Fern{\'a}ndez-Torreiro} M.,
   2019, \mn@doi [\mnras] {10.1093/mnrasl/slz072}, \href
  {https://ui.adsabs.harvard.edu/abs/2019MNRAS.487L..18R} {487, L18}

\bibitem[\protect\citeauthoryear{{Revalski}, {Crenshaw}, {Kraemer}, {Fischer},
  {Schmitt}  \& {Machuca}}{{Revalski} et~al.}{2018}]{Revalski2018}
{Revalski} M.,  {Crenshaw} D.~M.,  {Kraemer} S.~B.,  {Fischer} T.~C.,
  {Schmitt} H.~R.,   {Machuca} C.,  2018, \mn@doi [\apj]
  {10.3847/1538-4357/aab107}, \href
  {https://ui.adsabs.harvard.edu/abs/2018ApJ...856...46R} {856, 46}

\bibitem[\protect\citeauthoryear{{Revalski} et~al.,}{{Revalski}
  et~al.}{2021}]{Revalski2021}
{Revalski} M.,  et~al., 2021, \mn@doi [\apj] {10.3847/1538-4357/abdcad}, \href
  {https://ui.adsabs.harvard.edu/abs/2021ApJ...910..139R} {910, 139}

\bibitem[\protect\citeauthoryear{{Revalski} et~al.,}{{Revalski}
  et~al.}{2022}]{Revalski2022}
{Revalski} M.,  et~al., 2022, \mn@doi [\apj] {10.3847/1538-4357/ac5f3d}, \href
  {https://ui.adsabs.harvard.edu/abs/2022ApJ...930...14R} {930, 14}

\bibitem[\protect\citeauthoryear{{Riffel}}{{Riffel}}{2021}]{Riffel2021b}
{Riffel} R.~A.,  2021, \mn@doi [\mnras] {10.1093/mnras/stab1877}, \href
  {https://ui.adsabs.harvard.edu/abs/2021MNRAS.506.2950R} {506, 2950}

\bibitem[\protect\citeauthoryear{{Riffel}, {Rodr{\'\i}guez-Ardila}, {Aleman},
  {Brotherton}, {Pastoriza}, {Bonatto}  \& {Dors}}{{Riffel}
  et~al.}{2013}]{Riffel2013a}
{Riffel} R.,  {Rodr{\'\i}guez-Ardila} A.,  {Aleman} I.,  {Brotherton} M.~S.,
  {Pastoriza} M.~G.,  {Bonatto} C.,   {Dors} O.~L.,  2013, \mn@doi [\mnras]
  {10.1093/mnras/stt026}, \href
  {https://ui.adsabs.harvard.edu/abs/2013MNRAS.430.2002R} {430, 2002}

\bibitem[\protect\citeauthoryear{{Riffel}, {Vale}, {Storchi-Bergmann}  \&
  {McGregor}}{{Riffel} et~al.}{2014}]{Riffel2014}
{Riffel} R.~A.,  {Vale} T.~B.,  {Storchi-Bergmann} T.,   {McGregor} P.~J.,
  2014, \mn@doi [\mnras] {10.1093/mnras/stu843}, \href
  {https://ui.adsabs.harvard.edu/abs/2014MNRAS.442..656R} {442, 656}

\bibitem[\protect\citeauthoryear{{Riffel}, {Bianchin}, {Riffel},
  {Storchi-Bergmann}, {Sch{\"o}nell}, {Dahmer-Hahn}, {Dametto}  \&
  {Diniz}}{{Riffel} et~al.}{2021}]{Riffel2021}
{Riffel} R.~A.,  {Bianchin} M.,  {Riffel} R.,  {Storchi-Bergmann} T.,
  {Sch{\"o}nell} A.~J.,  {Dahmer-Hahn} L.~G.,  {Dametto} N.~Z.,   {Diniz}
  M.~R.,  2021, \mn@doi [\mnras] {10.1093/mnras/stab788}, \href
  {https://ui.adsabs.harvard.edu/abs/2021MNRAS.503.5161R} {503, 5161}

\bibitem[\protect\citeauthoryear{{Robinson}, {Binette}, {Fosbury}  \&
  {Tadhunter}}{{Robinson} et~al.}{1987}]{Robinson1987}
{Robinson} A.,  {Binette} L.,  {Fosbury} R.~A.~E.,   {Tadhunter} C.~N.,  1987,
  \mn@doi [\mnras] {10.1093/mnras/227.1.97}, \href
  {https://ui.adsabs.harvard.edu/abs/1987MNRAS.227...97R} {227, 97}

\bibitem[\protect\citeauthoryear{{Robinson} et~al.,}{{Robinson}
  et~al.}{1994}]{Robinson1994}
{Robinson} A.,  et~al., 1994, \aap, \href
  {https://ui.adsabs.harvard.edu/abs/1994A&A...291..351R} {291, 351}

\bibitem[\protect\citeauthoryear{{Rodr{\'\i}guez-Ardila}, {Riffel}  \&
  {Pastoriza}}{{Rodr{\'\i}guez-Ardila} et~al.}{2005}]{Rodriguez-Ardila2005}
{Rodr{\'\i}guez-Ardila} A.,  {Riffel} R.,   {Pastoriza} M.~G.,  2005, \mn@doi
  [\mnras] {10.1111/j.1365-2966.2005.09638.x}, \href
  {https://ui.adsabs.harvard.edu/abs/2005MNRAS.364.1041R} {364, 1041}

\bibitem[\protect\citeauthoryear{{Rose}, {Tadhunter}, {Holt}, {Ramos Almeida}
  \& {Littlefair}}{{Rose} et~al.}{2011}]{Rose2011}
{Rose} M.,  {Tadhunter} C.~N.,  {Holt} J.,  {Ramos Almeida} C.,   {Littlefair}
  S.~P.,  2011, \mn@doi [\mnras] {10.1111/j.1365-2966.2011.18639.x}, \href
  {https://ui.adsabs.harvard.edu/abs/2011MNRAS.414.3360R} {414, 3360}

\bibitem[\protect\citeauthoryear{{Rose}, {Tadhunter}, {Ramos Almeida},
  {Rodr{\'\i}guez Zaur{\'\i}n}, {Santoro}  \& {Spence}}{{Rose}
  et~al.}{2018}]{Rose2018}
{Rose} M.,  {Tadhunter} C.,  {Ramos Almeida} C.,  {Rodr{\'\i}guez Zaur{\'\i}n}
  J.,  {Santoro} F.,   {Spence} R.,  2018, \mn@doi [MNRAS]
  {10.1093/mnras/stx2590}, \href
  {https://ui.adsabs.harvard.edu/abs/2018MNRAS.474..128R} {474, 128}

\bibitem[\protect\citeauthoryear{{Santoro}, {Rose}, {Morganti}, {Tadhunter},
  {Oosterloo}  \& {Holt}}{{Santoro} et~al.}{2018}]{Santoro2018}
{Santoro} F.,  {Rose} M.,  {Morganti} R.,  {Tadhunter} C.,  {Oosterloo} T.~A.,
   {Holt} J.,  2018, \mn@doi [A\&A] {10.1051/0004-6361/201833248}, \href
  {https://ui.adsabs.harvard.edu/abs/2018A&A...617A.139S} {617, A139}

\bibitem[\protect\citeauthoryear{{Santoro}, {Tadhunter}, {Baron}, {Morganti}
  \& {Holt}}{{Santoro} et~al.}{2020}]{Santoro2020}
{Santoro} F.,  {Tadhunter} C.,  {Baron} D.,  {Morganti} R.,   {Holt} J.,  2020,
  \mn@doi [A\&A] {10.1051/0004-6361/202039077}, \href
  {https://ui.adsabs.harvard.edu/abs/2020A&A...644A..54S} {644, A54}

\bibitem[\protect\citeauthoryear{{Schaye} et~al.,}{{Schaye}
  et~al.}{2015}]{Schaye2015}
{Schaye} J.,  et~al., 2015, \mn@doi [\mnras] {10.1093/mnras/stu2058}, \href
  {https://ui.adsabs.harvard.edu/abs/2015MNRAS.446..521S} {446, 521}

\bibitem[\protect\citeauthoryear{{Schlafly} \& {Finkbeiner}}{{Schlafly} \&
  {Finkbeiner}}{2011}]{Schlafly2011}
{Schlafly} E.~F.,  {Finkbeiner} D.~P.,  2011, \mn@doi [ApJ]
  {10.1088/0004-637X/737/2/103}, \href
  {https://ui.adsabs.harvard.edu/abs/2011ApJ...737..103S} {737, 103}

\bibitem[\protect\citeauthoryear{{Schlegel}, {Finkbeiner}  \&
  {Davis}}{{Schlegel} et~al.}{1998}]{Schlegel1998}
{Schlegel} D.~J.,  {Finkbeiner} D.~P.,   {Davis} M.,  1998, \mn@doi [ApJ]
  {10.1086/305772}, \href
  {https://ui.adsabs.harvard.edu/abs/1998ApJ...500..525S} {500, 525}

\bibitem[\protect\citeauthoryear{{Seyfert}}{{Seyfert}}{1943}]{Seyfert1943}
{Seyfert} C.~K.,  1943, \mn@doi [\apj] {10.1086/144488}, \href
  {https://ui.adsabs.harvard.edu/abs/1943ApJ....97...28S} {97, 28}

\bibitem[\protect\citeauthoryear{{Sharp} \& {Bland-Hawthorn}}{{Sharp} \&
  {Bland-Hawthorn}}{2010}]{Sharp2010}
{Sharp} R.~G.,  {Bland-Hawthorn} J.,  2010, \mn@doi [\apj]
  {10.1088/0004-637X/711/2/818}, \href
  {https://ui.adsabs.harvard.edu/abs/2010ApJ...711..818S} {711, 818}

\bibitem[\protect\citeauthoryear{{Silk} \& {Rees}}{{Silk} \&
  {Rees}}{1998}]{Silk1998}
{Silk} J.,  {Rees} M.~J.,  1998, A\&A, \href
  {https://ui.adsabs.harvard.edu/abs/1998A&A...331L...1S} {331, L1}

\bibitem[\protect\citeauthoryear{{Somerville}, {Hopkins}, {Cox}, {Robertson}
  \& {Hernquist}}{{Somerville} et~al.}{2008}]{Somerville2008}
{Somerville} R.~S.,  {Hopkins} P.~F.,  {Cox} T.~J.,  {Robertson} B.~E.,
  {Hernquist} L.,  2008, \mn@doi [\mnras] {10.1111/j.1365-2966.2008.13805.x},
  \href {https://ui.adsabs.harvard.edu/abs/2008MNRAS.391..481S} {391, 481}

\bibitem[\protect\citeauthoryear{{Spence}}{{Spence}}{2018}]{SpenceThesis}
{Spence} R. A.~W.,  2018, PhD thesis, University of Sheffield, \url
  {https://etheses.whiterose.ac.uk/23431/}

\bibitem[\protect\citeauthoryear{{Spence}, {Tadhunter}, {Rose}  \&
  {Rodr{\'\i}guez Zaur{\'\i}n}}{{Spence} et~al.}{2018}]{Spence2018}
{Spence} R.~A.~W.,  {Tadhunter} C.~N.,  {Rose} M.,   {Rodr{\'\i}guez
  Zaur{\'\i}n} J.,  2018, \mn@doi [MNRAS] {10.1093/mnras/sty1046}, \href
  {https://ui.adsabs.harvard.edu/abs/2018MNRAS.478.2438S} {478, 2438}

\bibitem[\protect\citeauthoryear{{Speranza} et~al.,}{{Speranza}
  et~al.}{2022}]{Speranza2022}
{Speranza} G.,  et~al., 2022, \mn@doi [\aap] {10.1051/0004-6361/202243585},
  \href {https://ui.adsabs.harvard.edu/abs/2022A&A...665A..55S} {665, A55}

\bibitem[\protect\citeauthoryear{{Springel}, {Di Matteo}  \&
  {Hernquist}}{{Springel} et~al.}{2005}]{Springel2005}
{Springel} V.,  {Di Matteo} T.,   {Hernquist} L.,  2005, \mn@doi [MNRAS]
  {10.1111/j.1365-2966.2005.09238.x}, \href
  {https://ui.adsabs.harvard.edu/abs/2005MNRAS.361..776S} {361, 776}

\bibitem[\protect\citeauthoryear{{Storchi-Bergmann}, {McGregor}, {Riffel},
  {Sim{\~o}es Lopes}, {Beck}  \& {Dopita}}{{Storchi-Bergmann}
  et~al.}{2009}]{Storchi-Bergmann2009}
{Storchi-Bergmann} T.,  {McGregor} P.~J.,  {Riffel} R.~A.,  {Sim{\~o}es Lopes}
  R.,  {Beck} T.,   {Dopita} M.,  2009, \mn@doi [\mnras]
  {10.1111/j.1365-2966.2009.14388.x}, \href
  {https://ui.adsabs.harvard.edu/abs/2009MNRAS.394.1148S} {394, 1148}

\bibitem[\protect\citeauthoryear{{Storchi-Bergmann}, {Lopes}, {McGregor},
  {Riffel}, {Beck}  \& {Martini}}{{Storchi-Bergmann}
  et~al.}{2010}]{Storchi-Bergmann2010}
{Storchi-Bergmann} T.,  {Lopes} R.~D.~S.,  {McGregor} P.~J.,  {Riffel} R.~A.,
  {Beck} T.,   {Martini} P.,  2010, \mn@doi [\mnras]
  {10.1111/j.1365-2966.2009.15962.x}, \href
  {https://ui.adsabs.harvard.edu/abs/2010MNRAS.402..819S} {402, 819}

\bibitem[\protect\citeauthoryear{{Sun}, {Greene}  \& {Zakamska}}{{Sun}
  et~al.}{2017}]{Sun2017}
{Sun} A.-L.,  {Greene} J.~E.,   {Zakamska} N.~L.,  2017, \mn@doi [ApJ]
  {10.3847/1538-4357/835/2/222}, \href
  {https://ui.adsabs.harvard.edu/abs/2017ApJ...835..222S} {835, 222}

\bibitem[\protect\citeauthoryear{{Sutherland} \& {Dopita}}{{Sutherland} \&
  {Dopita}}{2017}]{Sutherland2017}
{Sutherland} R.~S.,  {Dopita} M.~A.,  2017, \mn@doi [\apjs]
  {10.3847/1538-4365/aa6541}, \href
  {https://ui.adsabs.harvard.edu/abs/2017ApJS..229...34S} {229, 34}

\bibitem[\protect\citeauthoryear{{Tadhunter}}{{Tadhunter}}{2016}]{Tadhunter2016}
{Tadhunter} C.,  2016, \mn@doi [\aapr] {10.1007/s00159-016-0094-x}, \href
  {https://ui.adsabs.harvard.edu/abs/2016A&ARv..24...10T} {24, 10}

\bibitem[\protect\citeauthoryear{{Tadhunter}, {Morganti}, {Rose}, {Oonk}  \&
  {Oosterloo}}{{Tadhunter} et~al.}{2014}]{Tadhunter2014}
{Tadhunter} C.,  {Morganti} R.,  {Rose} M.,  {Oonk} J.~B.~R.,   {Oosterloo} T.,
   2014, \mn@doi [Nat.] {10.1038/nature13520}, \href
  {https://ui.adsabs.harvard.edu/abs/2014Natur.511..440T} {511, 440}

\bibitem[\protect\citeauthoryear{{Ulvestad} \& {Wilson}}{{Ulvestad} \&
  {Wilson}}{1984}]{Ulvestad1984}
{Ulvestad} J.~S.,  {Wilson} A.~S.,  1984, \mn@doi [\apj] {10.1086/162520},
  \href {https://ui.adsabs.harvard.edu/abs/1984ApJ...285..439U} {285, 439}

\bibitem[\protect\citeauthoryear{{Venturi} et~al.,}{{Venturi}
  et~al.}{2021}]{Venturi2021}
{Venturi} G.,  et~al., 2021, \mn@doi [\aap] {10.1051/0004-6361/202039869},
  \href {https://ui.adsabs.harvard.edu/abs/2021A&A...648A..17V} {648, A17}

\bibitem[\protect\citeauthoryear{{Villar-Mart{\'\i}n}, {Tadhunter}, {Morganti},
  {Axon}  \& {Koekemoer}}{{Villar-Mart{\'\i}n} et~al.}{1999}]{VillarMartin1999}
{Villar-Mart{\'\i}n} M.,  {Tadhunter} C.,  {Morganti} R.,  {Axon} D.,
  {Koekemoer} A.,  1999, \mn@doi [MNRAS] {10.1046/j.1365-8711.1999.02603.x},
  \href {https://ui.adsabs.harvard.edu/abs/1999MNRAS.307...24V} {307, 24}

\bibitem[\protect\citeauthoryear{{Wagner} \& {Bicknell}}{{Wagner} \&
  {Bicknell}}{2011}]{Wagner2011}
{Wagner} A.~Y.,  {Bicknell} G.~V.,  2011, \mn@doi [ApJ]
  {10.1088/0004-637X/728/1/29}, \href
  {https://ui.adsabs.harvard.edu/abs/2011ApJ...728...29W} {728, 29}

\bibitem[\protect\citeauthoryear{{Wang}, {Fabbiano}, {Elvis}, {Risaliti},
  {Mundell}, {Karovska}  \& {Zezas}}{{Wang} et~al.}{2011a}]{Wang2011b}
{Wang} J.,  {Fabbiano} G.,  {Elvis} M.,  {Risaliti} G.,  {Mundell} C.~G.,
  {Karovska} M.,   {Zezas} A.,  2011a, \mn@doi [\apj]
  {10.1088/0004-637X/736/1/62}, \href
  {https://ui.adsabs.harvard.edu/abs/2011ApJ...736...62W} {736, 62}

\bibitem[\protect\citeauthoryear{{Wang} et~al.,}{{Wang}
  et~al.}{2011b}]{Wang2011}
{Wang} J.,  et~al., 2011b, \mn@doi [\apj] {10.1088/0004-637X/742/1/23}, \href
  {https://ui.adsabs.harvard.edu/abs/2011ApJ...742...23W} {742, 23}

\bibitem[\protect\citeauthoryear{{Williams} et~al.,}{{Williams}
  et~al.}{2017}]{Williams2017}
{Williams} D.~R.~A.,  et~al., 2017, \mn@doi [\mnras] {10.1093/mnras/stx2205},
  \href {https://ui.adsabs.harvard.edu/abs/2017MNRAS.472.3842W} {472, 3842}

\bibitem[\protect\citeauthoryear{{Wilson} \& {Ulvestad}}{{Wilson} \&
  {Ulvestad}}{1983}]{Wilson1983}
{Wilson} A.~S.,  {Ulvestad} J.~S.,  1983, \mn@doi [\apj] {10.1086/161507},
  \href {https://ui.adsabs.harvard.edu/abs/1983ApJ...275....8W} {275, 8}

\bibitem[\protect\citeauthoryear{{Wilson} \& {Ulvestad}}{{Wilson} \&
  {Ulvestad}}{1987}]{WilsonUlvestad1987}
{Wilson} A.~S.,  {Ulvestad} J.~S.,  1987, \mn@doi [\apj] {10.1086/165436},
  \href {https://ui.adsabs.harvard.edu/abs/1987ApJ...319..105W} {319, 105}

\bibitem[\protect\citeauthoryear{{Winge}, {Axon}, {Macchetto}  \&
  {Capetti}}{{Winge} et~al.}{1997}]{Winge1997}
{Winge} C.,  {Axon} D.~J.,  {Macchetto} F.~D.,   {Capetti} A.,  1997, \mn@doi
  [\apjl] {10.1086/310892}, \href
  {https://ui.adsabs.harvard.edu/abs/1997ApJ...487L.121W} {487, L121}

\bibitem[\protect\citeauthoryear{{Woo} \& {Urry}}{{Woo} \&
  {Urry}}{2002}]{Woo2002}
{Woo} J.-H.,  {Urry} C.~M.,  2002, \mn@doi [\apj] {10.1086/342878}, \href
  {https://ui.adsabs.harvard.edu/abs/2002ApJ...579..530W} {579, 530}

\bibitem[\protect\citeauthoryear{{Yuan} et~al.,}{{Yuan}
  et~al.}{2020}]{Yuan2020}
{Yuan} W.,  et~al., 2020, \mn@doi [\apj] {10.3847/1538-4357/abb377}, \href
  {https://ui.adsabs.harvard.edu/abs/2020ApJ...902...26Y} {902, 26}

\bibitem[\protect\citeauthoryear{{Zamorani} et~al.,}{{Zamorani}
  et~al.}{1981}]{Zamorani1981}
{Zamorani} G.,  et~al., 1981, \mn@doi [\apj] {10.1086/158815}, \href
  {https://ui.adsabs.harvard.edu/abs/1981ApJ...245..357Z} {245, 357}

\makeatother
\end{thebibliography}




\appendix

\section{Critical densities and ionisation energies for the emission lines used in our analysis}
\label{section: appendix_critical_densities}

In Table \ref{tab: critical_densities}, we present critical densities and ionisation energies for the lines used in our analysis, calculated using the \textsc{PyNeb Python} module for a gas of temperature T$_\mathrm{e}=15$,000\;K. 

There have previously been concerns that the transauroral lines --- used to derive electron densities and reddenings in Section \ref{section: tr_diagnostics} --- do not trace the same gas that is emitting other key diagnostic lines such as H$\mathrm{\beta}$ and [OIII]$\lambda\lambda$4959,5007 (see \citealt{Sun2017}, \citealt{Rose2018}, \citealt{Spence2018} and \citealt{Holden2022}). We note that, if the transauroral lines originate from denser clumps of gas within the same cloud complexes as the clouds emitting other lines \citep{Sun2017}, we would also expect those clumps to also radiate strongly in H$\mathrm{\beta}$ since the recombination line emissivity scales as $n^2$. Furthermore, we note that the transauroral lines have critical densities that are closer to the critical density of the [OIII]$\lambda\lambda$4959,5007 lines than the traditional [SII] and [OII] lines (Table \ref{tab: critical_densities}), so they are more likely to trace the [OIII]-emitting clouds than the traditional lines \citep{Rose2018, Spence2018}. Furthermore, the transauroral ratios involve emission lines that arise from transitions within the [OII] ion, which has an ionisation energy that is closer to the ionisation energy of [OIII] than [SII]. This highlights that the transauroral lines are likely better tracers of the [OIII]-emitting gas than the commonly used [SII](6717/6731) ratio.

\begin{table}
    \centering
    \begin{tabular}{ccc}
    \hline
    Emission line & n$_\mathrm{crit}$ (cm$^{-3}$) & E$_\mathrm{ion}$ (eV) \\ \hline
    {[}FeVII{]}$\lambda$6087 & 1.9$\times10^7$ & 125.0 \\  
    {[}FeVII{]}$\lambda$3759 & 3.5$\times10^8$ & 125.0 \\
    {[}NeV{]}$\lambda$3426 & 1.8$\times10^7$ & 126.2 \\  
    {[}NeIII{]}$\lambda$3869 & 1.3$\times10^7$  & 63.4 \\ \hline
    {[}OIII{]}$\lambda$4959,5007 & 7.8$\times10^5$ & 54.9 \\ \hline
    \multicolumn{3}{c}{Transauroral  {[}OII{]} and  {[}SII{]} lines} \\
    {[}OII{]}$\lambda$7320,7331 & 6.9$\times10^6$ & 35.1 \\
    {[}OII{]}$\lambda$7319,7330 & 4.7$\times10^6$ & 35.1 \\
    {[}SII{]}$\lambda$4069 & 3.2$\times10^6$ & 23.3 \\
    {[}SII{]}$\lambda$4076 & 1.6$\times10^6$ & 23.3 \\ \hline
    \multicolumn{3}{c}{Traditional  {[}OII{]} and  {[}SII{]} lines} \\
    {[}OII{]}$\lambda$3726 & 4.8$\times10^3$ & 35.1 \\
    {[}OII{]}$\lambda$3729 & 1.4$\times10^3$ & 35.1 \\ 
    {[}SII{]}$\lambda$6716 & 1.9$\times10^3$ & 23.3 \\
    {[}SII{]}$\lambda$6731 & 5.1$\times10^3$ & 23.3 \\ \hline
    \end{tabular}
    \caption{Critical densities and ionisation energies at T$_\mathrm{e}=15$,000\;K for several key diagnostic lines that trace the warm ionised outflow phase: the [NeV]$\lambda$3426 and [NeIII]$\lambda$3869 lines are used to investigate the presence of matter-bounded gas (Section \ref{section: mechanisms}); the lines in the [FeVII](6087/3759) ratio are used to determine densities for high ionisation gas (Section \ref{section: high_ionisation}); the [OIII]$\lambda$4959,5007 doublet is used for kinematics (Section \ref{section: kinematics}); the [SII](6717/6731) and [SII](3726/3729) are `traditional' density ratios, and the transauroral lines are used in the TR method to derive electron densities and reddenings (Section \ref{section: tr_diagnostics}). The \textsc{PyNeb Python} module was used to produce the values in this table.}
    \label{tab: critical_densities}
\end{table}

\section{Variation of the [FeVII](6086/3759) vs [NeV]$\lambda$3426/[FeVII]$\lambda$6086 diagnostic diagram with spectral index}
\label{section: appendix_nev_fevii}

\begin{figure}
  \includegraphics[width=\linewidth]{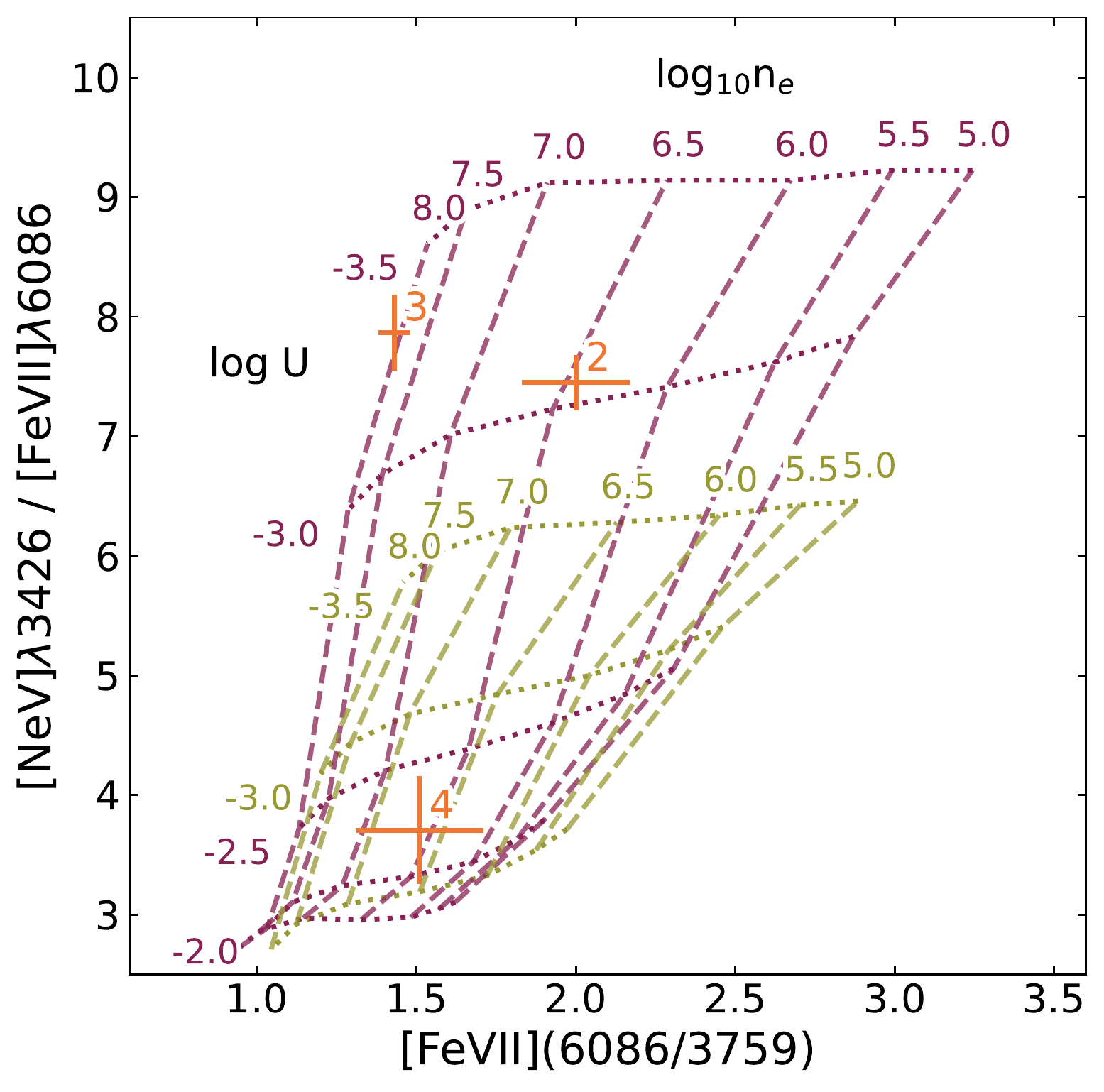}
  \caption{[FeVII](6087/3759) vs [NeV]$\lambda$3426/[FeVII]$\lambda$6086 diagnostic diagram, as in Figure \ref{fig: nev_fevii}, with radiation-bounded AGN-photoionisation grids generated with \textsc{CLOUDY} for two values of spectral index: $\alpha=1.0$ (green) and $\alpha=1.5$ (purple). Otherwise, the labelling, colour, marker and line scheme is the same as in Figure \ref{fig: nev_fevii}.}
  \label{fig: nev_fevii_a_vary}
 \end{figure}

The ratios used in the [FeVII](6086/3759) vs [NeV]$\lambda$3426/[FeVII]$\lambda$6086 diagnostic diagram (Section \ref{section: high_ionisation}; Figure \ref{fig: nev_fevii}) are sensitive to both electron density and temperature, and thus the position of the AGN-photoionisation grid on this diagram --- which we use to determine the density of the high ionisation gas --- changes with the assumed ionising source spectral index ($\alpha$) and ionisation parameter ($U$) of the gas. To further investigate this beyond only varying the ionisation parameter (as is shown in Figure \ref{fig: nev_fevii} for $\alpha=1.5$), here we show the effect of assuming a lower spectral index. We used the same \textsc{CLOUDY} model as described in Section \ref{section: high_ionisation}, but instead took the spectral index to be $\alpha=1.0$. We present the resulting grid (along with the grid for $\alpha=1.5$ for comparison purposes) in Figure \ref{fig: nev_fevii_a_vary}.

From Figure \ref{fig: nev_fevii_a_vary}, it can be seen that a flatter spectral index produces lower [NeV]$\lambda$3426/[FeVII]$\lambda$6086 ratios, with little effect on the [FeVII](6086/3759) ratios: for low values of [NeV]$\lambda$3426/[FeVII]$\lambda$6086 (i.e. as measured in Aperture 4 for NGC\;1068), the effect on derived density is small ($\sim$0.1\;dex). However, a shallower spectral index cannot reproduce higher values of [NeV]$\lambda$3426/[FeVII]$\lambda$6086 (as measured in NGC\;1068 apertures 2 and 3) without very low (log\;$U$\;\textless\;$-3.0$) ionisation parameters. Therefore, we use the \textsc{CLOUDY} grid with $\alpha=1.5$ to derive electron densities for the high ionisation gas in Section \ref{section: high_ionisation}.

\section{Modelling the effect of shock-ionisation on the transauroral line ratio density diagnostic grid}
\label{section: appendix_tr_shock}

\begin{figure}
  \includegraphics[width=\linewidth]{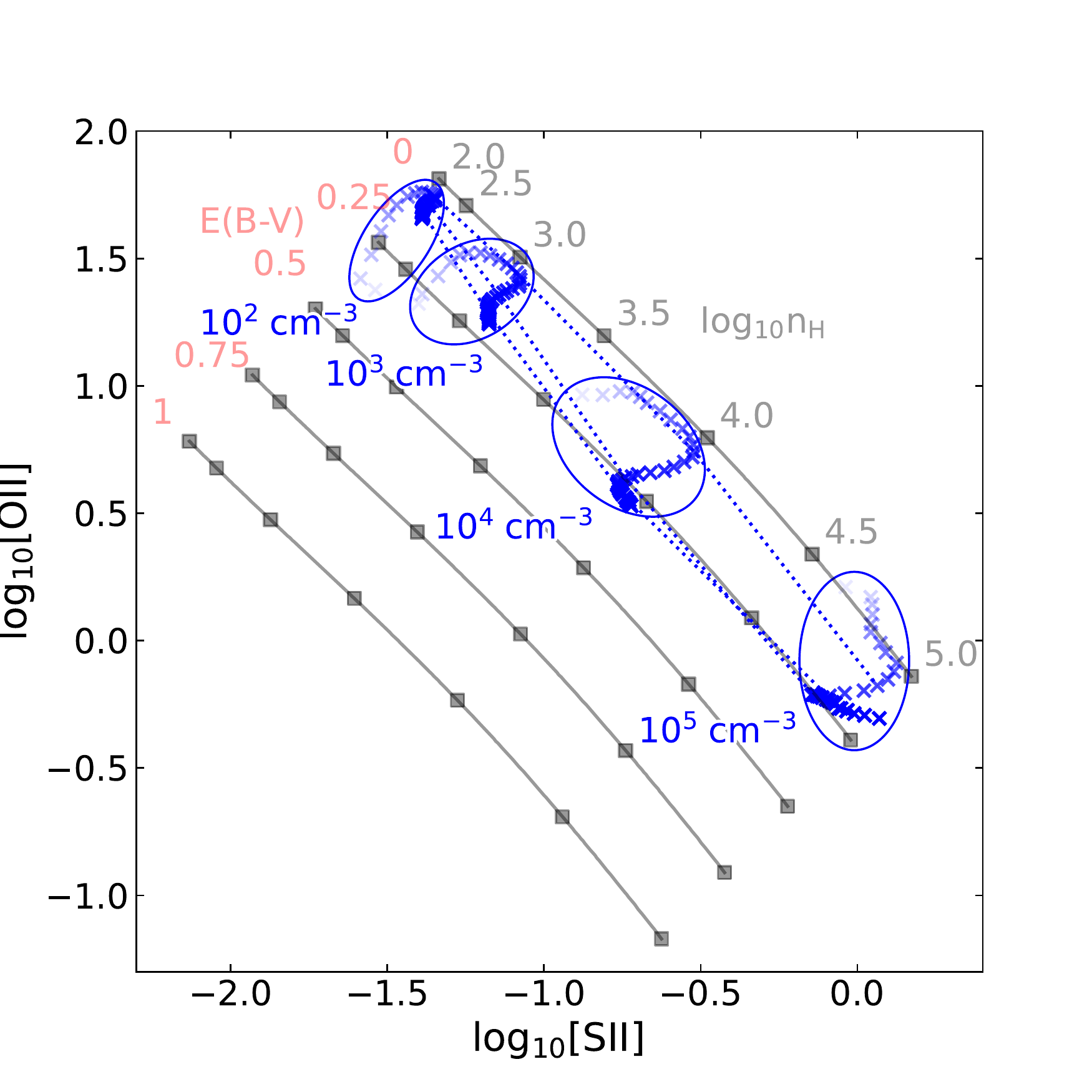}
  \caption{TR([SII]) vs TR([OII]) grid for radiation-bounded AGN-photoionised gas (black with red E(B-V) labels; as in Figure \ref{fig: tr_ddd}) with the line ratios expected from modelled shocks (blue crosses) of fixed magnetic parameter ($B/\sqrt{n}$=2\;$\mu$G\;cm$^{3/2}$) at different post-shock densities (labelled in blue), and for a range of shock velocities (0\;\textless\;$v_\mathrm{shock}$\;\textless\;1000\;km\;s$^{-1}$). Line ratios from shock modelling at a single post-shock density (assuming a compression factor of 100) are grouped by blue ellipses, and points of the same shock velocity (v$_\mathrm{shock}=400,600,800$\;km\;s$^{-1}$) are joined by dashed blue lines.}
  \label{fig: tr_shock_ddd_v_vary}
 \end{figure}

In order to investigate the effect of shock-ionisation on the TR technique\footnote{The effect of shock-ionisation on transauroral line density and reddening diagnostic was investigated in a preliminary fashion by \citet{SpenceThesis}.}, in Figure \ref{fig: tr_shock_ddd_v_vary} we plot the expected TR([OII]) and TR([SII]) line ratios with the radiation-bounded diagnostic grid for photoionised gas that we previously presented in Section \ref{section: tr_diagnostics} (Figure \ref{fig: tr_ddd}). The shock models shown on this grid were taken from the library presented by \citet{Allen2008} (generated with the \textsc{MAPPINGS III} code), and are for solar-composition gas.

We first investigate the effect of varying the gas velocity between 0\;\textless\;$v_\mathrm{shock}$\;\textless\;1000\;km\;s$^{-1}$ with a constant magnetic parameter of $B/\sqrt{n}$=2\;$\mu$G\;cm$^{3/2}$ and pre-shock densities of \mbox{$n=1$, 10, 100, 1000\;cm$^{-3}$}. Assuming a compression factor of $\sim100$ if the gas cools in pressure equilibrium behind the shock \citep{Sutherland2017, Santoro2018}, these correspond to post-shock densities of \mbox{$n=10^2$, 10$^3$, 10$^4$, 10$^5$\;cm$^{-3}$}. From Figure \ref{fig: tr_shock_ddd_v_vary}, it can be seen that for high shock velocities ($\gtrapprox$500\;km\;s$^{-1}$) at a given density, the modelled line ratios are similar to those predicted by photoionisation modelling. However, for lower shock velocities (a few hundred km\;s$^{-1}$), the predicted densities may differ by $\pm0.22$ orders of magnitude. Similarly, shock-ionisation may effect the derived color excesses by E(B-V)$_\mathrm{TR}\pm0.13$.

\begin{figure}
  \includegraphics[width=\linewidth]{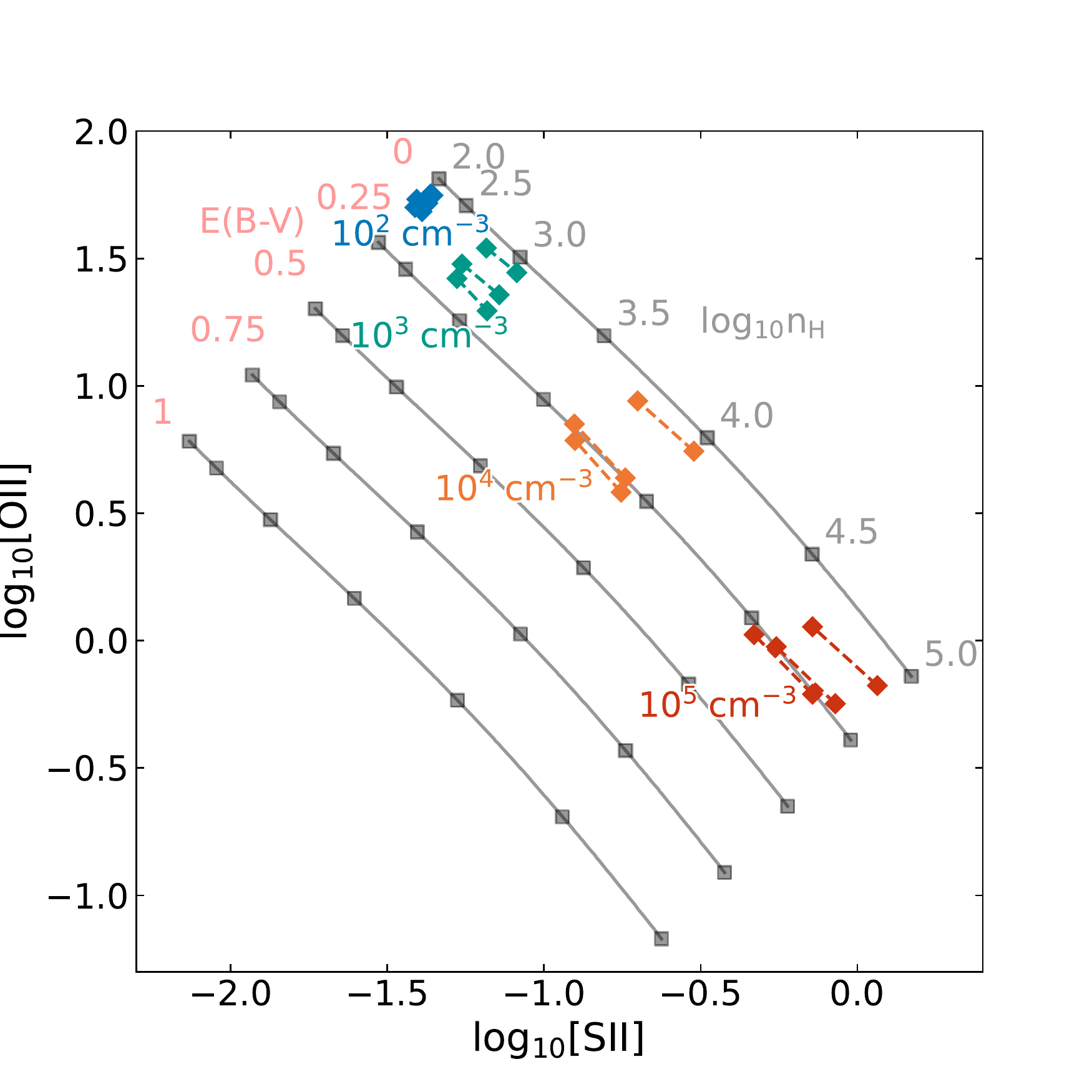}
  \caption{TR grid (as in Figure \ref{fig: tr_ddd}) with line ratios predicted by shock modelling for three values of shock velocity $v_\mathrm{shock}=$400, 600 and 800\;km\;s$^{-1}$ and two values of magnetic parameter (2\;\textless\;$B/\sqrt{n}$\;\textless\;4\;$\mu$G\;cm$^{3/2}$) which are typical of the ISM. Line ratios generated using the same post-shock density are show as crosses with the same color (labelled), and points of the same density and velocity are joined by a dashed line.}
  \label{fig: tr_shock_ddd_bn_vary}
\end{figure}

Secondly, we investigate the effect of varying the magnetic parameter between typical values for the ISM (2\;\textless\;$B/\sqrt{n}$\;\textless\;4\;$\mu$G\;cm$^{3/2}$: \citealt{Dopita1995, Allen2008}). We did this for three values of shock velocity \mbox{($v_\mathrm{shock}=400$, 600, 800\;km\;s$^{-1}$)}, which we show in Figure \ref{fig: tr_shock_ddd_bn_vary}. The impact on derived electron densities is greater at higher densities ($\pm0.25$ orders of magnitude) than at lower densities ($\pm0.10$ orders of magnitude), with little effect on the derived reddening value.

\begin{figure}
  \includegraphics[width=\linewidth]{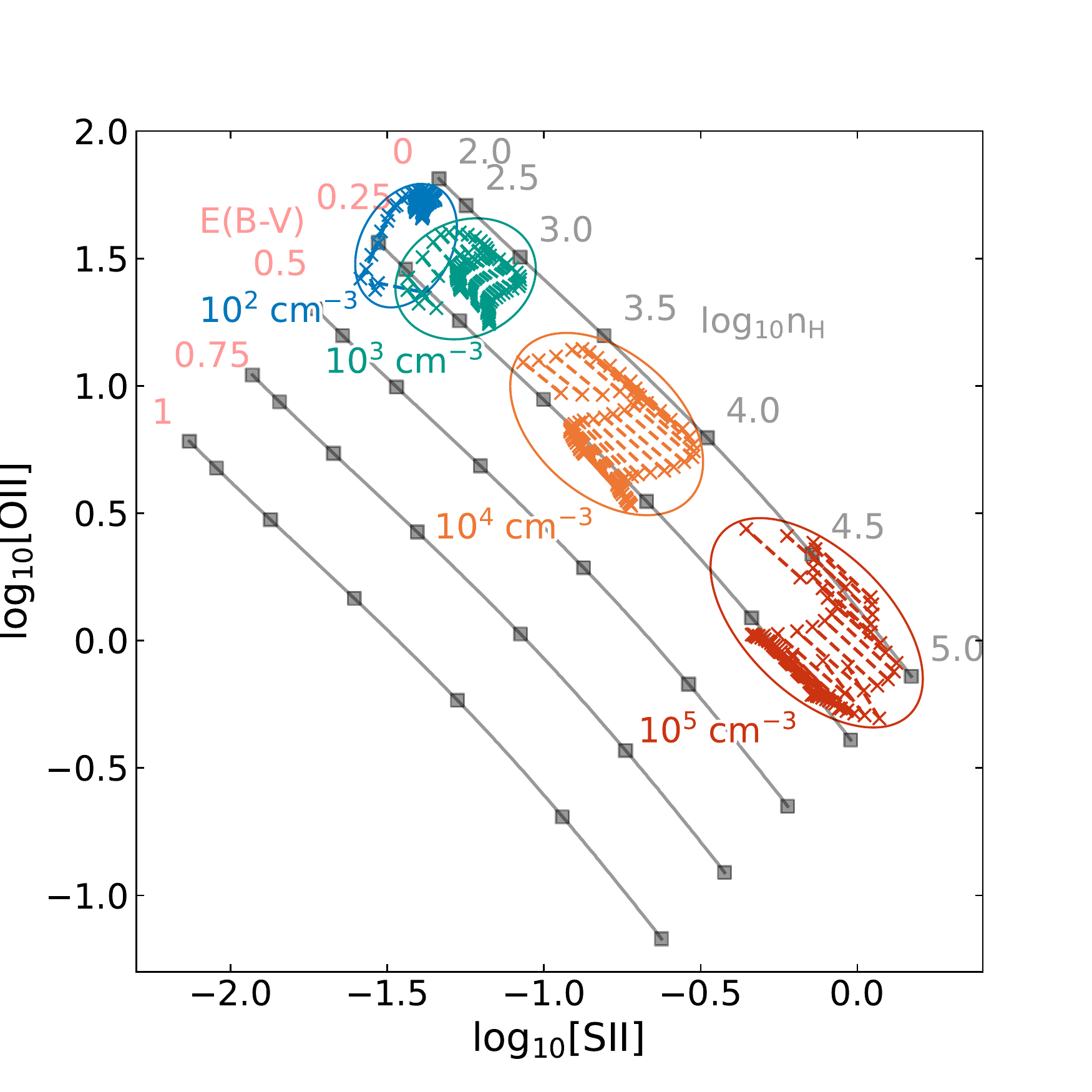}
  \caption{TR grid (as in Figure \ref{fig: tr_ddd}) with line ratios predicted by shock modelling for two magnetic parameters ($B/\sqrt{n}=$2,4\;$\mu$G\;cm$^{3/2}$) and a range of shock velocities (0\;\textless\;$v_\mathrm{shock}$\;\textless\;1000\;km\;s$^{-1}$) at each value of post-shock density. Predicted ratios with the same post-shock density are shown with crosses of a colour for each density (labelled) and are grouped with ellipses.}
  \label{fig: tr_shock_ddd_v_bn_vary}
\end{figure}

Finally, we quantify the effect of simultaneously varying the velocity \mbox{(between 0\;km\;s$^{-1}$\;\textless\;$v_\mathrm{shock}$\;\textless\;1000\;km\;s$^{-1}$)} and the magnetic parameter \mbox{(2\;\textless\;$B/\sqrt{n}$\;\textless\;4\;$\mu$G\;cm$^{3/2}$)} on the TR-derived electron densities and reddenings (Figure \ref{fig: tr_shock_ddd_v_bn_vary}). We find that the effect on the derived density is $\pm0.38$ orders of magnitude, regardless of the density of the modelled gas, and that the effect on derived reddening value is the same as varying the velocity (E(B-V)$\pm0.13$).

\section{The origin of the transauroral lines in NGC\;1068, NGC\;4151 and IC\;5063}
\label{appendix: tr-origin}

\begin{figure*}
    \includegraphics[width=\linewidth]{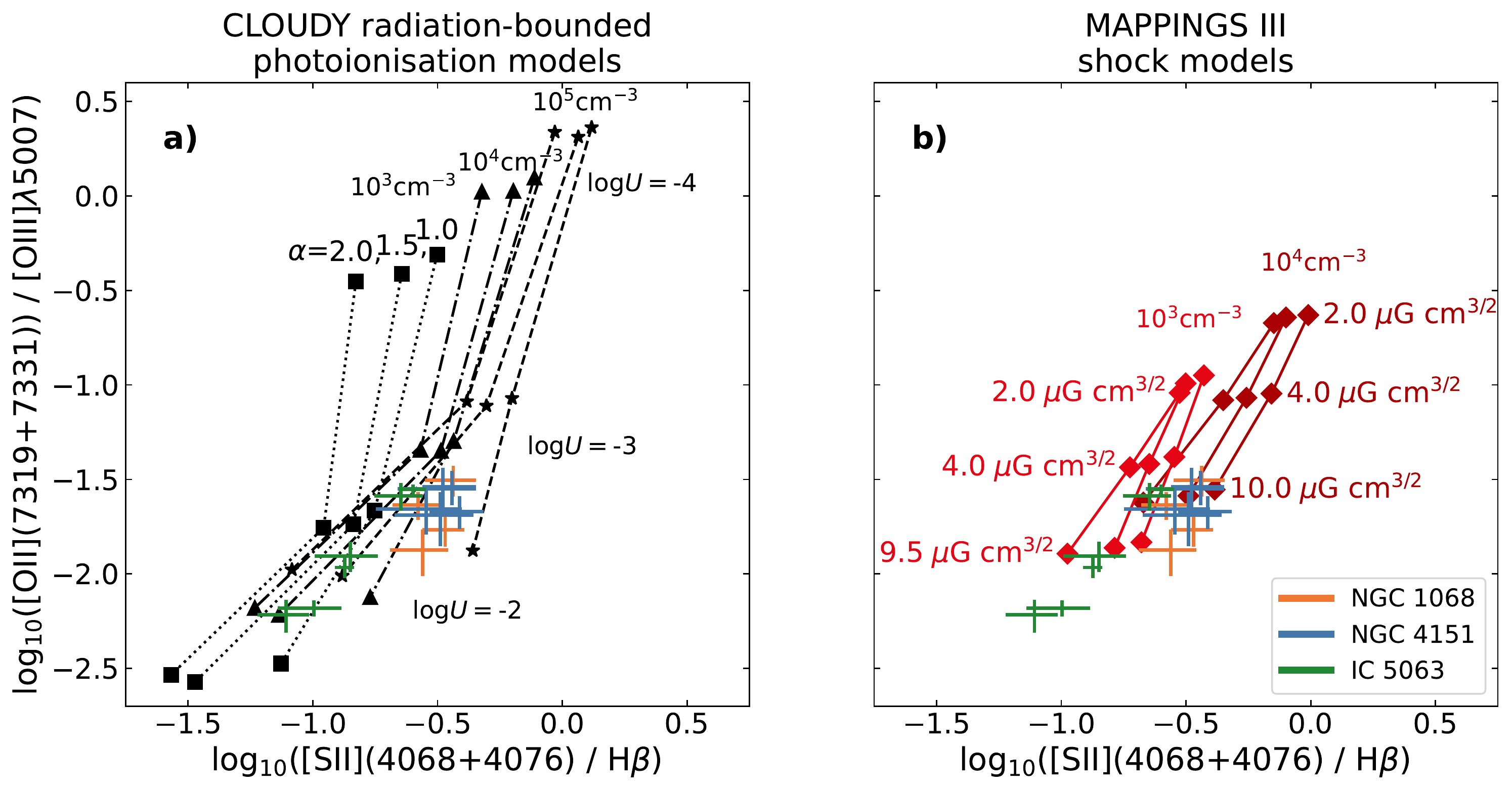}
    \caption{TR([SII])/H$\mathrm{\beta}$ vs TR([OII])/[OIII]$\lambda$5007 ratio grids from \textsc{CLOUDY} radiation-bounded photoionisation modelling (\textbf{a}) and the \citet{Allen2008} MAPPINGS III shock-ionisation models (\textbf{b}), along with measured values for each aperture in NGC\;1068 (orange crosses), NGC\;4151 (blue crosses) and IC\;5063 (green crosses; see \citealt{Holden2022}). The modelled photoionisation line ratio grid shown in \textbf{a)} is for three spectral indices ($\alpha=1.0, 1.5, 2.0$; labelled) at different electron densities (squares: $n_e$=10$^3$\;cm$^{-3}$; triangles: 10$^4$\;cm$^{-3}$; stars: 10$^5$\;cm$^{-3}$) and ionisation parameters (log\;$U$; labelled), while the shock-ionisation grids in \textbf{b)} are for post-shock densities of $n$=10$^3$\;cm$^{-3}$ and $n$=10$^{4}$\;cm$^{-3}$ (labelled; assuming a shock compression factor of 100), magnetic parameters of $B/\sqrt{n}=2$, 4, 10 \;$\mu$G\;cm$^{3/2}$. (labelled), and velocities of $v_\mathrm{shock}=400, 600, 800$\;km\;s$^{-1}$.}
    \label{fig: tr_oiii_hbeta}
\end{figure*}

In order to investigate the origin of the transauroral lines in the archetypal Seyfert galaxies (and the nearby Sey 2 IC\;5063: \citealt{Holden2022}), we plotted the measured [OII](7319+7331)/[OIII]$\lambda$5007 vs [SII](4068+4076)/H$\mathrm{\beta}$ ratios with overlaid grids from radiation-bounded photoionisation modelling (as first presented by \citealt{Spence2018}) and shock modelling, which we present in Figures \ref{fig: tr_oiii_hbeta}a and \ref{fig: tr_oiii_hbeta}b, respectively. 

The photoionisation models in Figure \ref{fig: tr_oiii_hbeta}a were generated using \textsc{CLOUDY} for a radiation-bounded cloud with varying values of density, ionisation parameter and spectral index. The measured ratios in our apertures for NGC\;1068 and NGC\;4151 are consistent with this grid; however, the corresponding ionisation parameters are a half-an-order-of-magnitude higher than that which was required to reproduce the measured transauroral line ratios (log\;$U=-3$; Figure \ref{fig: tr_ddd}). This is further evidence that radiation-bounded AGN photoionisation is not the dominant ionisation mechanism in our apertures, and in the case of NGC\;1068 can be explained as the [OIII] and H$\mathrm{\beta}$ emission being dominated by matter-bounded components (as indicated by the measured HeII/H$\mathrm{\beta}$ ratios; Figure \ref{fig: oiii_heii_hbeta_stis}). This is because matter-bounded emission will increase the relative strength of the [OIII] and H$\mathrm{\beta}$ lines, reducing the [OII](7319+7331)/[OIII]$\lambda$5007 and [SII](4068+4076)/H$\mathrm{\beta}$ ratios used in Figure \ref{fig: tr_oiii_hbeta}a and thus giving a higher corresponding ionisation parameter on the radiation-bounded grid.

In Figure \ref{fig: tr_oiii_hbeta}b, we present the same measured ratios, but with the expected ratios from shock-ionisation modelling. The shock models here are those presented by \citet{Allen2008}, and are for two values of pre-shock density (10$^1$\;cm$^{-3}$ and 10$^2$\;cm$^{-3}$), velocities of $v_\mathrm{shock}=400, 600, 800$\;km\;s$^{-1}$, and magnetic parameters in the range 2\;$\textless$\;$B/\sqrt{n}$\;$\textless$\;10\;$\mu$G\;cm$^{3/2}$. In all apertures for NGC\;4151, the measured line ratios are consistent with shock ionisation, albeit with relatively high magnetic parameters (4\;$\textless$\;$B/\sqrt{n}$\;$\textless$\;10\;$\mu$G\;cm$^{3/2}$). These magnetic parameters are higher than those typical for the ISM (2\;$\textless$\;$B/\sqrt{n}$\;$\textless$\;4\;$\mu$G\;cm$^{3/2}$; \citealt{Dopita1995, Allen2008}), potentially indicating higher magnetic fields associated with the shocked material. If the gas detected in our NGC\;4151 data is indeed shock-ionised, then the position of the shock model grid on the diagram would explain why the ionisation parameter deduced from the radiation-bounded photoionisation grid (Figure \ref{fig: tr_oiii_hbeta}a) for the NGC\;4151 apertures is higher than expected from the TR diagnostic grid (Figure \ref{fig: tr_ddd}): shock ionisation produces lower values of [OII](7319+7331)/[OIII]$\lambda$5007 and [SII](4068+4076)/H$\mathrm{\beta}$, corresponding to higher ionisation parameters on the photoionisation grid.

We also present [OII](7319+7331)/[OIII]$\lambda$5007 and [SII](4068+4076)/H$\mathrm{\beta}$ ratios for the Seyfert 2 galaxy IC\;5063, as measured from the dataset described in \citet{Holden2022}. In agreement with our previous findings \citep{Holden2022}, we find that the ratios are consistent with radiation-bounded AGN-photoionisation for an ionisation parameter of $-3$\;\textless\;log\;$U$\;\textless\;$-2$ and densities in the range 3\;\textless\;log$_{10}(n_e$[cm$^{-3}$])\;\textless\;4.

\bsp	
\label{lastpage}
\end{document}